\documentclass[fleqn,usenatbib]{mnras}

\usepackage{newtxtext,newtxmath}
\usepackage{xspace}

\usepackage[T1]{fontenc}
\DeclareRobustCommand{\VAN}[3]{#2}
\let\VANthebibliography\thebibliography
\def\thebibliography{\DeclareRobustCommand{\VAN}[3]{##3}\VANthebibliography}

\usepackage{graphicx}	
\usepackage{amsmath}	
\usepackage{subcaption}

\newcommand{\planck}{{\sl Planck}\xspace}

\title[SHAMe 2x2pt]{Cosmological constraints from galaxy clustering and galaxy-galaxy lensing with extended SubHalo Abundance Matching}

\author[C. Mahony et al.]{Constance Mahony $^{1,2}$\thanks{E-mail: constance.mahony@physics.ox.ac.uk},
Sergio Contreras$^{1,3}$,
Raul E. Angulo$^{1,4}$,
David Alonso$^{2}$,
Christos Georgiou$^{5,6}$,
\newauthor
Andrej Dvornik$^{7}$
\\
$^{1}$Donostia International Physics Center, Manuel Lardizabal Ibilbidea, 4, 20018 Donostia, Gipuzkoa, Spain\\
$^{2}$Department of Physics, University of Oxford, Denys Wilkinson Building, Keble Road, Oxford OX1 3RH, United Kingdom\\
$^{3}$Facultad de F\'isica. Universidad de Sevilla. Multidisciplinary Unit for Energy Science, Av. Reina Mercedes s/n 41012 Seville, Spain\\
$^{4}$IKERBASQUE, Basque Foundation for Science, 48013, Bilbao, Spain\\
$^{5}$Institute for Theoretical Physics, Utrecht University, Princetonplein 5, 3584 CC, Utrecht, The Netherlands\\
$^{6}$Institut de Física d’Altes Energies (IFAE), The Barcelona Institute of Science and Technology, Campus UAB, 08193 Bellaterra (Barcelona), Spain\\
$^{7}$Ruhr University Bochum, Faculty of Physics and Astronomy, Astronomical Institute (AIRUB), German Centre for Cosmological Lensing, \\ 44780 Bochum, Germany}

\date{Accepted XXX. Received YYY; in original form ZZZ}

\pubyear{\the\year{}}

\begin{document}
\label{firstpage}
\pagerange{\pageref{firstpage}--\pageref{lastpage}}
\maketitle

\begin{abstract}
We present the first cosmological constraints from a joint analysis of galaxy clustering and galaxy-galaxy lensing using extended SubHalo Abundance Matching (SHAMe). We analyse stellar mass-selected Galaxy And Mass Assembly (GAMA) galaxy clustering and Kilo-Degree Survey (KiDS-1000) galaxy-galaxy lensing and find constraints on $S_8\equiv\sigma_8\sqrt{\Omega_{\rm m}/0.3}=0.793^{+0.025}_{-0.024}$, in agreement with \planck at 1.7$\sigma$, with $\sigma_8$ the mass density fluctuation amplitude in 8 $h^{-1}{\rm Mpc}$ sphere at present and $\Omega_{\rm m}$ the density parameter in total matter. These results are in agreement with the Cosmic Microwave Background results from \planck. We are able to constrain all 5 SHAMe parameters, which describe the galaxy-subhalo connection. We validate our methodology by first applying it to simulated catalogues, generated from the TNG300 simulation, which mimic the stellar mass selection of our real data. We show that we are able to recover the input cosmology for both our fiducial and all-scale analyses. Our all-scale analysis extends to scales of galaxy-galaxy lensing below $r_\mathrm{p}<1.4\,\mathrm{Mpc}/h$, which we exclude in our fiducial analysis to avoid baryonic effects. When including all scales, we find a value of $S_8$, which is 1.26$\sigma$ higher than our fiducial result (against naive expectations where baryonic feedback should lead to small-scale power suppression), and in agreement with \planck at 0.9$\sigma$. We also find a 21\% tighter constraint on $S_8$ and a 29\% tighter constraint on $\Omega_\mathrm{m}$ compared to our fiducial analysis. This work shows the power and potential of joint small-scale galaxy clustering and galaxy-galaxy lensing analyses using SHAMe.
\end{abstract}




\section{Introduction}
Imaging galaxy surveys have been highly successful at constraining the current cosmological model \citep{Heymans:2020gsg,DES3x2pt,HSC3x2pt2023}. In particular, they have placed constraints on the matter density of the Universe $\Omega_\mathrm{m}$ and the amplitude of matter fluctuations\footnote{As is usual, we will quantify the amplitude of matter fluctuations in terms of $\sigma_8$: the standard deviation of the linear matter overdensity field on spheres of 8 $h^{-1}{\rm Mpc}$ radius, where $h$ is the Hubble constant in units of $100\,{\rm km}\,{\rm s}^{-1}\,{\rm Mpc}^{-1}$. We will also use the ``clumping'' parameter $S_8$, defined in terms of $\sigma_8$ and $\Omega_{\rm m}$ as $S_8\equiv\sigma_8\,(\Omega_m/0.3)^{0.5}$.} $S_8$ that are now competitive with Cosmic Microwave Background (CMB) experiments \citep{Planck}. The most powerful results generally come from combining measurements of the clustering of galaxies, how dark matter mildly distorts galaxy images via weak gravitational lensing, and how the structure traced by foreground galaxies distorts background galaxy images (galaxy-galaxy lensing). Recently, there has been much discussion regarding a potential tension between the value of $S_8$ recovered by galaxy surveys and that preferred by CMB data within the $\Lambda$CDM model, with one of the main proposed solutions being potential uncertainties in the modelling of these three observables on small scales \citep{Yoon2019,YoonJee2021,Amon:2022azi,Preston2023, AricoDES2023A&A...678A.109A,Garcia-Garcia2024JCAP...08..024G}. While recent weak lensing results from the Kilo-Degree Survey (KiDS) show no tension with the CMB \citep{Wright:2025xka, Stolzner:2025htz}, ensuring accurate modelling of all observables on small scales remains a priority.

A common conservative approach to mitigating small-scale uncertainties is to apply scale cuts to the data, removing data points where these biases are most likely to be relevant (e.g., \citealt{2021arXiv210513548K}). However, there are two problems with this. One is that this approach involves discarding a huge amount of data, which has a wealth of information about galaxies as well as cosmology. The second is that it is difficult to cleanly remove uncertain non-linear and astrophysical processes that are not fully localised in either Fourier or configuration space from common observables. Both of these issues will be more acute for the next generation of surveys, such as \textit{Euclid} \citep{Euclid2025} and the Vera C. Rubin Observatory Legacy Survey of Space and Time (LSST, \citealt{rubin2019}), which are already underway.

The difficulty in modelling small scales comes from linking the properties and spatial distribution of galaxies to the underlying matter inhomogeneities. On large scales, this relationship is linear and simple to model. However, on small scales, the relationship is non-linear, non-local, and non-deterministic. One approach to is to expand the bias perturbatively, which has been shown to be very successful at extending analyses to intermediate scales (e.g. \citealt{Beutler2017MNRAS.466.2242B,Amico2020JCAP...05..005D,Ivanov2020JCAP...05..042I, Nicola2024JCAP...02..015N}). To extend analyses to even smaller scales, whilst achieving the necessary accuracy, requires detailed modelling of the astrophysics of galaxies. Usually, this modelling is divided into two parts: how matter clusters into dark matter halos and how galaxies populate these halos. The distribution of dark matter halos is either calculated analytically through the halo model \citep{Smith:2002dz, Takahashi:2012em, HMCode}, taken directly from dark matter simulations \citep{Nishimichi:2018etk,Miyatake:2020uhg} or a combination of the two \citep{Mead:2020qdk,Mahony:2022emy}. Then these dark matter halos are populated with galaxies in a second step.

There are two main approaches to populating dark matter halos with galaxies. The first more popular approach is Halo Occupation Distribution (HOD) modelling, which provides a probabilistic description for this process. This approach is successful in predicting galaxy distributions (e.g., \citealt{contreras2024}), but it requires a large number of parameters and typically does not account for variable levels of assembly bias. Assembly bias refers to how the evolutionary history of halos impacts the clustering of their associated galaxies, not only their mass \citep{Gao2005,Wechsler2006,Dalal2008}. Some extensions to HOD modelling do include assembly bias \citep{Hearin2016MNRAS.460.2552H}, but these approaches further increase the parameter space. The second approach is SubHalo Abundance Matching (SHAM), which matches a dark matter subhalo property to a corresponding galaxy property. Usually, SHAM relates a property summarising the mass of a dark matter subhalo to the stellar mass or luminosity of a galaxy. Here, the assumption is that more massive subhalos tend to host more luminous galaxies. SHAM can reproduce both real observations of galaxies \citep{contreras_lensing_data2023} and results from hydrodynamic simulations \citep{2016MNRAS.460.3100C, contreras_lensing}. The main advantages of SHAM over HOD approaches is that SHAM utilises fewer parameters and includes a variable level of assembly bias. The main disadvantage is that SHAM requires subhalo information, which is computationally expensive and is a key factor in why SHAM has not previously been used for cosmological analyses of galaxy clustering and galaxy-galaxy lensing. However, it is possible to overcome these challenges by emulating the observables, and access to subhalo information aids in modelling the distribution of satellite galaxies, which is particularly useful for galaxy-galaxy lensing \citep{2023Chavesmontero, contreras_lensing}. 

Previous works have used HOD modelling to place tight constraints on $S_8$ and $\Omega_\mathrm{m}$ from combined analyses of small-scale galaxy clustering and galaxy-galaxy lensing \citep{Dvornik:2022xap,HSC3x2pt2023}. In this work, we obtain the first cosmological constraints from a combined analysis of galaxy clustering and galaxy-galaxy lensing using the SHAM method. In Section \ref{sec:method} we present our extended SubHalo Abundance Matching methodology, and detail the creation of our galaxy clustering and galaxy-galaxy lensing emulator. In Section \ref{sec:data} we present the galaxy clustering and galaxy-galaxy lensing data used in the analysis. In Section \ref{sec:validation}, we validate our methodology on a simulated galaxy sample before presenting our results for real data in Section \ref{sec:results}. We conclude in Section \ref{sec:conclusion}.
\section{Methodology} \label{sec:method}
\subsection{Extended abundance matching} \label{subsection: extended abundance matching}

In most implementations, the SHAM model has one free parameter, the scatter between a given subhalo property (e.g. mass or circular velocity) and the related galaxy property used to rank them (e.g. stellar mass or luminosity). However, these implementations provide limited precision and \citealt{2021MNRAS.508..175C} extended the basic SHAM model to include four additional parameters. They showed that by including these additional parameters, the extended SHAM (SHAMe) model was able to accurately reproduce galaxy clustering in both real and redshift space in a state-of-the-art hydrodynamical simulation. They then additionally showed that the SHAMe model was able to simultaneously describe galaxy-galaxy lensing and galaxy clustering in both hydrodynamic simulations \citep{contreras_lensing, contreras2024} and in real galaxy observations \citep{contreras_lensing_data2023}.

The 5 parameters included in the SHAMe model are,
\begin{equation}
  [\sigma, \ t_{\mathrm{merger}}, \ f_{\mathrm{k, c+s}}, \ f_{\mathrm{k, c-s}}, \ \beta].   
\end{equation}
In this case, $\sigma$ is the scatter between $V_{\mathrm{peak}}$, the peak maximum circular velocity of a dark matter subhalo during its evolution, and $M_{\star}$, the stellar mass or luminosity of the associated galaxy. $t_{\mathrm{merger}}$ regulates the number of orphan galaxies in the simulation by removing galaxies that have been orphans for a large number of dynamical times. Orphan galaxies are galaxies that are no longer associated with a dark matter subhalo, as the mass of the dark matter subhalo has fallen below the resolution of the simulation. Including orphan galaxies is important to reproduce the small-scale galaxy clustering pattern \citep{Guo:2013voa}. 

The parameters $f_{\mathrm{k, c+s}}$ and $f_{\mathrm{k, c-s}}$ regulate the amount of galaxy assembly bias in the sample. Galaxy assembly bias \citep{Croton:2007} is the impact of the evolutionary history of host haloes on the clustering of their associated galaxies. Specifically, $f_{\mathrm{k, c+s}}$ and $f_{\mathrm{k, c-s}}$ represent the sum and the difference between the amount of galaxy assembly bias added to central (c) and satellite (s) galaxies, and they re-order the $V_{\mathrm{peak}}$-$M_{\star}$ relation for the galaxies while preserving the original scatter $\sigma$. If these parameters are positive/negative, the model will assign a larger/smaller stellar mass or luminosity to galaxies with a larger linear galaxy bias, resulting in a positive/negative correlation. If they are equal to zero the stellar masses or luminosities are randomly re-assigned. The bias-per-object is calculated using the method from \cite{Paranjape2018} and is computed independently for centrals and satellites, so the satellite fraction is preserved. A deeper explanation of this implementation can be found in \cite{Contreras2021_AssemblyBias}.   

The final parameter is $\beta$, which sets the timescale for how long satellite galaxies survive within their host halo. Satellite galaxies progressively lose mass and luminosity over time due to dynamical friction, so eventually no longer have the minimum mass or luminosity required to enter our sample, or even completely disappear into the intra-cluster medium. Satellite galaxies should therefore be removed if they have been satellites for a long time. For further details about the SHAMe parameters, see \cite{2021MNRAS.508..175C}.

\subsection{Galaxy clustering and galaxy-galaxy lensing} \label{sec:method-clustering_lensing}

The SHAMe formalism can be applied to the subhalo catalogues of gravity-only simulations to populate them with galaxies. Summary statistics of the galaxy clustering and galaxy-galaxy lensing can then be computed from the distribution of these galaxies. 

 The galaxy clustering statistic we use in this work is the projected correlation function $ w_{\mathrm{p}}(r_{\mathrm{p}})$, which is found by integrating the three-dimensional galaxy correlation function $\xi_{\mathrm{gg}}(r_{\mathrm{p}},r_{\uppi})$ over the line of sight separation $r_{\uppi}$,
\begin{equation}
  w_{\mathrm{p}}(r_{\mathrm{p}}) = 2\int^{r_{\uppi, \mathrm{max}}}_0 \xi_{\mathrm{gg}}(r_{\mathrm{p}},r_{\uppi}) \, \mathrm{d}r_{\uppi} \ ,
  \label{eq:wp_theory}
\end{equation}
where $r_{\rm{p}}$ is the projected separation between two galaxies and $r_{\pi,\mathrm{max}}$ is the maximum integration range. We utilise \textsc{corrfunc}\footnote{\url{https://github.com/manodeep/Corrfunc}}\citep{Sinha:2019reo} to compute $ w_{\mathrm{p}}(r_{\mathrm{p}})$, and use $r_{\pi,\mathrm{max}}=60 \ \mathrm{Mpc}/h$ to match the measurements from GAMA used in this work (see section \ref{sec:data}). We include the effect of redshift-space distortions in the clustering, but they only have a small impact due to our value of $r_{\pi,\mathrm{max}}$.

The galaxy-galaxy lensing statistic we use in this work is the excess surface density (ESD) profile $\Delta\Sigma(r_{\mathrm{p}})$, where we subtract the projected surface mass density $\Sigma(r_{\mathrm{p}})$ from the mean surface mass density within a given projected radius $\overline{\Sigma}(\leq r_{\mathrm{p}})$,
\begin{equation}
    \Delta\Sigma(r_{\mathrm{p}}) = \overline{\Sigma}(\leq r_{\mathrm{p}}) - \Sigma(r_{\mathrm{p}}) \ .
\end{equation}
The projected surface mass density $\Sigma(r_{\mathrm{p}})$ is related to the galaxy-matter correlation function $\xi_\mathrm{gm} (r_{\mathrm{p}},r_{\uppi})$ by,
\begin{equation}
    \Sigma(r_{\mathrm{p}}) = 2\Omega_\mathrm{m} \rho_\mathrm{crit} 
    \int^{r_{\uppi, \mathrm{max}}}_0 \xi_\mathrm{gm} (r_{\mathrm{p}},r_{\uppi})\mathrm{d}r_\uppi \ ,
\end{equation}
where $\rho_\mathrm{crit}$ is the critical density of the Universe. The mean surface mass density within a given projected radius $\overline{\Sigma}(\leq r_{\mathrm{p}})$ is given by, 
\begin{equation}
  \overline{\Sigma}(\leq r_{\mathrm{p}}) = \frac{2}{r_{\rm{p}}^2}\int^{r_{\rm{p}}}_0 \Sigma(R')R'{\rm{d}} R' \ .
\end{equation}
We again utilise \textsc{corrfunc} to compute $\Delta\Sigma(r_{\mathrm{p}})$, and integrate out to $r_{\pi,\mathrm{max}}=233 \ \mathrm{Mpc}/h$ to match the measurements from KiDS-1000 (see section \ref{sec:data}). To compute $\xi_\mathrm{gm}$, we extract a subsampled version of the dark matter density field from the gravity-only simulation, which we then correlate with the distribution of galaxies produced by the SHAMe formalism. A dilution factor of $\sim1/3000$ was previously found to produce $\Delta\Sigma(r_{\mathrm{p}})$ measurements with sub-percent precision \citep{contreras_lensing}. 

\subsection{SHAMe emulator} \label{sec:shame_emulator}

We created an emulator to speed up the computation of galaxy clustering and galaxy-galaxy lensing using the procedure described in \citep{Contreras2023_MTNG}. Here, we briefly summarise the main steps in building the emulator.
\begin{itemize}
    \item We begin by running five gravity-only simulations with varying cosmology (as described in \citealt{Contreras2020_Scaling}). These simulations have $3072^3$ particles over a $(1024\ h^{-1}{\rm Mpc})^3$ volume, which is roughly a mass resolution of $3-4\times10^{9}h^{-1}M_{\odot}$\footnote{The mass resolution can vary slightly between the simulations, due to their differing cosmologies.}. The simulations were run using the fixed initial condition technique by \cite{Angulo2016_IC}, which significantly reduces the effect of cosmic variance on a simulation, at least for second-order statistics.
    \item We scaled these five simulations to 10,000 distinct cosmologies by employing the scaling technique of \cite{Angulo2010_scale}. The upper part of Table~\ref{tab:priors} shows the cosmological parameters we scaled and the range within which we varied them. For each cosmology, we generate an average of eight snapshots between $z=0$ and $z=2$. As demonstrated in \cite{Contreras2020_Scaling} and \cite{Contreras2023_MTNG}, the scaled simulations precisely reproduce the mass distribution of the desired cosmology, needing only 6 minutes to scale.
    \item We populate each of the $\sim 80,000$ scaled snapshots with two SHAMe mocks. The SHAMe parameters used, as well as the range within which they were varied, are shown in the lower part of Table~\ref{tab:priors}. These ranges were chosen based on the model's performance in fitting galaxy clustering and galaxy-galaxy lensing simulations \citep{contreras_lensing, Contreras2023_MTNG, contreras2024} and observations \citep{contreras_lensing_data2023}. For this work we have assumed the sum of the neutrino masses to be equal to zero.
    \item For each mock, we compute the galaxy correlation function $\xi_{\mathrm{gg}}(r_{\mathrm{p}},r_{\uppi})$ up to a scale of $r_{\rm \pi, max}=100\ h^{-1}{\rm Mpc}$, and the galaxy-matter correlation function $\xi_\mathrm{gm} (r_{\mathrm{p}},r_{\uppi})$ up to a scale of $r_{\rm \pi, max}=240\ h^{-1} {\rm Mpc}$ for 8 galaxy samples with number densities ranging from ${\rm n_{den}} = 10^{-4} h^{3}{\rm Mpc^{-3}}{\rm\ to\ }10^{-2.25} h^{3}{\rm Mpc^{-3}}$. We then compute $w_{\rm p}(r_{\rm p})$ and $\rm \Delta\Sigma(r_{\rm p})$ using the steps described in the previous section, up to a maximum scale of 60 and $233\,h^{-1} {\rm Mpc}$, respectively. We computed correlation functions with larger values of $\rm r_{\pi, max}$ so that we could use these calculations for future observational samples.
    \item Using over 1,000,000 projected correlation functions and galaxy-galaxy lensing calculations, we created an emulator that can predict these statistics as a function of the cosmological and SHAMe parameters, as well as the sample redshift and galaxy number density. The emulator was built using a feed-forward neural network, similar to the one described in \cite{Angulo2021_bacco}. The architecture used consists of two fully connected hidden layers, each with 200 neurons, and a rectified linear unit (ReLU) activation function, with each statistic represented by its own network. The neural networks were trained with the Keras front-end of the TensorFlow library \citep{tensorflow}. We applied the Adam optimisation algorithm with a learning rate of 0.001 and a mean squared error loss function. Evaluating one of these emulators takes $\sim 40$ milliseconds on a laptop and $\sim 1.6$ seconds to evaluate 100,000 samples (it is more efficient to evaluate the data in larger groups). 
\end{itemize}
\begin{table}
  \centering
  \caption{Parameter ranges of the SHAMe emulator, see section \ref{subsection: extended abundance matching} for details of the SHAMe parameters. These ranges are also the fiducial top-hat prior ranges used in our analysis of GAMA galaxy clustering and KiDS-1000 galaxy-galaxy lensing with SHAMe.}
  \begin{tabular}{lll}
  \textbf{Parameter} & \textbf{Prior}  \\
  \hline
  \hline
        $\sigma_8$ & [0.65, 0.9] \\
        $\Omega_{\mathrm{m}}$     & [0.23, 0.4] \\
        $\Omega_{\mathrm{b}}$     & [0.04, 0.06] \\
        $n_{\mathrm{s}}$     & [0.92, 1.01] \\
        $h$     & [0.64, 0.8] \\
  \hline
        $\sigma \log \mathrm{M}$     & [0.1, 1.8] \\
        $\log t_{\mathrm{merger}}$     & [-1.5, 1.2] \\
        $f_{\mathrm{k, c+s}}$     & [-1.5, 1.5] \\
        $f_{\mathrm{k, c-s}}$     & [-1.5, 1.5] \\
        $\beta$     & [0.0, 1.0] \\
  \end{tabular}
  \label{tab:priors}
  \end{table} 

The measurements of $w_{\mathrm{p}}$ and $\Delta\Sigma$ made from the real data depend implicitly on the fiducial cosmology used to transform redshifts into distances. This procedure is mimicked in the galaxy clustering and galaxy-galaxy lensing measurements made from the simulations used to construct the emulator, assuming that the observer uses a fiducial cosmological model (different from that of the simulation). We adopt the procedure described in \cite{Lange:2019}, where these effects are incorporated by modifying the box geometry and appropriately adjusting the coordinate system. In doing this, we assumed a fixed {\sl Planck}-like cosmology, with $\Omega_m=0.314$. This is slightly different from the cosmology assumed in the real data ($\Omega_m=0.3$) \citep{Dvornik:2022xap}, but we verified that the difference is negligible given the sensitivity of our measurements (the effect is less than 7\% of the statistical uncertainties).

\section{Data} \label{sec:data}
In this work we use galaxy clustering measurements from the final data release of the Galaxy And Mass Assembly survey (GAMA, \citealt{Driver:2022vyh}), and galaxy-galaxy lensing measurements from the  fourth data release of the Kilo-Degree Survey (KiDS-1000, \citealt{KiDS-DR4}). These two datasets are described below. The lens sample used for galaxy-galaxy lensing is based on the KiDS Bright sample \citep{Bilicki:2021hgn}, and was designed to match the selection of the GAMA sample. We will thus assume that these lenses and the GAMA sample used for galaxy clustering have the same properties and are described by the same SHAMe parameters. This allows us to avoid the significant impact of photometric redshift uncertainties in the galaxy clustering measurements.

\subsection{Kilo Degree Survey} \label{sec:KiDS}

The Kilo Degree Survey (KiDS) is an imaging survey carried out at the European Southern Observatory VLT Survey Telescope in the $ugri$ broad-band filters. The data cover a sky area of $1347$ deg$^2$ \citep{KiDS-DR4, Wright:2024qvd} and are complemented by fully overlapping infrared photometry in the $ZYJHK_s$ bands from the VISTA Kilo-degree INfrared Galaxy (VIKING) survey \citep{VIKING}. KiDS was specifically designed for weak gravitational lensing science and the imaging data reflect this in their excellent seeing and well-behaved point-spread function. The 9-band photometry also assists greatly with the determination of photometric redshift estimates (photo-$z$) for the galaxy sample. The $r$-band data, used for the estimation of galaxy shapes, were obtained within strict observing condition limits and reach a median $5\sigma$ limiting magnitude of 24.8 and median seeing of 0.7''.

We use measurements of the galaxy-galaxy lensing signal, based on KiDS-1000 data, as presented in \cite{Dvornik:2022xap} and detailed here as well. The sample of source galaxies used as gravitational shear tracers is detailed in \cite{Giblin}, with their photometric redshift estimates described in \cite{Hildebrandt}. The source sample number density is $6.17$ galaxies$/$arcmin$^2$ and spans a photo-$z$ range of $0.1<z_B<1.2$, where $z_B$ is the estimated photometric redshift. The lens galaxies used as density tracers are selected from the KiDS Bright sample \citep{Bilicki:2021hgn}. This sample is designed to match the selection of the GAMA galaxy sample (see Section \ref{sec:GAMA}) for which spectroscopic redshift information is available. The GAMA sample is then used as training data for estimating machine-learning-based photometric redshifts over the KiDS-1000 area. The resulting lens sample benefits from accurate photo-$z$ estimates, with a mean offset of $\delta z=5\times10^{-4}$ and a standard mean absolute deviation of $\sigma_z=0.018(1+z)$. Stellar mass estimates for this sample are also available using the spectral energy density template-fitting code \textsc{LePhare} \citep{LePhare1,LePhare2}. See \cite{Bilicki:2021hgn} for further details.

The lens galaxy sample is split into six volume-limited stellar mass bins, the details of this procedure are described in \cite{Dvornik:2022xap}. We only include the highest four stellar mass bins in our analysis, as the lowest two bins are currently outside the maximum number density of our SHAMe emulator (see Section \ref{sec:CSMF}). We plan to extend the number density range of the SHAMe emulator in the next version. The stellar mass bin boundaries used in this analysis are [9.95, 10.25, 10.5, 10.7, 11.3] in units of $\mathrm{{log}}(M_{{\star}}/h^{{-2}}M_{{\odot}})$ and the measured $\Delta\Sigma$ are shown in Figure \ref{fig:GAMA_KiDS_data}. 

\subsection{Galaxy And Mass Assembly Survey} \label{sec:GAMA}

To complement the galaxy-galaxy lensing measurements and break degeneracies between cosmological and SHAMe parameters, we make use of projected galaxy clustering measurements. However, the presence of un-modelled photometric redshift uncertainties makes attempting this measurement with the KiDS Bright sample difficult. This was previously identified in \cite{Dvornik:2022xap} and led to the introduction of an additional photometric dilution parameter. We also found it challenging to obtain a good fit to the KiDS Bright photometric clustering with the SHAMe model used in this study. We therefore decided to perform galaxy clustering measurements using the GAMA galaxy sample, which we treat as being described by the same SHAMe parameters as the KiDS Bright sample. Note that the galaxy-galaxy lensing measurements are significantly less affected by photometric errors \citep{Dvornik:2022xap}.

The Galaxy And Mass Assembly survey \citep{GAMA} is a spectroscopic survey carried out using the AAOmega spectrograph at the Anglo-Australian Telescope (AAT) covering approximately $250$ deg$^2$ of the sky in five patches. Of these, we use the three equatorial patches (G09, G12, and G15) that have a homogeneous target selection with a Petrosian $r$-band magnitude limit of $r^\mathrm{Petro} < 19.8$ mag and fully overlap with the KiDS survey. The survey's strategy prioritised completeness, involving multiple passes over the same sky area. We use data from the final data release \citep[GAMA III][]{GAMADR4}, which achieves a completeness of 95\% with a limiting magnitude of $r < 19.72$. We make use of stellar mass measurements from the \texttt{StellarMassesGKVv24} \citep{Taylor2011} Data Management Unit (DMU) and local-flow corrected redshifts (\texttt{Z\_TONRY} column) from the \texttt{DistancesFramesv14} DMU \citep{Baldry2012}.

The KiDS Bright sample, from which our lens galaxies are chosen, has been designed to match the GAMA galaxies as well as possible. An exact match is difficult due to GAMA having been selected on Petrosian $r$-band magnitude \citep[see][for details]{Bilicki:2021hgn}. The differences between these two samples were recently studied in \cite{Georgiou} and are present mostly at the tail end of the redshift distribution. Given that our lens sample is selected to be volume-limited and does not extend to redshifts beyond 0.4, this mismatch is not expected to be significant. Therefore, we treat these two samples as containing the same type of galaxies.

We apply the stellar mass and redshift cuts from the KiDS Bright sample to the GAMA sample and estimate the projected correlation function using equation \ref{eq:wp_theory}, where $\xi_{\mathrm{gg}}$ is now the estimated galaxy correlation function. We use 10 logarithmically spaced $r_{\mathrm{p}}$ bins between $0.15-25\,\mathrm{Mpc}/h$, and take $r_{\pi, \mathrm{max}}=60\,\mathrm{Mpc}/h$ binned in 30 linearly spaced bins. We chose the range of $r_p$ to mimic the range shown in \cite{Dvornik:2022xap} who previously study the clustering and lensing of KiDS-1000. Correlation functions are estimated using \textsc{treecorr}\footnote{\url{http://rmjarvis.github.io/TreeCorr}}\citep{treecorr} and the measured $w_{\mathrm{p}}$ are shown in Figure \ref{fig:GAMA_KiDS_data}. We use the GAMA \texttt{Randoms v02} DMU \citep{GAMARandoms} -- specifically designed for the GAMA survey\footnote{The random catalogue we used was built for GAMA II, but it should also be representative of the GAMA III sample since the two data releases have the same target selection in the equatorial regions.} -- as the random points catalogue. Given the bright magnitude cut and high completeness of GAMA, this random catalogue does not need to account for spatially varying observational systematics, but models carefully the radial selection function of the data.

\subsection{Covariance} \label{sec:Covariance}

We use the analytic covariance detailed in \cite{Dvornik:2022xap} (and previously validated by \citealt{2019A&A...624A..30J}) for the KiDS-1000 galaxy-galaxy lensing measurements, and compute a jackknife covariance using 12 patches for the GAMA galaxy clustering data. The jackknife patches, which are approximately 15 deg$^2$ in area, were defined using \textsc{kmeans-radec}\footnote{\url{https://github.com/esheldon/kmeans-radec}}. We do not include cross-correlations between the different galaxy clustering stellar mass bins, as they do not bring in significant additional information, and obtaining a sufficiently accurate covariance would have required a much larger set of jackknife realisations. Our covariance matrix also assumes no correlation between the KiDS-1000 $\Delta\Sigma$ measurements and the GAMA $w_\mathrm{p}$ measurements. \cite{Dvornik:2022xap} found the correlation between these observables within KiDS itself to be small, and we expect them to become negligible when switching to GAMA clustering, given GAMA only covers approximately 20\% of the KiDS sky area \citep{Wright:2024qvd}. 

\subsection{Cumulative number densities} \label{sec:CSMF}
In \cite{Dvornik:2022xap} the stellar mass function is an output of the galaxy-halo connection modeling, which is the analytic Conditional Stellar Mass Function formalism \citep{Yang:2007pg,Cacciato:2008hm,2013BoschCacciato,Cacciato2013MNRAS.430..767C,More2013,Wang2013}. This means that the stellar mass function can be jointly analysed, along with galaxy clustering and galaxy-galaxy lensing. In SHAMe, the stellar mass function acts as an input to the modelling. Normally, the specific form of the stellar mass function has very little effect \citep{Contreras:2020zgw}. However, when the galaxy sample is binned in stellar mass, it becomes more important because the SHAMe emulator requires a cumulative number density input.

To model the galaxy clustering in a particular stellar mass bin we first convert the stellar mass bin edges (${M_\star}_1$, ${M_\star}_2$ where ${M_\star}_1<{M_\star}_2$) to cumulative number densities ($n_1$, $n_2$ where $n_1>n_2$) using the cumulative stellar mass function. We are then able to use the SHAMe emulator to compute the galaxy clustering $w_{\mathrm{p}}$ for two galaxy samples where the minimum stellar mass corresponds to each of the bin edges,
\begin{equation}
\begin{split}
 &w_{\mathrm{p}}(M_\star > {M_\star}_1) = w_{\mathrm{p}}(n=n_1) \\
 &w_{\mathrm{p}}(M_\star > {M_\star}_2) = w_{\mathrm{p}}(n=n_2) \ .
\end{split}
\end{equation}
Then, from these, we estimate the galaxy clustering within a given stellar mass bin,
\begin{equation}
\begin{split}
    w_{\mathrm{p}}({M_\star}_1 < M_\star < {M_\star}_2) \simeq 
&w_{\mathrm{p}1} \left( \frac{n_1}{n_1 - n_2} \right)^2 + w_{\mathrm{p}2} \left( \frac{n_2}{n_1 - n_2} \right)^2 \\
&- 2 \sqrt{w_{\mathrm{p}1} w_{\mathrm{p}2} } \left( \frac{n_1 n_2}{(n_1 - n_2)^2} \right)
\end{split}
\end{equation}
where $w_{\mathrm{p}1}=w_{\mathrm{p}}(n=n_1)$, $w_{\mathrm{p}2}=w_{\mathrm{p}}(n=n_2)$ and $n_1>n_2$. The derivation of this expression can be found in Appendix~\ref{sec:derivation}.
For galaxy-galaxy lensing $\Delta\Sigma$, the computation is even simpler. We compute,  
\begin{equation}
\begin{split}
 &\Delta\Sigma(M_\star > {M_\star}_1) = \Delta\Sigma(n=n_1) \\
 &\Delta\Sigma(M_\star > {M_\star}_2) = \Delta\Sigma(n=n_2) \ ,
\end{split}
\end{equation}
using the SHAMe emulator. Then compute the galaxy-galaxy lensing within a given stellar mass bin, 
\begin{equation}
\Delta\Sigma({M_\star}_1 < M_\star < {M_\star}_2) = \left( \frac{n_1}{n_1 - n_2} \right) \Delta\Sigma_1 - \left( \frac{n_2}{n_1 - n_2} \right) \Delta\Sigma_2
\end{equation}
where $\Delta\Sigma_1=\Delta\Sigma(n=n_1)$, $\Delta\Sigma_2=\Delta\Sigma(n=n_2)$ and $n_1>n_2$. We validated these relations using the TNG300 hydrodynamic simulation \citep{Nelson:2018uso}, and found that for $w_{\mathrm{p}}$ the differences were less than 1\% across all scales and that for $\Delta\Sigma$ the relation was exact.   

\begin{figure}
    \centering
    \includegraphics[width=1.0\linewidth]{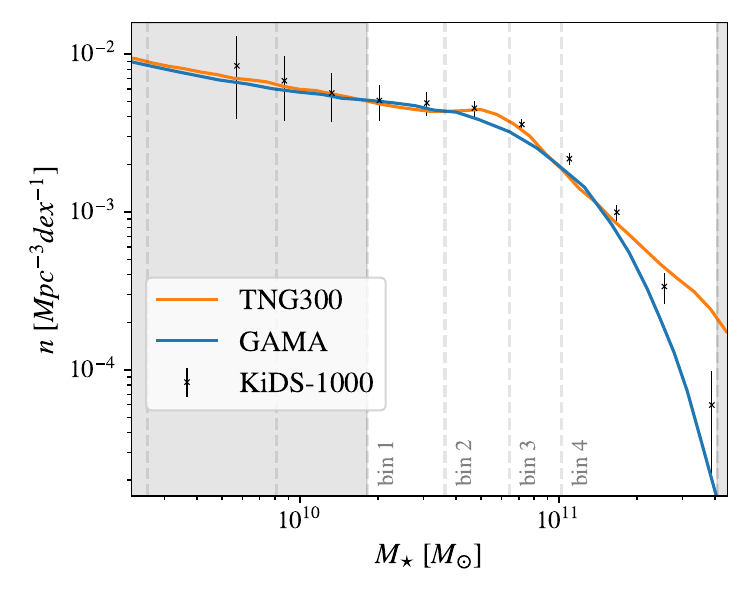}
    \caption{Differential stellar mass function from TNG300 compared to GAMA \citep{Driver:2022vyh} and KiDS-1000 \citep{Dvornik:2022xap}. The dashed lines indicate the stellar mass bin boundaries, which are required as inputs to the SHAMe emulator. We do not include the bins in the grey-shaded region, as they fall outside the range of the emulator.}
    \label{fig:differential stellar mass function}
\end{figure}

To convert the stellar mass bin edges to cumulative number densities, we require the cumulative stellar mass function of our galaxy sample. While this is easily accessible for simulated data, it is a lot more complex to obtain for real data, due to difficulties in accurately defining the volume of a galaxy survey. Therefore, in this work, we compute the cumulative number densities from the cumulative stellar mass function of TNG300, instead of from GAMA or KiDS-1000 data. Figure \ref{fig:differential stellar mass function} shows a comparison between the TNG300, GAMA \citep{Driver:2022vyh} and KiDS-1000 \citep{Dvornik:2022xap} differential stellar mass functions, where the grey dashed lines indicate the stellar mass bin boundaries used in \cite{Dvornik:2022xap} and the grey shaded region indicates the stellar mass bins outside the current number density range of our SHAMe emulator. We find good agreement between the three differential stellar mass functions, and in Section \ref{sec:fiducial analysis results} show that our fiducial analysis is resilient to our choice of cumulative number density inputs. A future extension of this work would be to adapt the SHAMe emulator to work with differential number densities, which are more accessible for real data, and then include these number densities with an associated uncertainty in the analysis.

\subsection{Likelihood} \label{sec:likelihood}
  To obtain constraints on the free parameters of our model, we assume a Gaussian likelihood of the form
  \begin{equation}
    -2\log p({\bf d}|\vec{\theta})\equiv\chi^2=({\bf d}-{\bf t}(\vec{\theta}))^T{\sf C}^{-1}({\bf d}-{\bf t}(\vec{\theta})),
  \end{equation}
  where ${\bf d}$ is our data vector, ${\bf t}$ is the theoretical prediction for ${\bf d}$, dependent on the free model parameters $\vec{\theta}$, and ${\sf C}$ is the covariance matrix of our measurements. The measurements and covariance are described above.
  
  Specifically, we use measurements of the projected correlation function $w_\mathrm{p}$ of galaxies in four stellar mass bins, as well as their cross-correlation with gravitational shear $\Delta\Sigma$, from a single sample of source galaxies. The data vector ${\bf d}$ thus comprises four different estimates of $w_\mathrm{p}$ and $\Delta\Sigma$, one per stellar mass bin. We do not consider cross-correlations between mass bins. The measurements span transverse separations in the range $0.15\,{\rm Mpc}/h<r_p<25\,{\rm Mpc}/h$. In our fiducial analysis we use this full range for $w_\mathrm{p}$, but impose a scale cut of $r_\mathrm{p}<1.4\,\mathrm{Mpc}/h$ for the $\Delta\Sigma$ data, to avoid the potential impact of baryonic feedback, which is not currently included in our model. 
  We determine this value by looking at the expected impact of baryonic effects following \cite{contreras2024}.
  We study the dependence of our results on this choice (see Section \ref{sec:allscales}). Our fiducial data vector contains a total of 64 elements.

  The free parameters of the model are the 5 $\Lambda$CDM cosmological parameters $(\sigma_8,\Omega_{\rm m},\Omega_{\rm b},n_s,h)$, as well as 5 SHAMe parameters, which are described in Section \ref{subsection: extended abundance matching}. These parameters and their associated priors are shown in Table \ref{tab:priors}. We evaluate the emulator (see Section \ref{sec:shame_emulator}) at the median redshift of each stellar mass bin [0.18, 0.22, 0.27, 0.32], and assume that the SHAMe parameters do not evolve with redshift or stellar mass. This assumption could be explored further in future work. We do not include intrinsic alignments or lens galaxy magnification in our model as \cite{Dvornik:2022xap} found them to have a negligible impact on both $w_{\mathrm{p}}$ and $\Delta\Sigma$. We sample the parameter space using \textsc{nautilus}\footnote{\url{https://github.com/johannesulf/nautilus}} \citep{nautilus} with 3000 live points.

\section{Validation on simulated sample} \label{sec:validation}
Before applying our methodology to KiDS-1000 and GAMA data, we validated it on a simulated sample from the TNG300 hydrodynamic simulation. This was to determine whether we were able to obtain robust cosmological constraints. 

We constructed our simulated sample by selecting galaxies from the redshift zero snapshot of TNG300 with stellar masses within the stellar mass ranges of the real data. This gave us four simulated galaxy samples, from which we computed the galaxy clustering $w_{\mathrm{p}}$ and galaxy-galaxy lensing $\Delta\Sigma$ using the same methodology described in section \ref{sec:method-clustering_lensing}, except for one difference. When creating the simulated $\Delta\Sigma$ from TNG300, instead of extracting only the dark matter field, as we do for the emulator in Section \ref{sec:method-clustering_lensing}, which uses gravity-only simulations, we also extracted the density fields for stars and gas. This means that the simulated $\Delta\Sigma$ from TNG300 includes the impact of baryons, but that the emulator for $\Delta\Sigma$ does not include them. A future extension to this work would be to include baryonic effects in the SHAMe emulator following the techniques introduced in \citealt{Arico2021a} and \citealt{Zennaro2024}.

\begin{figure}
    \centering
    \begin{subfigure}[b]{0.48\textwidth}
        \centering
        \includegraphics[width=\textwidth]{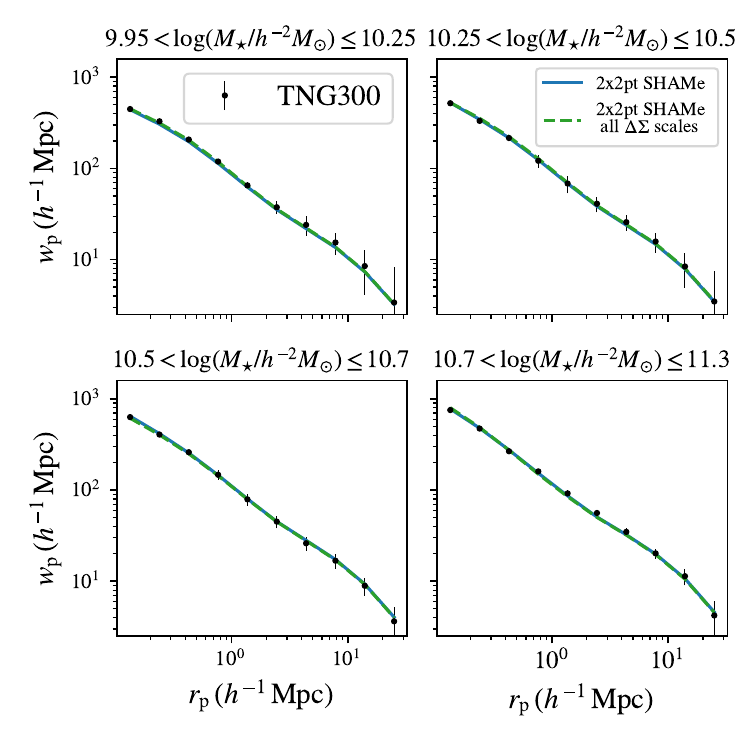}
    \end{subfigure}
    \begin{subfigure}[b]{0.48\textwidth}
        \centering
        \includegraphics[width=\textwidth]{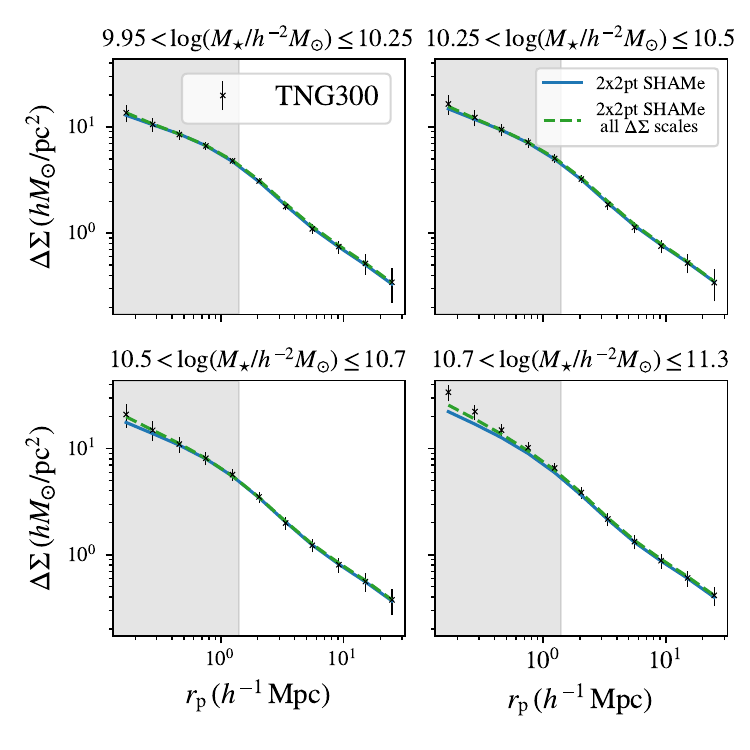}
    \end{subfigure}
    \caption{TNG300 galaxy clustering (top) and TNG300 galaxy-galaxy lensing (bottom) for each of the four stellar mass bins used in this work. The solid blue line shows the best fit when analysing the joint data vector with SHAMe, where the galaxy-galaxy lensing data in the grey shaded region is not included, and the green dashed line shows the best fit when all scales of the galaxy-galaxy lensing are included in the data vector.}
    \label{fig:TNG300_data}
\end{figure}

The simulated data can be seen in Figure \ref{fig:TNG300_data}. It is very similar to the real GAMA and KiDS-1000 data, but does not match exactly for a few reasons. Firstly, the simulated data is computed at redshift zero instead of the median redshift of each stellar mass bin, which are [0.18, 0.22, 0.27, 0.32]. This results in a slight discrepancy between the real and simulated data. Secondly, the underlying cosmology of the TNG300 sample may be slightly different from that of the real Universe. Thirdly, there may be some small differences in how stellar masses are computed for the real and simulated data, as well as some possible incompleteness at low stellar masses. Finally, the galaxy formation physics in TNG300 is likely not representative of the real Universe, and the clustering data was not used to calibrate the sub-grid physics. Despite these differences, the simulated data vector is very close to the real data vector, so it provides a good test for our methodology.

\begin{figure}
    \centering
    \includegraphics[width=1.0\linewidth]{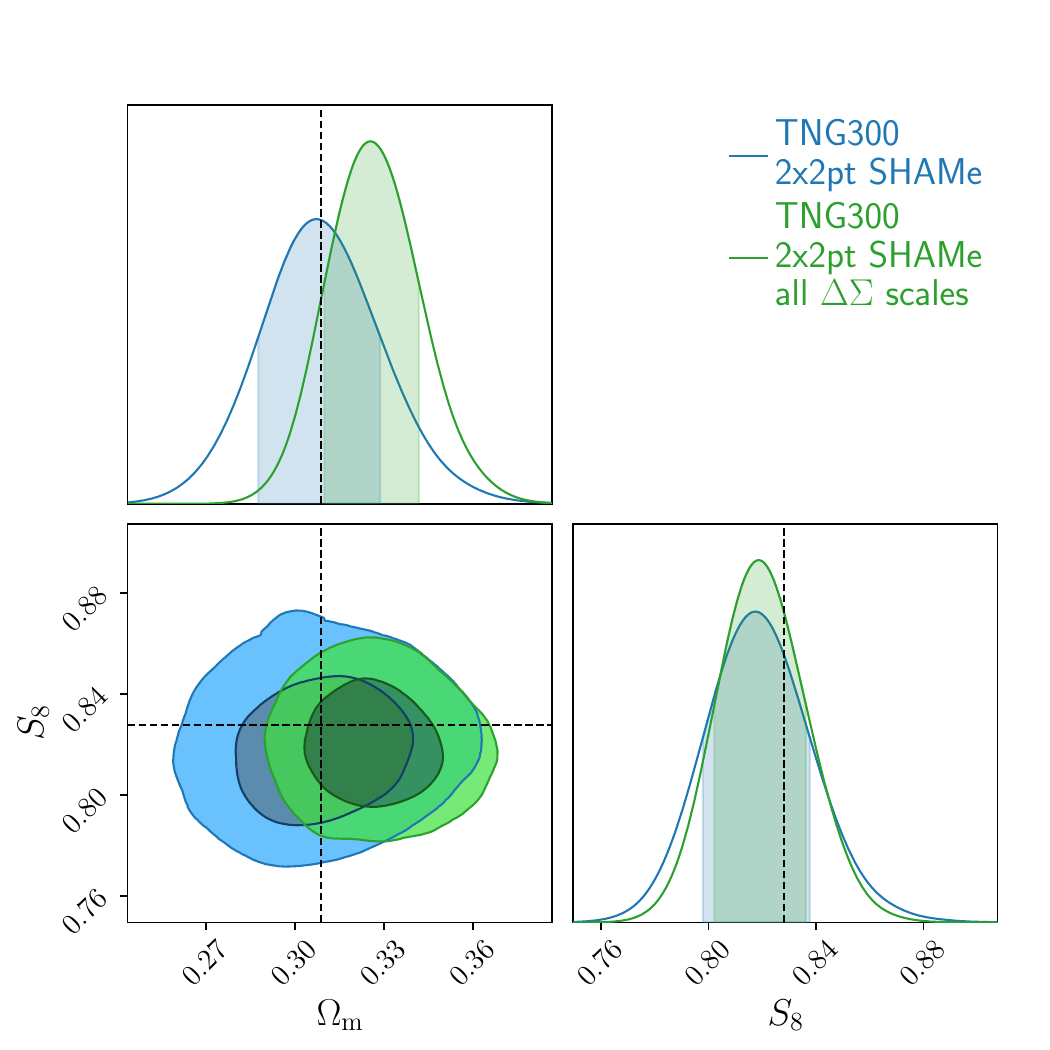}
    \caption{Marginalised posterior constraints on $S_8$ and $\Omega_\mathrm{m}$ for TNG300 galaxy clustering and galaxy-galaxy lensing using SHAMe (blue) and the same analysis without scale cuts in the galaxy-galaxy lensing (green). We recover the TNG300 input cosmology (dashed lines) within $1\sigma$, where the covariance matches the covariance used in the real data analysis.}
    \label{fig:TNG300_sample}
\end{figure}

After constructing the simulated galaxy clustering and galaxy-galaxy lensing data vectors from TNG300 we analysed them in exactly the same way as the real data; using the same covariance (Section \ref{sec:Covariance}), cumulative number density inputs (Section \ref{sec:CSMF}), and likelihood (Section \ref{sec:likelihood}). Figure \ref{fig:TNG300_sample} shows our marginal posterior constraints on $S_8$ and $\Omega_\mathrm{m}$ for two variations of our analysis. The blue contour shows our fiducial case where we remove scales of $\Delta\Sigma$ where $r_\mathrm{p}<1.4\,\mathrm{Mpc}/h$, as we do not model baryonic effects, and the green contour shows an extended analysis where we include all scales of $\Delta\Sigma$. For a detailed discussion of the impact of baryons on our analyses, see Section \ref{sec:allscales}. The dashed lines indicate the input cosmology of TNG300, and we are able to recover the input cosmology of the simulation within 1$\sigma$ in both cases.

Figure \ref{fig:TNG300_data} shows the best-fit galaxy-galaxy lensing and clustering predictions for our fiducial analysis (blue), and for the extended analysis including smaller scales in $\Delta\Sigma$ (green). These predictions were estimated for the maximum a posteriori parameter values, which align with the marginal distributions in Figure \ref{fig:TNG300_sample}. Both cases provide a good fit to the data, with the extended analysis providing a better fit for the small-scale galaxy-galaxy lensing, as expected. Since we use the real data covariance when analysing our simulated data vectors, the corresponding chi-squared values are meaningless as an absolute goodness-of-fit metric. 

\section{Results} \label{sec:results}
\begin{figure}
    \centering
    \includegraphics[width=1.0\linewidth]{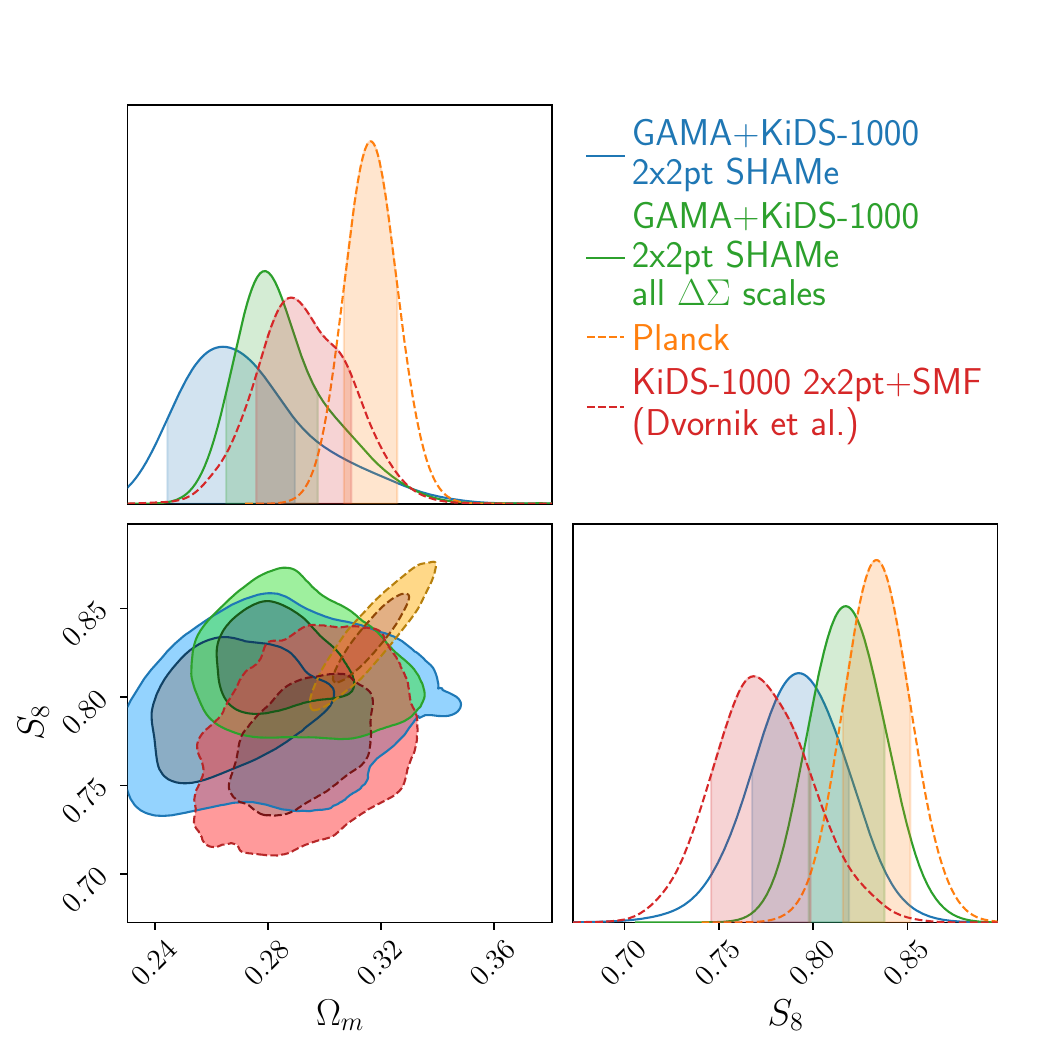}
    \caption{Marginalised posterior constraints on $S_8$ and $\Omega_\mathrm{m}$ for GAMA galaxy clustering and KiDS-1000 galaxy-galaxy lensing using SHAMe (blue) and the same analysis without scale cuts in the galaxy-galaxy lensing (green). For comparison, we show CMB results from \planck (\citealt{Planck}, orange dashed) and results for KiDS-1000 galaxy clustering and galaxy-galaxy lensing using an analytic conditional stellar mass function to describe the galaxy-halo connection (\citealt{Dvornik:2022xap}, red dashed). The results are consistent, with SHAMe (this work) giving a slightly higher value of $S_8$, particularly when all scales are included in the galaxy-galaxy lensing.}
    \label{fig:S8_omegam_all_scales_kids}
\end{figure}

We present cosmological constraints for a combined analysis of GAMA galaxy clustering and KiDS-1000 galaxy-galaxy lensing analysed using SHAMe to model the galaxy-halo connection. Figure \ref{fig:S8_omegam_all_scales_kids} shows our main result, constraints on $\Omega_\mathrm{m}$ and $S_8 \equiv \sigma_8\sqrt{\Omega_\mathrm{m}/0.3}$. We show our fiducial result (blue contour), as well as an extended analysis that includes all scales of galaxy-galaxy lensing (green contour). Our results are compared to the \planck TT,TE,EE+lowE CMB result (orange contour, \citealt{Planck}), and a precursor analysis with KiDS-1000 (red contour, \citealt{Dvornik:2022xap}). 

\subsection{Fiducial Analysis} \label{sec:fiducial analysis results}
\begin{figure}
    \centering
    \begin{subfigure}[b]{0.48\textwidth}
        \centering
        \includegraphics[width=\textwidth]{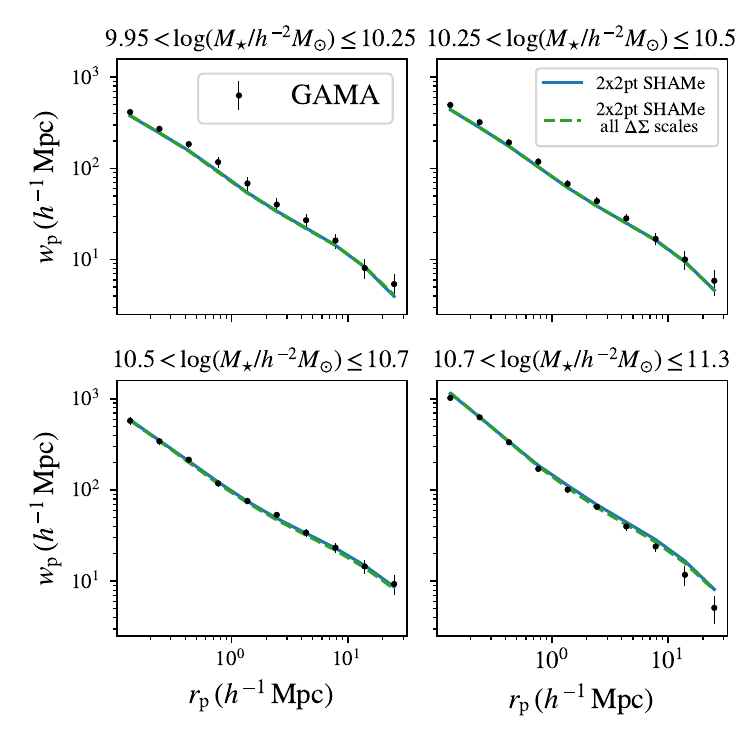}
    \end{subfigure}
    \begin{subfigure}[b]{0.48\textwidth}
        \centering
        \includegraphics[width=\textwidth]{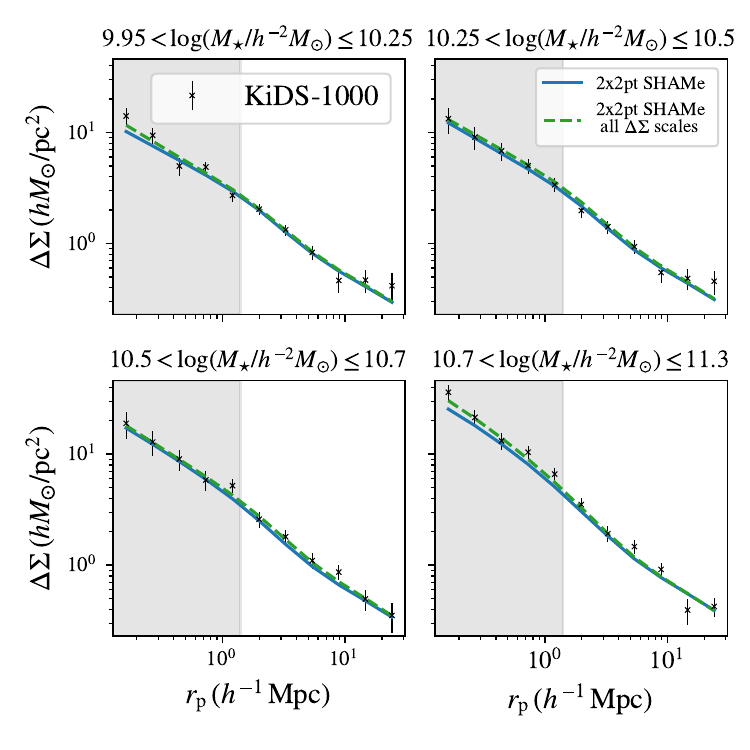}
    \end{subfigure}
    \caption{GAMA galaxy clustering (top) and KiDS-1000 galaxy-galaxy lensing (bottom) for each of the four stellar mass bins used in this work. The solid blue line shows the best fit when analysing the joint data vector with SHAMe, where the galaxy-galaxy lensing data in the grey shaded region is not included in order to avoid the impact of baryonic effects (Figure \ref{fig:S8_omegam_all_scales_kids}, blue contour). The green dashed line shows the best fit when all scales of the galaxy-galaxy lensing are included in the data vector (Figure \ref{fig:S8_omegam_all_scales_kids}, green contour).}
    \label{fig:GAMA_KiDS_data}
\end{figure}

In our fiducial analysis (Figure \ref{fig:S8_omegam_all_scales_kids}, blue contour) we find $S_8=0.793^{+0.025}_{-0.024}$ and $\Omega_\mathrm{m}=0.260^{+0.027}_{-0.015}$. These results are consistent with \planck, with a lower value of $S_8$ that is in agreement at the 1.7$\sigma$ level (computed as the 1D difference between the two predictions), and a slightly larger difference for $\Omega_\mathrm{m}$ values (also shifted to lower values).
We are also able to constrain all five SHAMe parameters (see section \ref{sec:shame_param_results} for details). We find a good fit to the data with a chi-squared value of 74.2 and a chi-squared per degree of freedom of 1.37, where we use a simple estimate of the number of degrees of freedom as the number of data points minus the number of constrained parameters, in this case 64-10. The associated $p$-value is 0.036, which we consider to be adequate (the model is able to explain the data at less than 2$\sigma$). {These values are similar to the ones obtained using the small scales of the galaxy-galaxy lensing (1.28 and 0.051, receptively), which is particularly impressive, since we are not fitting those scales.} Our best-fit clustering and galaxy-galaxy lensing, calculated using the maximum posterior values of our fiducial result, can be seen in Figure \ref{fig:GAMA_KiDS_data} (blue line). We show the best fit galaxy-galaxy lensing extrapolated to the full range of $r_\mathrm{p}$ scales, but we only compute this best fit using the data outside of the grey shaded region in order to remove baryonic effects. The current version of the SHAMe emulator does not model baryonic effects in $\Delta\Sigma(r_{\rm p})$, so in our fiducial case, we remove scales where $r_\mathrm{p}<1.4\,\mathrm{Mpc}/h$. See section \ref{sec:allscales} for further details about the impact of baryons. In our fiducial case, the model is able to describe the data on all scales, even those not used to constrain it, which emphasises the physically motivated nature of the SHAMe model. We are also able to simultaneously reproduce the galaxy clustering and galaxy-galaxy lensing measurements across all the stellar mass bins, which SHAMe was not originally designed to do.

\begin{figure}
    \centering
    \includegraphics[width=1.0\linewidth]{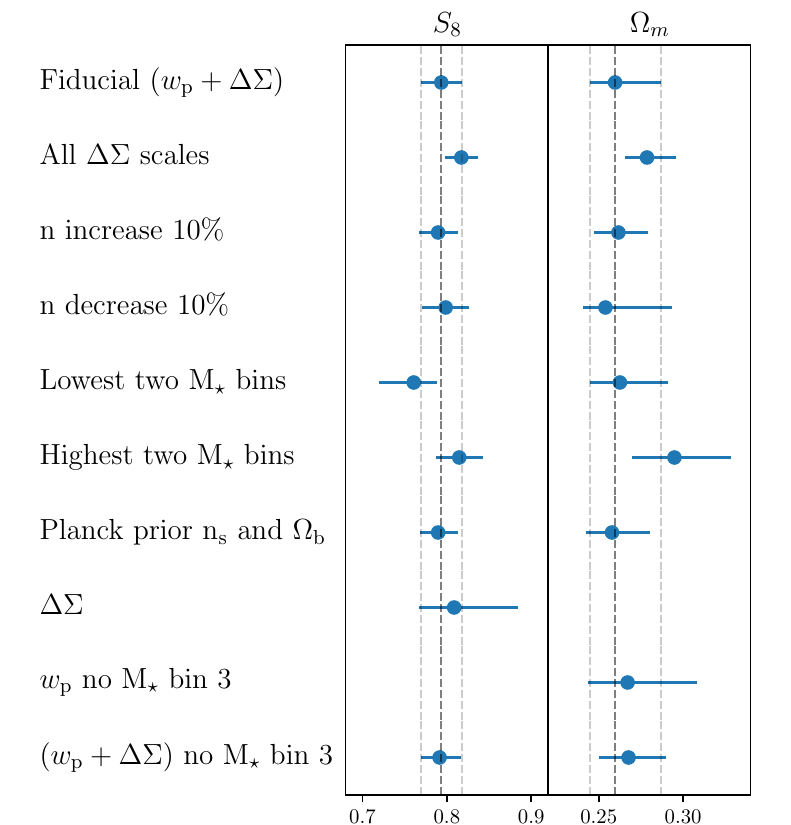}
    \caption{$S_8$ and $\Omega_\mathrm{m}$ marginal constraints for variations of our fiducial joint analysis of GAMA galaxy clustering and KiDS-1000 galaxy-galaxy lensing using SHAMe. We find consistency between our analysis and all of the variations shown.}
    \label{fig:S8_summary_analysis_variations}
\end{figure}

In Figure \ref{fig:S8_summary_analysis_variations} we quantify the impact of various analysis choices on our fiducial constraints. Any differences in the recovered value of $S_8$ compared to the fiducial result are quantified in terms of the metric $(S_8^{\rm fid}-S_8^{\rm var}) / \sigma_{S_8^{\rm var}}$. We discuss extending the analysis to all $\Delta\Sigma$ scales and the impact of baryons in detail in Section \ref{sec:allscales}.

In our fiducial analysis, we use the cumulative number densities from the TNG300 simulation as inputs for the SHAMe emulator, instead of those from GAMA or KiDS-1000 (see Section \ref{sec:CSMF}). We assess the impact of this choice by increasing/decreasing the SHAMe number density inputs by 10\%, motivated by Figure \ref{fig:differential stellar mass function}, and find shifts of 0.16$\sigma$/0.19$\sigma$ in $S_8$, demonstrating that our analysis is robust to this choice. We check that our results are consistent across the stellar mass bins by analysing the two lowest and two highest stellar mass bins separately. We find broad consistency between the two sets of bins, with shifts of 1.16$\sigma$/0.77$\sigma$ in $S_8$ for the two lowest/highest bins, respectively. However, these variations are limited by the number of datapoints compared to the number of free parameters, and the majority of the constraining power for the 5 SHAMe parameters comes from the two highest stellar mass bins. Therefore, even though we find broad consistency across the stellar mass bins, the power of the analysis comes from combining a range of stellar masses. We also check the impact of adding a \planck prior to the unconstrained cosmological parameters, $n_\mathrm{s}$ and $\Omega_\mathrm{b}$, and find a small shift of 0.16$\sigma$ in $S_8$. 

In the last three rows of Figure \ref{fig:S8_summary_analysis_variations}, we illustrate the power of combined analyses of galaxy clustering and galaxy-galaxy lensing. When analysing $\Delta\Sigma$ alone, we are only able to place a wide constraint on $S_8$, and when analysing $w_{\mathrm{p}}$ alone, we are only able to place a wide constraint on $\Omega_\mathrm{m}$. However, combining the two, we can constrain both $S_8$ and $\Omega_\mathrm{m}$ with greater precision. For a discussion of the power of combining $w_{\mathrm{p}}$ and $\Delta\Sigma$ in the context of the SHAMe parameters, see Section \ref{sec:shame_param_results}. When initially analysing $w_{\mathrm{p}}$ alone we found a significant constraint on $S_8$, which was inconsistent with our fiducial result, and is not expected using only galaxy clustering data. We found that this constraint disappeared when we removed stellar mass bin 3, $10.5 < \mathrm{{log}}(M_{{\star}}/h^{{-2}}M_{{\odot}}) \leq 10.7$, and even if we just removed the large scales of bin 3, where $r_{\mathrm{max}}>1.5 \ h^{-1}\mathrm{Mpc}$ (see Appendix \ref{app:clustering_inconsistency}). We therefore suspect the presence of an uncontrolled systematic effect in the large scales of bin 3. To avoid confirmation bias, we did not remove $w_{\mathrm{p}}$ bin 3 from our fiducial joint data vector. However, in the final row of Figure \ref{fig:S8_summary_analysis_variations} we show that our fiducial analysis is only negligibly affected when stellar mass bin 3 is removed from the clustering, resulting in a shift of 0.08$\sigma$ in $S_8$. We also test removing mass bin 4, since the stellar mass function can not correctly reproduce this mass regime, as shown in Figure \ref{fig:differential stellar mass function}. For this case, we find similar constraints as when using the full sample (not shown here); this occurs because we actually select the galaxies based on their number density, which mitigates the impact of the different mass functions on the galaxy selection (see Appendix \ref{sec:NoBin4} for more details).

\subsection{Extension to all scales of galaxy-galaxy lensing} \label{sec:allscales}

In Figure \ref{fig:S8_omegam_all_scales_kids} we present cosmological constraints for an extension to our fiducial analysis where we include all scales in the galaxy-galaxy lensing data (green contour). In our fiducial analysis, we remove scales of $\Delta\Sigma$ where $r_\mathrm{p}<1.4\,\mathrm{Mpc}/h$ because the current version of the SHAMe emulator does not include baryonic effects. However, the model works well on small, intra-halo scales in the absence of baryons, and in Section \ref{sec:validation} we showed that for a simulated analysis, in which the simulated data incorporated the impact of baryonic effects, we were still able to recover the input cosmology when including all scales. It will also be possible to extend the emulator in the future \citep{Arico2021a, Zennaro2024}. We therefore explore the consistency of our results with the minimum scale used in the galaxy-galaxy lensing and any potential gains in constraining power.

When we include all scales we find $S_8=0.817^{+0.020}_{-0.019}$ and $\Omega_\mathrm{m}=0.279^{+0.017}_{-0.013}$. The $S_8$ value is 1.26$\sigma$ higher than our fiducial result and in agreement with \planck at the 0.9$\sigma$ level. We also find a 21\% tighter constraint on $S_8$ and a 29\% tighter constraint on $\Omega_\mathrm{m}$. This result is consistent with our fiducial result, with the upwards shift in the value of both parameters driven by preference of the small-scale data for the region of high $S_8$ and $\Omega_{\rm m}$ allowed by the fiducial analysis. An increase in the value of $S_8$, towards the \planck value, is contrary to what might be naively expected when adding small scales without modelling baryonic feedback. In most scenarios, baryonic feedback leads to a suppression of clustering on small scales, which would be interpreted as a lower $S_8$ if unmodelled  \citep{Yoon2019,YoonJee2021,Amon:2022azi,Preston2023, AricoDES2023A&A...678A.109A,Garcia-Garcia2024JCAP...08..024G}. This is thus an interesting result in the context of the $S_8$ tension observed between galaxy surveys and CMB data. Since our data approximately corresponds to halo masses of $10^{11.5}-10^{13}\, h^{-1}M_{\odot}$ \citep{Dvornik:2022xap}, one possible explanation could be that less gas is ejected from low mass halos.

In Figure \ref{fig:GAMA_KiDS_data} we show our best-fit clustering and galaxy-galaxy lensing calculated using the maximum posterior values of our all-scale result (green line). We find a good fit to the data with a chi-squared value of 85.5 and a chi-squared per degree of freedom of 1.15, where we simply estimate the degrees of freedom as 84-10. The improvement in the chi-squared value per degree of freedom compared to our fiducial analysis is likely driven by the increase in data points resulting from including all scales. The associated p-value is 0.17, showing that the model is able to describe the small-scale galaxy-galaxy lensing data well.

\begin{figure*}
    \centering
    \begin{subfigure}[b]{0.48\textwidth}
        \centering
        \includegraphics[width=\textwidth]{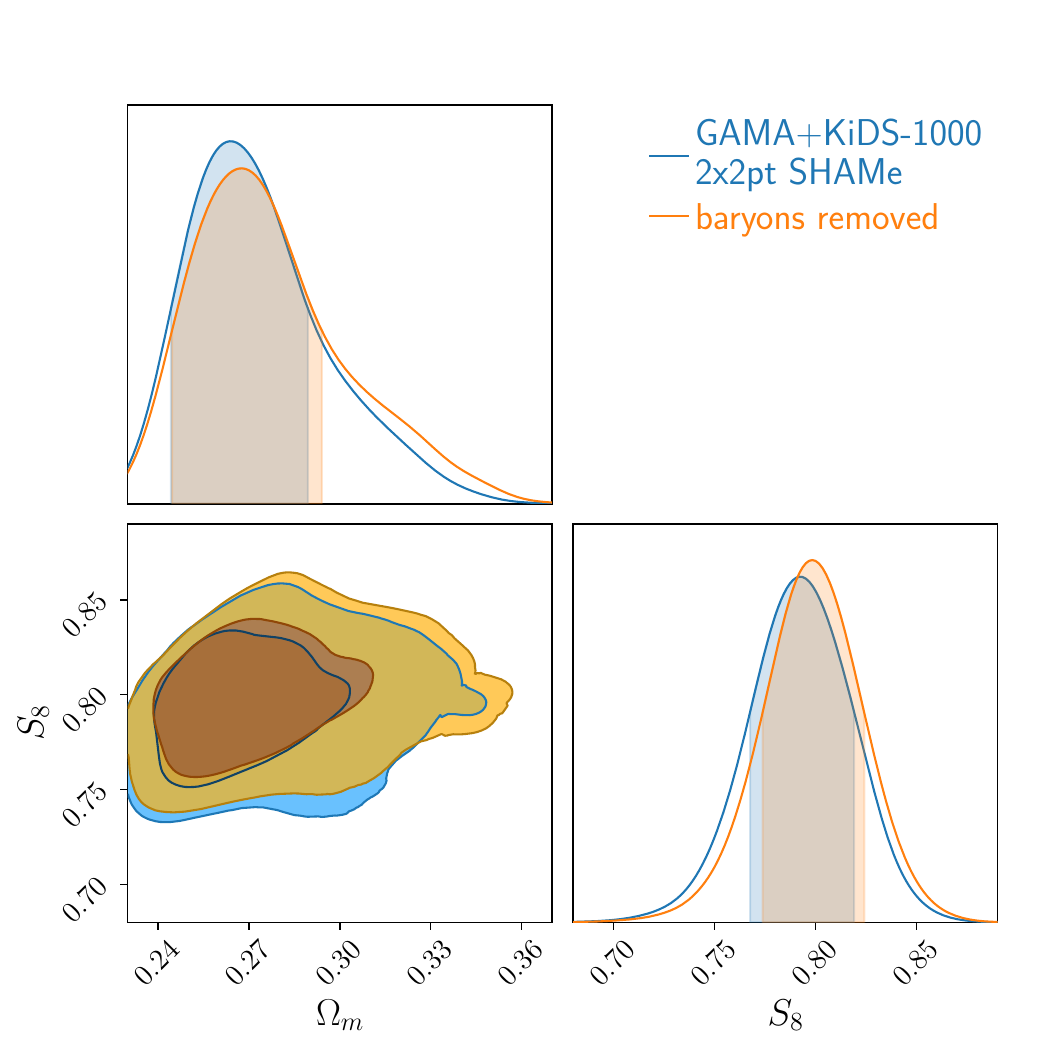}
    \end{subfigure}
    \begin{subfigure}[b]{0.48\textwidth}
        \centering
        \includegraphics[width=\textwidth]{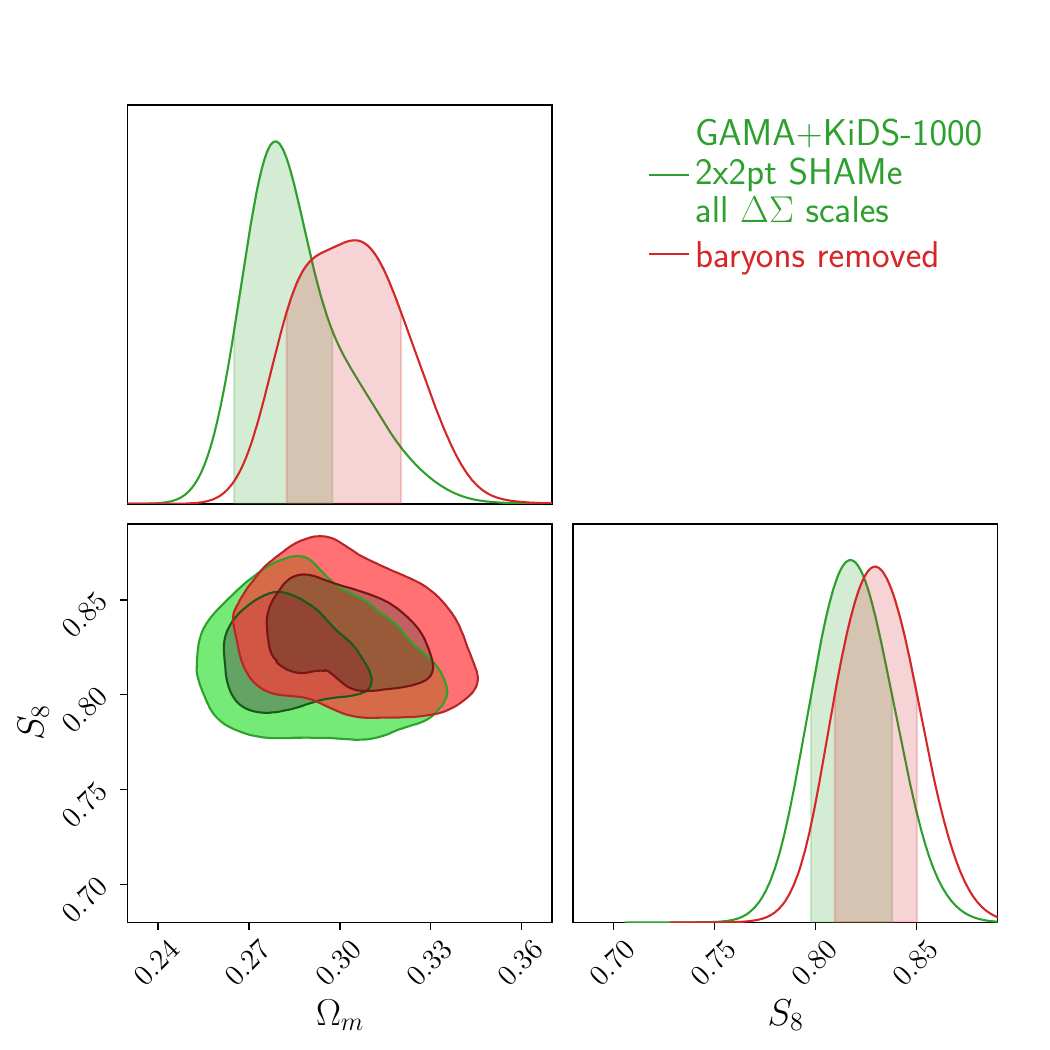}
    \end{subfigure}
    \caption{Marginalised posterior constraints on $S_8$ and $\Omega_\mathrm{m}$ for GAMA galaxy clustering and KiDS-1000 galaxy-galaxy lensing using SHAMe (left) and the same analysis without scale cuts in the galaxy-galaxy lensing (right). The blue and green contours match those in Figure \ref{fig:S8_omegam_all_scales_kids} and show the analysis with the real data vectors. The orange and red contours show the impact of removing baryons from the data vector, using an estimate from simulations. Our fiducial analysis is robust to baryonic effects, and even our less conservative all-scales analysis is only slightly impacted.}
    \label{fig:no baryons}
\end{figure*}

In Figure \ref{fig:no baryons} we estimate the impact of not modelling baryonic effects present in the galaxy-galaxy lensing measurements, for both our fiducial (left) and all-scale analysis (right). While it is not possible to fully remove the effect of baryons from our galaxy-galaxy lensing measurements, we can mitigate the impact by subtracting the expected signal predicted by hydrodynamic simulations. We measure this signal by computing the galaxy-galaxy lensing from a SHAMe mock catalogue calibrated to the best-fitting parameters of our all-scale analysis. We then apply the baryonification model from \cite{Arico2021a}, which displaces the particles of a dark matter-only simulation to mimic the effect of baryons. The baryonification model reproduces the matter distribution expected for the standard Bahamas simulations \citep{McCarthy:2016mry}, which are thought to provide the most accurate estimate of baryonic effects in the Universe since they are calibrated using the observed stellar and gas fractions in galaxy clusters. We then measure the difference in the galaxy-galaxy lensing signal between the mock with and without the baryonification implementation, and finally subtract the additional signal from our KiDS-1000 galaxy-galaxy lensing measurements. This allows us to obtain an order-of-magnitude estimate of the impact of baryonic effects. 

As expected, in both cases we find that removing baryonic effects increases the value of $S_8$ to be more consistent with \planck. This emphasises the surprising increase in $S_8$ found when we move from our fiducial to all-scale analysis (see Figure \ref{fig:S8_omegam_all_scales_kids}). For our fiducial analysis, we find a shift of 0.2$\sigma$ in $S_8$ when we remove baryonic effects. Without scale cuts, we observe a shift of 0.6$\sigma$. We therefore conclude that our fiducial analysis is robust to baryonic effects, and while our all-scale analysis is less conservative, we still find a relatively small impact. These results ultimately depend on the level of feedback in the Universe, and we would expect to find a smaller shift if we were to assume a level of feedback that was smaller than from the Bahamas simulations, or larger, as the one recently predicted by \cite{Bigwood:2024}.
While these results are promising, we would like to remind the reader that without a breakthrough on the constraints of baryons in the galaxy-galaxy lensing signal, we need to continue being agnostic to their effect, and therefore, we will refer to the results of the fiducial model (i.e., with the limited range of scales in the galaxy-galaxy lensing measurement) as the main result of this work.

\subsection{SHAMe parameters} \label{sec:shame_param_results}

\begin{figure*}
    \centering
    \includegraphics[width=1.0\linewidth]{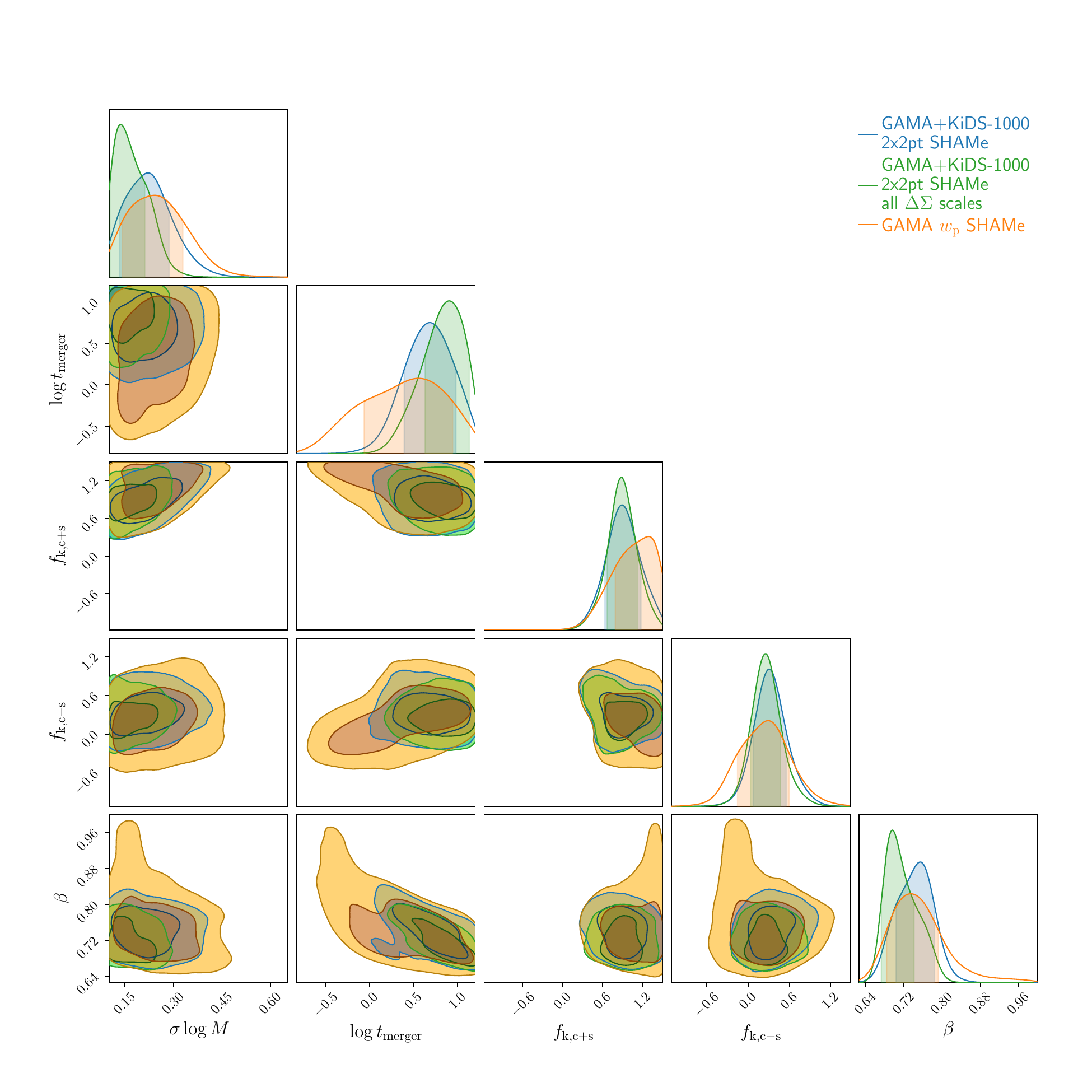}
    \caption{Marginalised posterior constraints on the 5 SHAMe parameters ($\sigma$, $t_{\mathrm{merger}}$, $f_{\mathrm{k, c+s}}$, $f_{\mathrm{k, c-s}}$ and $\beta$) for GAMA galaxy clustering and KiDS-1000 galaxy-galaxy lensing using SHAMe (blue), the same analysis without scale cuts in the galaxy-galaxy lensing (green) and for GAMA galaxy clustering alone (orange). We are able to constrain all 5 SHAMe parameters, as well as the cosmological parameters $S_8$ and $\Omega_\mathrm{m}$ (see Figure \ref{fig:S8_omegam_all_scales_kids}). The SHAMe parameter constraints predominantly come from the galaxy-clustering, but adding small-scale galaxy-galaxy lensing is also helpful. Note these contours do not include clustering stellar mass bin 3 (see Section \ref{sec:fiducial analysis results} and Appendix \ref{app:clustering_inconsistency} for details).}
    \label{fig:SHAMe params}
\end{figure*}

In Figure \ref{fig:SHAMe params} we present constraints on the SHAMe parameters ($\sigma$, $t_{\mathrm{merger}}$, $f_{\mathrm{k, c+s}}$, $f_{\mathrm{k, c-s}}$ and $\beta$) for our fiducial analysis of GAMA galaxy clustering and KiDS-1000 galaxy-galaxy lensing (blue contour) and our extended analysis where we include all scales of the galaxy-galaxy lensing (green contour). We are able to constrain all of the SHAMe parameters in both cases, but we find that the inclusion of small-scale $\Delta\Sigma$ data leads to a significant improvement in the constraints on $\sigma \log M$, $\log t_{\mathrm{merger}}$ and $\beta$. In turn, this results in the cosmological constraints becoming more precise, highlighting the benefits of including small-scale lensing information when using non-perturbative approaches to describe the clustering of galaxies. 

We also find evidence of strong assembly bias for both central and satellite galaxies in the sample. Since $f_{\mathrm{k, c+s}}$ and $f_{\mathrm{k, c-s}}$ are both positive, there is a positive correlation between large-scale linear bias and the stellar mass or luminosity of a galaxy in the sample (see Section \ref{subsection: extended abundance matching}). If we then populate a mock galaxy catalogue using the best-fitting SHAMe parameters from Figure \ref{fig:SHAMe params}, we find that this translates to strong assembly bias for both central and satellite galaxies. This implies it is important to model assembly bias in these types of analyses. 

In Figure \ref{fig:SHAMe params} we additionally show the constraints on the SHAMe parameters for GAMA galaxy clustering alone (orange contour). This illustrates that most of the constraining power on the SHAMe parameters comes from $w_{\mathrm{p}}$. $\Delta\Sigma$ alone is only able to constrain $\sigma$ and $\beta$, but when combined with $w_{\mathrm{p}}$ in a 2x2pt analysis we are able to constrain all of the SHAMe parameters, as well as $S_8$ and $\Omega_\mathrm{m}$. This again illustrates the power of combined analyses of galaxy clustering and galaxy-galaxy lensing\footnote{The contours in Figure \ref{fig:SHAMe params} are the versions without clustering stellar mass bin 3, where $10.5 < \mathrm{{log}}(M_{{\star}}/h^{{-2}}M_{{\odot}}) \leq 10.7$. See Section \ref{sec:fiducial analysis results} and Appendix \ref{app:clustering_inconsistency} for details.}.

\subsection{Comparison with precursor analysis}

In Figure \ref{fig:S8_omegam_all_scales_kids} we also include the $\Omega_\mathrm{m}$ and $S_8$ constraints from a precursor analysis of KiDS-1000 galaxy-galaxy lensing, galaxy clustering and galaxy abundance (2x2pt+SMF, \citealt{Dvornik:2022xap})\footnote{The contours look slightly different due to the choice of smoothing scale.}. The main difference between the two analyses is the modelling of the galaxy-halo connection. \cite{Dvornik:2022xap} use an analytical halo model-based description incorporating an effective Halo Occupation Distribution linked to the conditional stellar mass function. Instead, we use an abundance matching scheme to describe the galaxy-halo connection, extended to include the impact of assembly bias, and with predictions built from the application of this model to $N$-body simulations (instead of using an analytical approach). A key benefit of SHAMe, in addition to the use of simulations and the inclusion of assembly bias, is its dependence on a smaller number of parameters (5 instead of 11). The main disadvantage is that SHAMe requires subhalo information, which is computationally expensive. However, this can be addressed through emulation, and having subhalo information aids in modelling the distribution of satellite galaxies, as it is not necessary to rely on ad-hoc parametrisations of the satellite profile. This is particularly useful for galaxy-galaxy lensing \citep{2023Chavesmontero, contreras_lensing}. 

The data used in the two analyses is also slightly different. We do not include the lowest two stellar mass bins used in \cite{Dvornik:2022xap}, resulting in four stellar mass bins instead of six. The lowest two stellar mass bins are currently outside the maximum number density of our SHAMe emulator; however, this limitation will be addressed in future versions. We also use GAMA spectroscopic galaxy clustering instead of KiDS-1000 photometric galaxy clustering (see Section \ref{sec:data} for details). In our fiducial analysis, we do not include small-scale galaxy-galaxy lensing, where $r_\mathrm{p}<1.4\,\mathrm{Mpc}/h$, as we do not explicitly model baryonic effects (see Section \ref{sec:allscales}). \cite{Dvornik:2022xap} also do not explicitly model baryonic effects but they allow the concentration of both halos and satellites to vary, which has been shown to provide enough flexibility to account for the impact of baryons \citep{Debackere2021MNRAS.505..593D, Cacciato2013MNRAS.430..767C, Viola2015MNRAS.452.3529V, Edo2018MNRAS.476.4662V, Dvornik2018MNRAS.479.1240D, Amon2023MNRAS.518..477A}. Finally, we include galaxy abundance through the emulator cumulative number density inputs at the stellar mass bin edges (see Section \ref{sec:CSMF}), instead of jointly modelling the galaxy stellar mass function.

For our fiducial analysis, we find a value of $S_8$ that is 1.07$\sigma$ higher than \cite{Dvornik:2022xap}. This difference increases to 2.65$\sigma$ when including all scales in $\Delta\Sigma$. We find similar uncertainties on $S_8$ for our fiducial analysis compared to \cite{Dvornik:2022xap}, with only a 5\% increase in constraining power. This is despite using two fewer stellar mass bins and not including small-scale galaxy-galaxy lensing. For our all-scale analysis, we find our uncertainty on $S_8$ is improved by 25\% compared to \cite{Dvornik:2022xap}. The consistency between the cosmological constraints found in these two analyses, particularly in the fiducial case, is reassuring, given the vastly different approaches used to model the small-scale clustering of galaxies.

\section{Conclusion} \label{sec:conclusion}

In this paper we analysed GAMA galaxy clustering and KiDS-1000 galaxy-galaxy lensing using extended SubHalo Abundance Matching (SHAMe) to model the galaxy-halo connection. SHAMe has previously been shown to be capable of simultaneously modelling both galaxy clustering and galaxy-galaxy lensing \citep{contreras_lensing,contreras_lensing_data2023}, but has not been used to obtain cosmological constraints. In this work we built an emulator to simultaneously model both the SHAMe and cosmological parameters, and here highlight the most important results: 
\begin{itemize}
    \item We validated our analysis by constructing stellar mass-selected samples from the TNG300 hydrodynamic simulation with similar characteristics to those from GAMA and KiDS. We then computed simulated galaxy clustering and galaxy-galaxy lensing data vectors, which we analysed using our SHAMe emulator. We found that we could successfully reproduce both the galaxy clustering and galaxy-galaxy lensing measurements (Figure \ref{fig:TNG300_data}), as well as correctly recover the input cosmology of the simulation for both our fiducial and all-scale analysis (Figure \ref{fig:TNG300_sample}). 
    \item We jointly analysed four stellar mass bins ($9.95 < \mathrm{{log}}(M_{{\star}}/h^{{-2}}M_{{\odot}}) \leq 11.3$) of GAMA galaxy clustering and KiDS-1000 galaxy-galaxy lensing data using our SHAMe emulator. In our fiducial analysis we found $S_8=0.793^{+0.025}_{-0.024}$ and $\Omega_\mathrm{m}=0.260^{+0.027}_{-0.015}$ (Figure \ref{fig:S8_omegam_all_scales_kids}), and were also able to constrain all 5 SHAMe parameters (Figure \ref{fig:SHAMe params}). These constraints are consistent with those found by \planck within {\bf 1.7$\sigma$}, and we found a good fit to both the galaxy clustering and galaxy-galaxy lensing data (Figure \ref{fig:GAMA_KiDS_data}).
    \item We extended our fiducial analysis to include scales below $r_\mathrm{p}<1.4\,\mathrm{Mpc}/h$ in the galaxy-galaxy lensing. We removed these scales in our fiducial analysis to avoid the impact of baryonic effects, which while protecting our results from bias (Figure \ref{fig:no baryons}) limits the constraining power of our approach. When extending our analysis to all scales we found a 21\% improvement in our $S_8$ constraint and a 29\% improvement in our $\Omega_\mathrm{m}$ constraint. We also found a value of $S_8$, which is 1.26$\sigma$ higher than our fiducial result, bringing the result into closer agreement with \planck $(\sim1\sigma)$ (Figure \ref{fig:S8_omegam_all_scales_kids}). Although the statistical evidence of this shift is small, it is interesting to note that such an upwards shift is the opposite of what would be expected in the presence of baryonic feedback, which would normally lead to a suppression in the small-scale clustering of matter (Figure \ref{fig:no baryons}). 
    These results should be interpreted with caution, since our constraints on the impact of baryons on the galaxy-galaxy lensing are still limited, and we therefore cannot make strong statements about these constraints as we can do with the ones where we remove the small scales of the galaxy-galaxy lensing signal, which are the main result of this work. Nonetheless, these results show the power of including smaller scales in the modelling of galaxy clustering observables and the potential of this approach for future analysis.
\end{itemize}

The results of this paper highlight the potential of joint small-scale galaxy clustering and galaxy-galaxy lensing analyses using SHAMe. We find competitive constraints, which reduce the previously seen tension with \planck. The SHAMe model has only 5 free parameters and, when combined with a scaling technique capable of producing a 1 $h^{-1}$ Gpc dark matter simulation in a matter of minutes \citep{Angulo2010_scale}, we were able to build an emulator using a limited amount of computational resources. These analyses are capable of placing constraints on both cosmological and galaxy formation parameters, including galaxy assembly bias, and provide an exciting prospect for analysing upcoming data from \textit{Euclid} and LSST.

\section*{Acknowledgements}

We thank Marika Asgari for comments on the manuscript, and Jon\'{a}s Chaves-Montero for writing the code used to compute the galaxy-galaxy lensing from the simulated galaxy catalogues. CM, SC, and REA acknowledge support under the grant number PID2021-128338NB-I00 from the Spanish Ministry of Science. CM and DA acknowledge support from the Beecroft Trust. SC acknowledges the support of the `Juan de la Cierva Incorporac\'ion' fellowship (IJC2020-045705-I) and ``Ram\'on y Cajal'' fellowship (RYC2023-043783-I). CG is funded by the MICINN project PID2022-141079NB-C32. IFAE is partially funded by the CERCA program of the Generalitat de Catalunya. AD acknowledges support from an ERC Consolidator Grant (No. 770935).

The authors also acknowledge the computing resources at MareNostrum and the technical support provided by Barcelona Supercomputing Center (RES-AECT-2024-2-0022). Technical and human support provided by DIPC Supercomputing Center is gratefully acknowledged.

GAMA is a joint European-Australasian project based around a spectroscopic campaign using the Anglo-Australian Telescope. The GAMA input catalogue is based on data taken from the Sloan Digital Sky Survey and the UKIRT Infrared Deep Sky Survey. Complementary imaging of the GAMA regions is being obtained by a number of independent survey programmes including GALEX MIS, VST KiDS, VISTA VIKING, WISE, Herschel-ATLAS, GMRT and ASKAP providing UV to radio coverage. GAMA is funded by the STFC (UK), the ARC (Australia), the AAO, and the participating institutions. The GAMA website is https://www.gama-survey.org/.

Based on observations made with ESO Telescopes at the La Silla Paranal Observatory under programme IDs 177.A-3016, 177.A-3017, 177.A-3018 and 179.A-2004, and on data products produced by the KiDS consortium. The KiDS production team acknowledges support from: Deutsche Forschungsgemeinschaft, ERC, NOVA and NWO-M grants; Target; the University of Padova, and the University Federico II (Naples).

\section*{Data Availability}

The data underlying this article will be shared on reasonable request to the corresponding authors.



\bibliographystyle{mnras}
\bibliography{Shame2x2pt} 

@ARTICLE{Bigwood:2024,
       author = {{Bigwood}, L. and {Amon}, A. and {Schneider}, A. and {Salcido}, J. and {McCarthy}, I.~G. and {Preston}, C. and {Sanchez}, D. and {Sijacki}, D. and {Schaan}, E. and {Ferraro}, S. and {Battaglia}, N. and {Chen}, A. and {Dodelson}, S. and {Roodman}, A. and {Pieres}, A. and {Fert{\'e}}, A. and {Alarcon}, A. and {Drlica-Wagner}, A. and {Choi}, A. and {Navarro-Alsina}, A. and {Campos}, A. and {Ross}, A.~J. and {Carnero Rosell}, A. and {Yin}, B. and {Yanny}, B. and {S{\'a}nchez}, C. and {Chang}, C. and {Davis}, C. and {Doux}, C. and {Gruen}, D. and {Rykoff}, E.~S. and {Huff}, E.~M. and {Sheldon}, E. and {Tarsitano}, F. and {Andrade-Oliveira}, F. and {Bernstein}, G.~M. and {Giannini}, G. and {Diehl}, H.~T. and {Huang}, H. and {Harrison}, I. and {Sevilla-Noarbe}, I. and {Tutusaus}, I. and {Elvin-Poole}, J. and {McCullough}, J. and {Zuntz}, J. and {Blazek}, J. and {DeRose}, J. and {Cordero}, J. and {Prat}, J. and {Myles}, J. and {Eckert}, K. and {Bechtol}, K. and {Herner}, K. and {Secco}, L.~F. and {Gatti}, M. and {Raveri}, M. and {Kind}, M. Carrasco and {Becker}, M.~R. and {Troxel}, M.~A. and {Jarvis}, M. and {MacCrann}, N. and {Friedrich}, O. and {Alves}, O. and {Leget}, P. -F. and {Chen}, R. and {Rollins}, R.~P. and {Wechsler}, R.~H. and {Gruendl}, R.~A. and {Cawthon}, R. and {Allam}, S. and {Bridle}, S.~L. and {Pandey}, S. and {Everett}, S. and {Shin}, T. and {Hartley}, W.~G. and {Fang}, X. and {Zhang}, Y. and {Aguena}, M. and {Annis}, J. and {Bacon}, D. and {Bertin}, E. and {Bocquet}, S. and {Brooks}, D. and {Carretero}, J. and {Castander}, F.~J. and {da Costa}, L.~N. and {Pereira}, M.~E.~S. and {De Vicente}, J. and {Desai}, S. and {Doel}, P. and {Ferrero}, I. and {Flaugher}, B. and {Frieman}, J. and {Garc{\'\i}a-Bellido}, J. and {Gaztanaga}, E. and {Gutierrez}, G. and {Hinton}, S.~R. and {Hollowood}, D.~L. and {Honscheid}, K. and {Huterer}, D. and {James}, D.~J. and {Kuehn}, K. and {Lahav}, O. and {Lee}, S. and {Marshall}, J.~L. and {Mena-Fern{\'a}ndez}, J. and {Miquel}, R. and {Muir}, J. and {Paterno}, M. and {Plazas Malag{\'o}n}, A.~A. and {Porredon}, A. and {Romer}, A.~K. and {Samuroff}, S. and {Sanchez}, E. and {Sanchez Cid}, D. and {Smith}, M. and {Soares-Santos}, M. and {Suchyta}, E. and {Swanson}, M.~E.~C. and {Tarle}, G. and {To}, C. and {Weaverdyck}, N. and {Weller}, J. and {Wiseman}, P. and {Yamamoto}, M.},
        title = "{Weak lensing combined with the kinetic Sunyaev-Zel'dovich effect: a study of baryonic feedback}",
      journal = {\mnras},
         year = 2024,
        month = oct,
       volume = {534},
       number = {1},
        pages = {655-682},
          doi = {10.1093/mnras/stae2100},
archivePrefix = {arXiv},
       eprint = {2404.06098},
 primaryClass = {astro-ph.CO},
       adsurl = {https://ui.adsabs.harvard.edu/abs/2024MNRAS.534..655B}
}

@ARTICLE{2019A&A...624A..30J,
       author = {{Johnston}, Harry and {Georgiou}, Christos and {Joachimi}, Benjamin and {Hoekstra}, Henk and {Chisari}, Nora Elisa and {Farrow}, Daniel and {Fortuna}, Maria Cristina and {Heymans}, Catherine and {Joudaki}, Shahab and {Kuijken}, Konrad and {Wright}, Angus},
        title = "{KiDS+GAMA: Intrinsic alignment model constraints for current and future weak lensing cosmology}",
      journal = {\aap},
         year = 2019,
        month = apr,
       volume = {624},
          eid = {A30},
        pages = {A30},
          doi = {10.1051/0004-6361/201834714},
archivePrefix = {arXiv},
       eprint = {1811.09598},
 primaryClass = {astro-ph.CO},
       adsurl = {https://ui.adsabs.harvard.edu/abs/2019A&A...624A..30J}
}

@ARTICLE{Lange:2019,
       author = {{Lange}, Johannes U. and {van den Bosch}, Frank C. and {Zentner}, Andrew R. and {Wang}, Kuan and {Hearin}, Andrew P. and {Guo}, Hong},
        title = "{Cosmological Evidence Modelling: a new simulation-based approach to constrain cosmology on non-linear scales}",
      journal = {\mnras},
         year = 2019,
        month = dec,
       volume = {490},
       number = {2},
        pages = {1870-1878},
          doi = {10.1093/mnras/stz2664},
archivePrefix = {arXiv},
       eprint = {1909.03107},
 primaryClass = {astro-ph.CO},
       adsurl = {https://ui.adsabs.harvard.edu/abs/2019MNRAS.490.1870L}
}

@ARTICLE{2021arXiv210513548K,
       author = {{Krause}, E. and {Fang}, X. and {Pandey}, S. and {Secco}, L.~F. and {Alves}, O. and {Huang}, H. and {Blazek}, J. and {Prat}, J. and {Zuntz}, J. and {Eifler}, T.~F. and {MacCrann}, N. and {DeRose}, J. and {Crocce}, M. and {Porredon}, A. and {Jain}, B. and {Troxel}, M.~A. and {Dodelson}, S. and {Huterer}, D. and {Liddle}, A.~R. and {Leonard}, C.~D. and {Amon}, A. and {Chen}, A. and {Elvin-Poole}, J. and {Fert{\'e}}, A. and {Muir}, J. and {Park}, Y. and {Samuroff}, S. and {Brandao-Souza}, A. and {Weaverdyck}, N. and {Zacharegkas}, G. and {Rosenfeld}, R. and {Campos}, A. and {Chintalapati}, P. and {Choi}, A. and {Di Valentino}, E. and {Doux}, C. and {Herner}, K. and {Lemos}, P. and {Mena-Fern{\'a}ndez}, J. and {Omori}, Y. and {Paterno}, M. and {Rodriguez-Monroy}, M. and {Rogozenski}, P. and {Rollins}, R.~P. and {Troja}, A. and {Tutusaus}, I. and {Wechsler}, R.~H. and {Abbott}, T.~M.~C. and {Aguena}, M. and {Allam}, S. and {Andrade-Oliveira}, F. and {Annis}, J. and {Bacon}, D. and {Baxter}, E. and {Bechtol}, K. and {Bernstein}, G.~M. and {Brooks}, D. and {Buckley-Geer}, E. and {Burke}, D.~L. and {Carnero Rosell}, A. and {Carrasco Kind}, M. and {Carretero}, J. and {Castander}, F.~J. and {Cawthon}, R. and {Chang}, C. and {Costanzi}, M. and {da Costa}, L.~N. and {Pereira}, M.~E.~S. and {De Vicente}, J. and {Desai}, S. and {Diehl}, H.~T. and {Doel}, P. and {Everett}, S. and {Evrard}, A.~E. and {Ferrero}, I. and {Flaugher}, B. and {Fosalba}, P. and {Frieman}, J. and {Garc{\'\i}a-Bellido}, J. and {Gaztanaga}, E. and {Gerdes}, D.~W. and {Giannantonio}, T. and {Gruen}, D. and {Gruendl}, R.~A. and {Gschwend}, J. and {Gutierrez}, G. and {Hartley}, W.~G. and {Hinton}, S.~R. and {Hollowood}, D.~L. and {Honscheid}, K. and {Hoyle}, B. and {Huff}, E.~M. and {James}, D.~J. and {Kuehn}, K. and {Kuropatkin}, N. and {Lahav}, O. and {Lima}, M. and {Maia}, M.~A.~G. and {Marshall}, J.~L. and {Martini}, P. and {Melchior}, P. and {Menanteau}, F. and {Miquel}, R. and {Mohr}, J.~J. and {Morgan}, R. and {Myles}, J. and {Palmese}, A. and {Paz-Chinch{\'o}n}, F. and {Petravick}, D. and {Pieres}, A. and {Plazas Malag{\'o}n}, A.~A. and {Sanchez}, E. and {Scarpine}, V. and {Schubnell}, M. and {Serrano}, S. and {Sevilla-Noarbe}, I. and {Smith}, M. and {Soares-Santos}, M. and {Suchyta}, E. and {Tarle}, G. and {Thomas}, D. and {To}, C. and {Varga}, T.~N. and {Weller}, J.},
        title = "{Dark Energy Survey Year 3 Results: Multi-Probe Modeling Strategy and Validation}",
      journal = {arXiv e-prints},
         year = 2021,
        month = may,
          eid = {arXiv:2105.13548},
        pages = {arXiv:2105.13548},
          doi = {10.48550/arXiv.2105.13548},
archivePrefix = {arXiv},
       eprint = {2105.13548},
 primaryClass = {astro-ph.CO},
       adsurl = {https://ui.adsabs.harvard.edu/abs/2021arXiv210513548K}
}

@ARTICLE{2016MNRAS.460.3100C,
       author = {{Chaves-Montero}, Jon{\'a}s and {Angulo}, Raul E. and {Schaye}, Joop and {Schaller}, Matthieu and {Crain}, Robert A. and {Furlong}, Michelle and {Theuns}, Tom},
        title = "{Subhalo abundance matching and assembly bias in the EAGLE simulation}",
      journal = {\mnras},
         year = 2016,
        month = aug,
       volume = {460},
       number = {3},
        pages = {3100-3118},
          doi = {10.1093/mnras/stw1225},
archivePrefix = {arXiv},
       eprint = {1507.01948},
 primaryClass = {astro-ph.GA},
       adsurl = {https://ui.adsabs.harvard.edu/abs/2016MNRAS.460.3100C}
}

@ARTICLE{2021MNRAS.508..175C,
       author = {{Contreras}, S. and {Angulo}, R.~E. and {Zennaro}, M.},
        title = "{A flexible subhalo abundance matching model for galaxy clustering in redshift space}",
      journal = {\mnras},
         year = 2021,
        month = nov,
       volume = {508},
       number = {1},
        pages = {175-189},
          doi = {10.1093/mnras/stab2560},
archivePrefix = {arXiv},
       eprint = {2012.06596},
 primaryClass = {astro-ph.CO},
       adsurl = {https://ui.adsabs.harvard.edu/abs/2021MNRAS.508..175C}
}

@ARTICLE{contreras_lensing,
       author = {{Contreras}, Sergio and {Angulo}, Raul E. and {Chaves-Montero}, Jon{\'a}s and {White}, Simon D.~M. and {Aric{\`o}}, Giovanni},
        title = "{Consistent and simultaneous modelling of galaxy clustering and galaxy-galaxy lensing with subhalo abundance matching}",
      journal = {\mnras},
         year = 2023,
        month = mar,
       volume = {520},
       number = {1},
        pages = {489-502},
          doi = {10.1093/mnras/stad122},
archivePrefix = {arXiv},
       eprint = {2211.11745},
 primaryClass = {astro-ph.CO},
       adsurl = {https://ui.adsabs.harvard.edu/abs/2023MNRAS.520..489C}
}

@ARTICLE{contreras_lensing_data2023,
       author = {{Contreras}, Sergio and {Chaves-Montero}, Jon{\'a}s and {Angulo}, Raul E.},
        title = "{Consistent clustering and lensing of SDSS-III BOSS galaxies with an extended abundance matching formalism}",
      journal = {\mnras},
         year = 2023,
        month = oct,
       volume = {525},
       number = {2},
        pages = {3149-3161},
          doi = {10.1093/mnras/stad2434},
archivePrefix = {arXiv},
       eprint = {2305.09637},
 primaryClass = {astro-ph.CO},
       adsurl = {https://ui.adsabs.harvard.edu/abs/2023MNRAS.525.3149C}
}

@article{Guo:2013voa,
    author = "Guo, Qi and White, Simon",
    title = "{Numerical resolution limits on subhalo abundance matching}",
    eprint = "1303.3586",
    archivePrefix = "arXiv",
    primaryClass = "astro-ph.CO",
    doi = "10.1093/mnras/stt2116",
    journal = "Mon. Not. Roy. Astron. Soc.",
    volume = "437",
    number = "4",
    pages = "3228--3235",
    year = "2014"
}

@article{Contreras:2020zgw,
       author = {{Contreras}, S. and {Angulo}, R.~E. and {Zennaro}, M.},
        title = "{A flexible modelling of galaxy assembly bias}",
      journal = {\mnras},
         year = 2021,
        month = jul,
       volume = {504},
       number = {4},
        pages = {5205-5220},
          doi = {10.1093/mnras/stab1170},
archivePrefix = {arXiv},
       eprint = {2005.03672},
 primaryClass = {astro-ph.GA},
       adsurl = {https://ui.adsabs.harvard.edu/abs/2021MNRAS.504.5205C}
}

@article{Dvornik:2022xap,
    author = "Dvornik, Andrej and others",
    title = "{KiDS-1000: Combined halo-model cosmology constraints from galaxy abundance, galaxy clustering and galaxy-galaxy lensing}",
    eprint = "2210.03110",
    archivePrefix = "arXiv",
    primaryClass = "astro-ph.CO",
    doi = "10.1051/0004-6361/202245158",
    journal = "Astron. Astrophys.",
    volume = "675",
    pages = "A189",
    year = "2023",
    note = "[Erratum: Astron.Astrophys. 688, C3 (2024)]"
}

@article{Driver:2022vyh,
    author = "Driver, Simon P. and others",
    title = "{Galaxy And Mass Assembly (GAMA): Data Release 4 and the z \ensuremath{<} 0.1 total and z \ensuremath{<} 0.08 morphological galaxy stellar mass functions}",
    eprint = "2203.08539",
    archivePrefix = "arXiv",
    primaryClass = "astro-ph.GA",
    doi = "10.1093/mnras/stac472",
    journal = "Mon. Not. Roy. Astron. Soc.",
    volume = "513",
    number = "1",
    pages = "439--467",
    year = "2022"
}

@ARTICLE{Planck,
       author = {{Planck Collaboration} and {Aghanim}, N. and {Akrami}, Y. and {Ashdown}, M. and {Aumont}, J. and {Baccigalupi}, C. and {Ballardini}, M. and {Banday}, A.~J. and {Barreiro}, R.~B. and {Bartolo}, N. and {Basak}, S. and {Battye}, R. and {Benabed}, K. and {Bernard}, J. -P. and {Bersanelli}, M. and {Bielewicz}, P. and {Bock}, J.~J. and {Bond}, J.~R. and {Borrill}, J. and {Bouchet}, F.~R. and {Boulanger}, F. and {Bucher}, M. and {Burigana}, C. and {Butler}, R.~C. and {Calabrese}, E. and {Cardoso}, J. -F. and {Carron}, J. and {Challinor}, A. and {Chiang}, H.~C. and {Chluba}, J. and {Colombo}, L.~P.~L. and {Combet}, C. and {Contreras}, D. and {Crill}, B.~P. and {Cuttaia}, F. and {de Bernardis}, P. and {de Zotti}, G. and {Delabrouille}, J. and {Delouis}, J. -M. and {Di Valentino}, E. and {Diego}, J.~M. and {Dor{\'e}}, O. and {Douspis}, M. and {Ducout}, A. and {Dupac}, X. and {Dusini}, S. and {Efstathiou}, G. and {Elsner}, F. and {En{\ss}lin}, T.~A. and {Eriksen}, H.~K. and {Fantaye}, Y. and {Farhang}, M. and {Fergusson}, J. and {Fernandez-Cobos}, R. and {Finelli}, F. and {Forastieri}, F. and {Frailis}, M. and {Fraisse}, A.~A. and {Franceschi}, E. and {Frolov}, A. and {Galeotta}, S. and {Galli}, S. and {Ganga}, K. and {G{\'e}nova-Santos}, R.~T. and {Gerbino}, M. and {Ghosh}, T. and {Gonz{\'a}lez-Nuevo}, J. and {G{\'o}rski}, K.~M. and {Gratton}, S. and {Gruppuso}, A. and {Gudmundsson}, J.~E. and {Hamann}, J. and {Handley}, W. and {Hansen}, F.~K. and {Herranz}, D. and {Hildebrandt}, S.~R. and {Hivon}, E. and {Huang}, Z. and {Jaffe}, A.~H. and {Jones}, W.~C. and {Karakci}, A. and {Keih{\"a}nen}, E. and {Keskitalo}, R. and {Kiiveri}, K. and {Kim}, J. and {Kisner}, T.~S. and {Knox}, L. and {Krachmalnicoff}, N. and {Kunz}, M. and {Kurki-Suonio}, H. and {Lagache}, G. and {Lamarre}, J. -M. and {Lasenby}, A. and {Lattanzi}, M. and {Lawrence}, C.~R. and {Le Jeune}, M. and {Lemos}, P. and {Lesgourgues}, J. and {Levrier}, F. and {Lewis}, A. and {Liguori}, M. and {Lilje}, P.~B. and {Lilley}, M. and {Lindholm}, V. and {L{\'o}pez-Caniego}, M. and {Lubin}, P.~M. and {Ma}, Y. -Z. and {Mac{\'\i}as-P{\'e}rez}, J.~F. and {Maggio}, G. and {Maino}, D. and {Mandolesi}, N. and {Mangilli}, A. and {Marcos-Caballero}, A. and {Maris}, M. and {Martin}, P.~G. and {Martinelli}, M. and {Mart{\'\i}nez-Gonz{\'a}lez}, E. and {Matarrese}, S. and {Mauri}, N. and {McEwen}, J.~D. and {Meinhold}, P.~R. and {Melchiorri}, A. and {Mennella}, A. and {Migliaccio}, M. and {Millea}, M. and {Mitra}, S. and {Miville-Desch{\^e}nes}, M. -A. and {Molinari}, D. and {Montier}, L. and {Morgante}, G. and {Moss}, A. and {Natoli}, P. and {N{\o}rgaard-Nielsen}, H.~U. and {Pagano}, L. and {Paoletti}, D. and {Partridge}, B. and {Patanchon}, G. and {Peiris}, H.~V. and {Perrotta}, F. and {Pettorino}, V. and {Piacentini}, F. and {Polastri}, L. and {Polenta}, G. and {Puget}, J. -L. and {Rachen}, J.~P. and {Reinecke}, M. and {Remazeilles}, M. and {Renzi}, A. and {Rocha}, G. and {Rosset}, C. and {Roudier}, G. and {Rubi{\~n}o-Mart{\'\i}n}, J.~A. and {Ruiz-Granados}, B. and {Salvati}, L. and {Sandri}, M. and {Savelainen}, M. and {Scott}, D. and {Shellard}, E.~P.~S. and {Sirignano}, C. and {Sirri}, G. and {Spencer}, L.~D. and {Sunyaev}, R. and {Suur-Uski}, A. -S. and {Tauber}, J.~A. and {Tavagnacco}, D. and {Tenti}, M. and {Toffolatti}, L. and {Tomasi}, M. and {Trombetti}, T. and {Valenziano}, L. and {Valiviita}, J. and {Van Tent}, B. and {Vibert}, L. and {Vielva}, P. and {Villa}, F. and {Vittorio}, N. and {Wandelt}, B.~D. and {Wehus}, I.~K. and {White}, M. and {White}, S.~D.~M. and {Zacchei}, A. and {Zonca}, A.},
        title = "{Planck 2018 results. VI. Cosmological parameters}",
      journal = {\aap},
         year = 2020,
        month = sep,
       volume = {641},
          eid = {A6},
        pages = {A6},
          doi = {10.1051/0004-6361/201833910},
archivePrefix = {arXiv},
       eprint = {1807.06209},
 primaryClass = {astro-ph.CO},
       adsurl = {https://ui.adsabs.harvard.edu/abs/2020A&A...641A...6P}
}

@article{Amon:2022azi,
    author = "Amon, Alexandra and Efstathiou, George",
    title = "{A non-linear solution to the $S_8$ tension?}",
    eprint = "2206.11794",
    archivePrefix = "arXiv",
    primaryClass = "astro-ph.CO",
    doi = "10.1093/mnras/stac2429",
    journal = "Mon. Not. Roy. Astron. Soc.",
    volume = "516",
    number = "4",
    pages = "5355--5366",
    year = "2022"
}

@ARTICLE{KiDS-DR4,
       author = {{Kuijken}, K. and {Heymans}, C. and {Dvornik}, A. and {Hildebrandt}, H. and {de Jong}, J.~T.~A. and {Wright}, A.~H. and {Erben}, T. and {Bilicki}, M. and {Giblin}, B. and {Shan}, H. -Y. and {Getman}, F. and {Grado}, A. and {Hoekstra}, H. and {Miller}, L. and {Napolitano}, N. and {Paolilo}, M. and {Radovich}, M. and {Schneider}, P. and {Sutherland}, W. and {Tewes}, M. and {Tortora}, C. and {Valentijn}, E.~A. and {Verdoes Kleijn}, G.~A.},
        title = "{The fourth data release of the Kilo-Degree Survey: ugri imaging and nine-band optical-IR photometry over 1000 square degrees}",
      journal = {\aap},
         year = 2019,
        month = may,
       volume = {625},
          eid = {A2},
        pages = {A2},
          doi = {10.1051/0004-6361/201834918},
archivePrefix = {arXiv},
       eprint = {1902.11265},
 primaryClass = {astro-ph.GA},
       adsurl = {https://ui.adsabs.harvard.edu/abs/2019A&A...625A...2K}
}

@article{McCarthy:2016mry,
    author = "McCarthy, Ian G. and Schaye, Joop and Bird, Simeon and Le Brun, Amandine M. C.",
    title = "{The BAHAMAS project: Calibrated hydrodynamical simulations for large-scale structure cosmology}",
    eprint = "1603.02702",
    archivePrefix = "arXiv",
    primaryClass = "astro-ph.CO",
    doi = "10.1093/mnras/stw2792",
    journal = "Mon. Not. Roy. Astron. Soc.",
    volume = "465",
    number = "3",
    pages = "2936--2965",
    year = "2017"
}

@article{Bilicki:2021hgn,
    author = "Bilicki, M. and others",
    title = "{Bright galaxy sample in the Kilo-Degree Survey Data Release 4 - Selection, photometric redshifts, and physical properties}",
    eprint = "2101.06010",
    archivePrefix = "arXiv",
    primaryClass = "astro-ph.GA",
    doi = "10.1051/0004-6361/202140352",
    journal = "Astron. Astrophys.",
    volume = "653",
    pages = "A82",
    year = "2021"
}

@article{Yang:2007pg,
    author = "Yang, Xiaohu and Mo, H. J. and Bosch, Frank C. van den",
    title = "{Galaxy Groups in the SDSS DR4. 2. Halo occupation statistics}",
    eprint = "0710.5096",
    archivePrefix = "arXiv",
    primaryClass = "astro-ph",
    doi = "10.1086/528954",
    journal = "Astrophys. J.",
    volume = "676",
    pages = "248--261",
    year = "2008"
}

@ARTICLE{2013BoschCacciato,
       author = {\VAN{Bosch}{Van}{van} {den Bosch}, Frank C. and {More}, Surhud and {Cacciato}, Marcello and
         {Mo}, Houjun and {Yang}, Xiaohu},
        title = "{Cosmological constraints from a combination of galaxy clustering and lensing - I. Theoretical framework}",
      journal = {\mnras},
         year = 2013,
        month = apr,
       volume = {430},
       number = {2},
        pages = {725-746},
archivePrefix = {arXiv},
       eprint = {1206.6890},
 primaryClass = {astro-ph.CO},
       adsurl = {https://ui.adsabs.harvard.edu/abs/2013MNRAS.430..725V}
}

@article{Cacciato2013MNRAS.430..767C,
       author = {{Cacciato}, Marcello and {van den Bosch}, Frank C. and {More}, Surhud and
         {Mo}, Houjun and {Yang}, Xiaohu},
        title = "{Cosmological constraints from a combination of galaxy clustering and lensing - III. Application to SDSS data}",
      journal = {\mnras},
         year = "2013",
        month = "Apr",
       volume = {430},
       number = {2},
        pages = {767-786},
archivePrefix = {arXiv},
       eprint = {1207.0503},
 primaryClass = {astro-ph.CO},
       adsurl = {https://ui.adsabs.harvard.edu/abs/2013MNRAS.430..767C}
}

@article{Cacciato:2008hm,
    author = "Cacciato, Marcello and Bosch, Frank C. van den and More, Surhud and Li, Ran and Mo, H. J. and Yang, Xiaohu",
    title = "{Galaxy Clustering \& Galaxy-Galaxy Lensing: A Promising Union to Constrain Cosmological Parameters}",
    eprint = "0807.4932",
    archivePrefix = "arXiv",
    primaryClass = "astro-ph",
    doi = "10.1111/j.1365-2966.2008.14362.x",
    journal = "Mon. Not. Roy. Astron. Soc.",
    volume = "394",
    pages = "929--946",
    year = "2009"
}

@ARTICLE{More2013,
       author = {{More}, Surhud and {van den Bosch}, Frank C. and {Cacciato}, Marcello and {More}, Anupreeta and {Mo}, Houjun and {Yang}, Xiaohu},
        title = "{Cosmological constraints from a combination of galaxy clustering and lensing - II. Fisher matrix analysis}",
      journal = {\mnras},
         year = 2013,
        month = apr,
       volume = {430},
       number = {2},
        pages = {747-766},
          doi = {10.1093/mnras/sts697},
archivePrefix = {arXiv},
       eprint = {1207.0004},
 primaryClass = {astro-ph.CO},
       adsurl = {https://ui.adsabs.harvard.edu/abs/2013MNRAS.430..747M}
}

@ARTICLE{Wang2013,
       author = {{Wang}, L. and {Farrah}, D. and {Oliver}, S.~J. and {Amblard}, A. and {B{\'e}thermin}, M. and {Bock}, J. and {Conley}, A. and {Cooray}, A. and {Halpern}, M. and {Heinis}, S. and {Ibar}, E. and {Ilbert}, O. and {Ivison}, R.~J. and {Marsden}, G. and {Roseboom}, I.~G. and {Rowan-Robinson}, M. and {Schulz}, B. and {Smith}, A.~J. and {Viero}, M. and {Zemcov}, M.},
        title = "{Connecting stellar mass and star-formation rate to dark matter halo mass out to z {\textasciitilde} 2}",
      journal = {\mnras},
         year = 2013,
        month = may,
       volume = {431},
       number = {1},
        pages = {648-661},
          doi = {10.1093/mnras/stt190},
archivePrefix = {arXiv},
       eprint = {1203.5828},
 primaryClass = {astro-ph.CO},
       adsurl = {https://ui.adsabs.harvard.edu/abs/2013MNRAS.431..648W}
}

@article{Wright:2024qvd,
    author = "Wright, Angus H. and others",
    title = "{The fifth data release of the Kilo Degree Survey: Multi-epoch optical/NIR imaging covering wide and legacy-calibration fields}",
    eprint = "2503.19439",
    archivePrefix = "arXiv",
    primaryClass = "astro-ph.GA",
    doi = "10.1051/0004-6361/202346730",
    journal = "Astron. Astrophys.",
    volume = "686",
    pages = "A170",
    year = "2024"
}

@article{Nelson:2018uso,
    author = "Nelson, Dylan and others",
    title = "{The IllustrisTNG simulations: public data release}",
    eprint = "1812.05609",
    archivePrefix = "arXiv",
    primaryClass = "astro-ph.GA",
    doi = "10.1186/s40668-019-0028-x",
    journal = "Comput. Astrophys. Cosmol.",
    volume = "6",
    number = "1",
    pages = "2",
    year = "2019"
}

@article{nautilus,
    author = {Lange, Johannes U},
    title = "{nautilus: boosting Bayesian importance nested sampling with deep learning}",
    journal = {Monthly Notices of the Royal Astronomical Society},
    volume = {525},
    number = {2},
    pages = {3181-3194},
    year = {2023},
    month = {08},
    doi = {10.1093/mnras/stad2441},
    url = {https://doi.org/10.1093/mnras/stad2441},
    eprint = {https://academic.oup.com/mnras/article-pdf/525/2/3181/51331635/stad2441.pdf},
}

@article{Mahony:2022emy,
    author = "Mahony, Constance and Dvornik, Andrej and Mead, Alexander and Heymans, Catherine and Asgari, Marika and Hildebrandt, Hendrik and Miyatake, Hironao and Nishimichi, Takahiro and Reischke, Robert",
    title = "{The halo model with beyond-linear halo bias: unbiasing cosmological constraints from galaxy\textendash{}galaxy lensing and clustering}",
    eprint = "2202.01790",
    archivePrefix = "arXiv",
    primaryClass = "astro-ph.CO",
    reportNumber = "YITP-22-07",
    doi = "10.1093/mnras/stac1858",
    journal = "Mon. Not. Roy. Astron. Soc.",
    volume = "515",
    number = "2",
    pages = "2612--2623",
    year = "2022"
}

@article{Wright:2025xka,
       author = {{Wright}, Angus H. and {St{\"o}lzner}, Benjamin and {Asgari}, Marika and {Bilicki}, Maciej and {Giblin}, Benjamin and {Heymans}, Catherine and {Hildebrandt}, Hendrik and {Hoekstra}, Henk and {Joachimi}, Benjamin and {Kuijken}, Konrad and {Li}, Shun-Sheng and {Reischke}, Robert and {von Wietersheim-Kramsta}, Maximilian and {Yoon}, Mijin and {Burger}, Pierre and {Chisari}, Nora Elisa and {de Jong}, Jelte and {Dvornik}, Andrej and {Georgiou}, Christos and {Harnois-D{\'e}raps}, Joachim and {Jalan}, Priyanka and {William}, Anjitha John and {Joudaki}, Shahab and {Lesci}, Giorgio Francesco and {Linke}, Laila and {Loureiro}, Arthur and {Mahony}, Constance and {Maturi}, Matteo and {Miller}, Lance and {Moscardini}, Lauro and {Napolitano}, Nicola R. and {Porth}, Lucas and {Radovich}, Mario and {Schneider}, Peter and {Tr{\"o}ster}, Tilman and {Wittje}, Anna and {Yan}, Ziang and {Zhang}, Yun-Hao},
        title = "{KiDS-Legacy: Cosmological constraints from cosmic shear with the complete Kilo-Degree Survey}",
      journal = {arXiv e-prints},
         year = 2025,
        month = mar,
          eid = {arXiv:2503.19441},
        pages = {arXiv:2503.19441},
          doi = {10.48550/arXiv.2503.19441},
archivePrefix = {arXiv},
       eprint = {2503.19441},
 primaryClass = {astro-ph.CO},
       adsurl = {https://ui.adsabs.harvard.edu/abs/2025arXiv250319441W}
}

@article{Stolzner:2025htz,
       author = {{St{\"o}lzner}, Benjamin and {Wright}, Angus H. and {Asgari}, Marika and {Heymans}, Catherine and {Hildebrandt}, Hendrik and {Hoekstra}, Henk and {Joachimi}, Benjamin and {Kuijken}, Konrad and {Li}, Shun-Sheng and {Mahony}, Constance and {Reischke}, Robert and {Yoon}, Mijin and {Bilicki}, Maciej and {Burger}, Pierre and {Chisari}, Nora Elisa and {Dvornik}, Andrej and {Georgiou}, Christos and {Giblin}, Benjamin and {Harnois-D{\'e}raps}, Joachim and {Jalan}, Priyanka and {William}, Anjitha John and {Joudaki}, Shahab and {Lesci}, Giorgio Francesco and {Linke}, Laila and {Loureiro}, Arthur and {Maturi}, Matteo and {Moscardini}, Lauro and {Napolitano}, Nicola R. and {Porth}, Lucas and {Radovich}, Mario and {Tr{\"o}ster}, Tilman and {von Wietersheim-Kramsta}, Maximilian and {Wittje}, Anna and {Yan}, Ziang and {Zhang}, Yun-Hao},
        title = "{KiDS-Legacy: Consistency of cosmic shear measurements and joint cosmological constraints with external probes}",
      journal = {arXiv e-prints},
         year = 2025,
        month = mar,
          eid = {arXiv:2503.19442},
        pages = {arXiv:2503.19442},
          doi = {10.48550/arXiv.2503.19442},
archivePrefix = {arXiv},
       eprint = {2503.19442},
 primaryClass = {astro-ph.CO},
       adsurl = {https://ui.adsabs.harvard.edu/abs/2025arXiv250319442S}
}

@ARTICLE{Yoon2019,
       author = {{Yoon}, Mijin and {Jee}, M. James and {Tyson}, J. Anthony and {Schmidt}, Samuel and {Wittman}, David and {Choi}, Ami},
        title = "{Constraints on Cosmology and Baryonic Feedback with the Deep Lens Survey Using Galaxy-Galaxy and Galaxy-Mass Power Spectra}",
      journal = {\apj},
         year = 2019,
        month = jan,
       volume = {870},
       number = {2},
          eid = {111},
        pages = {111},
          doi = {10.3847/1538-4357/aaf3a9},
archivePrefix = {arXiv},
       eprint = {1807.09195},
 primaryClass = {astro-ph.CO},
       adsurl = {https://ui.adsabs.harvard.edu/abs/2019ApJ...870..111Y}
}

@ARTICLE{YoonJee2021,
       author = {{Yoon}, Mijin and {Jee}, M. James},
        title = "{Baryonic Feedback Measurement From KV450 Cosmic Shear Analysis}",
      journal = {\apj},
         year = 2021,
        month = feb,
       volume = {908},
       number = {1},
          eid = {13},
        pages = {13},
          doi = {10.3847/1538-4357/abcd9e},
archivePrefix = {arXiv},
       eprint = {2007.16166},
 primaryClass = {astro-ph.CO},
       adsurl = {https://ui.adsabs.harvard.edu/abs/2021ApJ...908...13Y}
}

@ARTICLE{Preston2023,
       author = {{Preston}, Calvin and {Amon}, Alexandra and {Efstathiou}, George},
        title = "{A non-linear solution to the S$_{8}$ tension - II. Analysis of DES Year 3 cosmic shear}",
      journal = {\mnras},
         year = 2023,
        month = nov,
       volume = {525},
       number = {4},
        pages = {5554-5564},
          doi = {10.1093/mnras/stad2573},
archivePrefix = {arXiv},
       eprint = {2305.09827},
 primaryClass = {astro-ph.CO},
       adsurl = {https://ui.adsabs.harvard.edu/abs/2023MNRAS.525.5554P}
}

@article{Heymans:2020gsg,
    author = "Heymans, Catherine and others",
    title = "{KiDS-1000 Cosmology: Multi-probe weak gravitational lensing and spectroscopic galaxy clustering constraints}",
    eprint = "2007.15632",
    archivePrefix = "arXiv",
    primaryClass = "astro-ph.CO",
    doi = "10.1051/0004-6361/202039063",
    journal = "Astron. Astrophys.",
    volume = "646",
    pages = "A140",
    year = "2021"
}

@ARTICLE{DES3x2pt,
       author = {{Abbott}, T.~M.~C. and {Aguena}, M. and {Alarcon}, A. and {Allam}, S. and {Alves}, O. and {Amon}, A. and {Andrade-Oliveira}, F. and {Annis}, J. and {Avila}, S. and {Bacon}, D. and {Baxter}, E. and {Bechtol}, K. and {Becker}, M.~R. and {Bernstein}, G.~M. and {Bhargava}, S. and {Birrer}, S. and {Blazek}, J. and {Brandao-Souza}, A. and {Bridle}, S.~L. and {Brooks}, D. and {Buckley-Geer}, E. and {Burke}, D.~L. and {Camacho}, H. and {Campos}, A. and {Carnero Rosell}, A. and {Carrasco Kind}, M. and {Carretero}, J. and {Castander}, F.~J. and {Cawthon}, R. and {Chang}, C. and {Chen}, A. and {Chen}, R. and {Choi}, A. and {Conselice}, C. and {Cordero}, J. and {Costanzi}, M. and {Crocce}, M. and {da Costa}, L.~N. and {da Silva Pereira}, M.~E. and {Davis}, C. and {Davis}, T.~M. and {De Vicente}, J. and {DeRose}, J. and {Desai}, S. and {Di Valentino}, E. and {Diehl}, H.~T. and {Dietrich}, J.~P. and {Dodelson}, S. and {Doel}, P. and {Doux}, C. and {Drlica-Wagner}, A. and {Eckert}, K. and {Eifler}, T.~F. and {Elsner}, F. and {Elvin-Poole}, J. and {Everett}, S. and {Evrard}, A.~E. and {Fang}, X. and {Farahi}, A. and {Fernandez}, E. and {Ferrero}, I. and {Fert{\'e}}, A. and {Fosalba}, P. and {Friedrich}, O. and {Frieman}, J. and {Garc{\'\i}a-Bellido}, J. and {Gatti}, M. and {Gaztanaga}, E. and {Gerdes}, D.~W. and {Giannantonio}, T. and {Giannini}, G. and {Gruen}, D. and {Gruendl}, R.~A. and {Gschwend}, J. and {Gutierrez}, G. and {Harrison}, I. and {Hartley}, W.~G. and {Herner}, K. and {Hinton}, S.~R. and {Hollowood}, D.~L. and {Honscheid}, K. and {Hoyle}, B. and {Huff}, E.~M. and {Huterer}, D. and {Jain}, B. and {James}, D.~J. and {Jarvis}, M. and {Jeffrey}, N. and {Jeltema}, T. and {Kovacs}, A. and {Krause}, E. and {Kron}, R. and {Kuehn}, K. and {Kuropatkin}, N. and {Lahav}, O. and {Leget}, P. -F. and {Lemos}, P. and {Liddle}, A.~R. and {Lidman}, C. and {Lima}, M. and {Lin}, H. and {MacCrann}, N. and {Maia}, M.~A.~G. and {Marshall}, J.~L. and {Martini}, P. and {McCullough}, J. and {Melchior}, P. and {Mena-Fern{\'a}ndez}, J. and {Menanteau}, F. and {Miquel}, R. and {Mohr}, J.~J. and {Morgan}, R. and {Muir}, J. and {Myles}, J. and {Nadathur}, S. and {Navarro-Alsina}, A. and {Nichol}, R.~C. and {Ogando}, R.~L.~C. and {Omori}, Y. and {Palmese}, A. and {Pandey}, S. and {Park}, Y. and {Paz-Chinch{\'o}n}, F. and {Petravick}, D. and {Pieres}, A. and {Plazas Malag{\'o}n}, A.~A. and {Porredon}, A. and {Prat}, J. and {Raveri}, M. and {Rodriguez-Monroy}, M. and {Rollins}, R.~P. and {Romer}, A.~K. and {Roodman}, A. and {Rosenfeld}, R. and {Ross}, A.~J. and {Rykoff}, E.~S. and {Samuroff}, S. and {S{\'a}nchez}, C. and {Sanchez}, E. and {Sanchez}, J. and {Sanchez Cid}, D. and {Scarpine}, V. and {Schubnell}, M. and {Scolnic}, D. and {Secco}, L.~F. and {Serrano}, S. and {Sevilla-Noarbe}, I. and {Sheldon}, E. and {Shin}, T. and {Smith}, M. and {Soares-Santos}, M. and {Suchyta}, E. and {Swanson}, M.~E.~C. and {Tabbutt}, M. and {Tarle}, G. and {Thomas}, D. and {To}, C. and {Troja}, A. and {Troxel}, M.~A. and {Tucker}, D.~L. and {Tutusaus}, I. and {Varga}, T.~N. and {Walker}, A.~R. and {Weaverdyck}, N. and {Wechsler}, R. and {Weller}, J. and {Yanny}, B. and {Yin}, B. and {Zhang}, Y. and {Zuntz}, J. and {DES Collaboration}},
        title = "{Dark Energy Survey Year 3 results: Cosmological constraints from galaxy clustering and weak lensing}",
      journal = {\prd},
         year = 2022,
        month = jan,
       volume = {105},
       number = {2},
          eid = {023520},
        pages = {023520},
          doi = {10.1103/PhysRevD.105.023520},
archivePrefix = {arXiv},
       eprint = {2105.13549},
 primaryClass = {astro-ph.CO},
       adsurl = {https://ui.adsabs.harvard.edu/abs/2022PhRvD.105b3520A}
}

@ARTICLE{HSC3x2pt2023,
       author = {{Miyatake}, Hironao and {Sugiyama}, Sunao and {Takada}, Masahiro and {Nishimichi}, Takahiro and {Li}, Xiangchong and {Shirasaki}, Masato and {More}, Surhud and {Kobayashi}, Yosuke and {Nishizawa}, Atsushi J. and {Rau}, Markus M. and {Zhang}, Tianqing and {Takahashi}, Ryuichi and {Dalal}, Roohi and {Mandelbaum}, Rachel and {Strauss}, Michael A. and {Hamana}, Takashi and {Oguri}, Masamune and {Osato}, Ken and {Luo}, Wentao and {Kannawadi}, Arun and {Hsieh}, Bau-Ching and {Armstrong}, Robert and {Bosch}, James and {Komiyama}, Yutaka and {Lupton}, Robert H. and {Lust}, Nate B. and {MacArthur}, Lauren A. and {Miyazaki}, Satoshi and {Murayama}, Hitoshi and {Okura}, Yuki and {Price}, Paul A. and {Sunayama}, Tomomi and {Tait}, Philip J. and {Tanaka}, Masayuki and {Wang}, Shiang-Yu},
        title = "{Hyper Suprime-Cam Year 3 results: Cosmology from galaxy clustering and weak lensing with HSC and SDSS using the emulator based halo model}",
      journal = {\prd},
         year = 2023,
        month = dec,
       volume = {108},
       number = {12},
          eid = {123517},
        pages = {123517},
          doi = {10.1103/PhysRevD.108.123517},
archivePrefix = {arXiv},
       eprint = {2304.00704},
 primaryClass = {astro-ph.CO},
       adsurl = {https://ui.adsabs.harvard.edu/abs/2023PhRvD.108l3517M}
}

@ARTICLE{Euclid2025,
       author = {{Euclid Collaboration} and {Mellier}, Y. and {Abdurro'uf} and {Acevedo Barroso}, J.~A. and {Ach{\'u}carro}, A. and {Adamek}, J. and {Adam}, R. and {Addison}, G.~E. and {Aghanim}, N. and {Aguena}, M. and {Ajani}, V. and {Akrami}, Y. and {Al-Bahlawan}, A. and {Alavi}, A. and {Albuquerque}, I.~S. and {Alestas}, G. and {Alguero}, G. and {Allaoui}, A. and {Allen}, S.~W. and {Allevato}, V. and {Alonso-Tetilla}, A.~V. and {Altieri}, B. and {Alvarez-Candal}, A. and {Alvi}, S. and {Amara}, A. and {Amendola}, L. and {Amiaux}, J. and {Andika}, I.~T. and {Andreon}, S. and {Andrews}, A. and {Angora}, G. and {Angulo}, R.~E. and {Annibali}, F. and {Anselmi}, A. and {Anselmi}, S. and {Arcari}, S. and {Archidiacono}, M. and {Aric{\`o}}, G. and {Arnaud}, M. and {Arnouts}, S. and {Asgari}, M. and {Asorey}, J. and {Atayde}, L. and {Atek}, H. and {Atrio-Barandela}, F. and {Aubert}, M. and {Aubourg}, E. and {Auphan}, T. and {Auricchio}, N. and {Aussel}, B. and {Aussel}, H. and {Avelino}, P.~P. and {Avgoustidis}, A. and {Avila}, S. and {Awan}, S. and {Azzollini}, R. and {Baccigalupi}, C. and {Bachelet}, E. and {Bacon}, D. and {Baes}, M. and {Bagley}, M.~B. and {Bahr-Kalus}, B. and {Balaguera-Antolinez}, A. and {Balbinot}, E. and {Balcells}, M. and {Baldi}, M. and {Baldry}, I. and {Balestra}, A. and {Ballardini}, M. and {Ballester}, O. and {Balogh}, M. and {Ba{\~n}ados}, E. and {Barbier}, R. and {Bardelli}, S. and {Baron}, M. and {Barreiro}, T. and {Barrena}, R. and {Barriere}, J. -C. and {Barros}, B.~J. and {Barthelemy}, A. and {Bartolo}, N. and {Basset}, A. and {Battaglia}, P. and {Battisti}, A.~J. and {Baugh}, C.~M. and {Baumont}, L. and {Bazzanini}, L. and {Beaulieu}, J. -P. and {Beckmann}, V. and {Belikov}, A.~N. and {Bel}, J. and {Bellagamba}, F. and {Bella}, M. and {Bellini}, E. and {Benabed}, K. and {Bender}, R. and {Benevento}, G. and {Bennett}, C.~L. and {Benson}, K. and {Bergamini}, P. and {Bermejo-Climent}, J.~R. and {Bernardeau}, F. and {Bertacca}, D. and {Berthe}, M. and {Berthier}, J. and {Bethermin}, M. and {Beutler}, F. and {Bevillon}, C. and {Bhargava}, S. and {Bhatawdekar}, R. and {Bianchi}, D. and {Bisigello}, L. and {Biviano}, A. and {Blake}, R.~P. and {Blanchard}, A. and {Blazek}, J. and {Blot}, L. and {Bosco}, A. and {Bodendorf}, C. and {Boenke}, T. and {B{\"o}hringer}, H. and {Boldrini}, P. and {Bolzonella}, M. and {Bonchi}, A. and {Bonici}, M. and {Bonino}, D. and {Bonino}, L. and {Bonvin}, C. and {Bon}, W. and {Booth}, J.~T. and {Borgani}, S. and {Borlaff}, A.~S. and {Borsato}, E. and {Bose}, B. and {Botticella}, M.~T. and {Boucaud}, A. and {Bouche}, F. and {Boucher}, J.~S. and {Boutigny}, D. and {Bouvard}, T. and {Bouwens}, R. and {Bouy}, H. and {Bowler}, R.~A.~A. and {Bozza}, V. and {Bozzo}, E. and {Branchini}, E. and {Brando}, G. and {Brau-Nogue}, S. and {Brekke}, P. and {Bremer}, M.~N. and {Brescia}, M. and {Breton}, M. -A. and {Brinchmann}, J. and {Brinckmann}, T. and {Brockley-Blatt}, C. and {Brodwin}, M. and {Brouard}, L. and {Brown}, M.~L. and {Bruton}, S. and {Bucko}, J. and {Buddelmeijer}, H. and {Buenadicha}, G. and {Buitrago}, F. and {Burger}, P. and {Burigana}, C. and {Busillo}, V. and {Busonero}, D. and {Cabanac}, R. and {Cabayol-Garcia}, L. and {Cagliari}, M.~S. and {Caillat}, A. and {Caillat}, L. and {Calabrese}, M. and {Calabro}, A. and {Calderone}, G. and {Calura}, F. and {Camacho Quevedo}, B. and {Camera}, S. and {Campos}, L. and {Ca{\~n}as-Herrera}, G. and {Candini}, G.~P. and {Cantiello}, M. and {Capobianco}, V. and {Cappellaro}, E. and {Cappelluti}, N. and {Cappi}, A. and {Caputi}, K.~I. and {Cara}, C. and {Carbone}, C. and {Cardone}, V.~F. and {Carella}, E. and {Carlberg}, R.~G. and {Carle}, M. and {Carminati}, L. and {Caro}, F. and {Carrasco}, J.~M. and {Carretero}, J. and {Carrilho}, P. and {Carron Duque}, J. and {Carry}, B.},
        title = "{Euclid: I. Overview of the Euclid mission}",
      journal = {\aap},
         year = 2025,
        month = may,
       volume = {697},
          eid = {A1},
        pages = {A1},
          doi = {10.1051/0004-6361/202450810},
archivePrefix = {arXiv},
       eprint = {2405.13491},
 primaryClass = {astro-ph.CO},
       adsurl = {https://ui.adsabs.harvard.edu/abs/2025A&A...697A...1E}
}

@ARTICLE{rubin2019,
       author = {{Ivezi{\'c}}, {\v{Z}}eljko and {Kahn}, Steven M. and {Tyson}, J. Anthony and {Abel}, Bob and {Acosta}, Emily and {Allsman}, Robyn and {Alonso}, David and {AlSayyad}, Yusra and {Anderson}, Scott F. and {Andrew}, John and {Angel}, James Roger P. and {Angeli}, George Z. and {Ansari}, Reza and {Antilogus}, Pierre and {Araujo}, Constanza and {Armstrong}, Robert and {Arndt}, Kirk T. and {Astier}, Pierre and {Aubourg}, {\'E}ric and {Auza}, Nicole and {Axelrod}, Tim S. and {Bard}, Deborah J. and {Barr}, Jeff D. and {Barrau}, Aurelian and {Bartlett}, James G. and {Bauer}, Amanda E. and {Bauman}, Brian J. and {Baumont}, Sylvain and {Bechtol}, Ellen and {Bechtol}, Keith and {Becker}, Andrew C. and {Becla}, Jacek and {Beldica}, Cristina and {Bellavia}, Steve and {Bianco}, Federica B. and {Biswas}, Rahul and {Blanc}, Guillaume and {Blazek}, Jonathan and {Blandford}, Roger D. and {Bloom}, Josh S. and {Bogart}, Joanne and {Bond}, Tim W. and {Booth}, Michael T. and {Borgland}, Anders W. and {Borne}, Kirk and {Bosch}, James F. and {Boutigny}, Dominique and {Brackett}, Craig A. and {Bradshaw}, Andrew and {Brandt}, William Nielsen and {Brown}, Michael E. and {Bullock}, James S. and {Burchat}, Patricia and {Burke}, David L. and {Cagnoli}, Gianpietro and {Calabrese}, Daniel and {Callahan}, Shawn and {Callen}, Alice L. and {Carlin}, Jeffrey L. and {Carlson}, Erin L. and {Chandrasekharan}, Srinivasan and {Charles-Emerson}, Glenaver and {Chesley}, Steve and {Cheu}, Elliott C. and {Chiang}, Hsin-Fang and {Chiang}, James and {Chirino}, Carol and {Chow}, Derek and {Ciardi}, David R. and {Claver}, Charles F. and {Cohen-Tanugi}, Johann and {Cockrum}, Joseph J. and {Coles}, Rebecca and {Connolly}, Andrew J. and {Cook}, Kem H. and {Cooray}, Asantha and {Covey}, Kevin R. and {Cribbs}, Chris and {Cui}, Wei and {Cutri}, Roc and {Daly}, Philip N. and {Daniel}, Scott F. and {Daruich}, Felipe and {Daubard}, Guillaume and {Daues}, Greg and {Dawson}, William and {Delgado}, Francisco and {Dellapenna}, Alfred and {de Peyster}, Robert and {de Val-Borro}, Miguel and {Digel}, Seth W. and {Doherty}, Peter and {Dubois}, Richard and {Dubois-Felsmann}, Gregory P. and {Durech}, Josef and {Economou}, Frossie and {Eifler}, Tim and {Eracleous}, Michael and {Emmons}, Benjamin L. and {Fausti Neto}, Angelo and {Ferguson}, Henry and {Figueroa}, Enrique and {Fisher-Levine}, Merlin and {Focke}, Warren and {Foss}, Michael D. and {Frank}, James and {Freemon}, Michael D. and {Gangler}, Emmanuel and {Gawiser}, Eric and {Geary}, John C. and {Gee}, Perry and {Geha}, Marla and {Gessner}, Charles J.~B. and {Gibson}, Robert R. and {Gilmore}, D. Kirk and {Glanzman}, Thomas and {Glick}, William and {Goldina}, Tatiana and {Goldstein}, Daniel A. and {Goodenow}, Iain and {Graham}, Melissa L. and {Gressler}, William J. and {Gris}, Philippe and {Guy}, Leanne P. and {Guyonnet}, Augustin and {Haller}, Gunther and {Harris}, Ron and {Hascall}, Patrick A. and {Haupt}, Justine and {Hernandez}, Fabio and {Herrmann}, Sven and {Hileman}, Edward and {Hoblitt}, Joshua and {Hodgson}, John A. and {Hogan}, Craig and {Howard}, James D. and {Huang}, Dajun and {Huffer}, Michael E. and {Ingraham}, Patrick and {Innes}, Walter R. and {Jacoby}, Suzanne H. and {Jain}, Bhuvnesh and {Jammes}, Fabrice and {Jee}, M. James and {Jenness}, Tim and {Jernigan}, Garrett and {Jevremovi{\'c}}, Darko and {Johns}, Kenneth and {Johnson}, Anthony S. and {Johnson}, Margaret W.~G. and {Jones}, R. Lynne and {Juramy-Gilles}, Claire and {Juri{\'c}}, Mario and {Kalirai}, Jason S. and {Kallivayalil}, Nitya J. and {Kalmbach}, Bryce and {Kantor}, Jeffrey P. and {Karst}, Pierre and {Kasliwal}, Mansi M. and {Kelly}, Heather and {Kessler}, Richard and {Kinnison}, Veronica and {Kirkby}, David and {Knox}, Lloyd and {Kotov}, Ivan V. and {Krabbendam}, Victor L. and {Krughoff}, K. Simon and {Kub{\'a}nek}, Petr and {Kuczewski}, John and {Kulkarni}, Shri and {Ku}, John and {Kurita}, Nadine R. and {Lage}, Craig S. and {Lambert}, Ron and {Lange}, Travis and {Langton}, J. Brian and {Le Guillou}, Laurent and {Levine}, Deborah and {Liang}, Ming and {Lim}, Kian-Tat and {Lintott}, Chris J. and {Long}, Kevin E. and {Lopez}, Margaux and {Lotz}, Paul J. and {Lupton}, Robert H. and {Lust}, Nate B. and {MacArthur}, Lauren A. and {Mahabal}, Ashish and {Mandelbaum}, Rachel and {Markiewicz}, Thomas W. and {Marsh}, Darren S. and {Marshall}, Philip J. and {Marshall}, Stuart and {May}, Morgan and {McKercher}, Robert and {McQueen}, Michelle and {Meyers}, Joshua and {Migliore}, Myriam and {Miller}, Michelle and {Mills}, David J.},
        title = "{LSST: From Science Drivers to Reference Design and Anticipated Data Products}",
      journal = {\apj},
         year = 2019,
        month = mar,
       volume = {873},
       number = {2},
          eid = {111},
        pages = {111},
          doi = {10.3847/1538-4357/ab042c},
archivePrefix = {arXiv},
       eprint = {0805.2366},
 primaryClass = {astro-ph},
       adsurl = {https://ui.adsabs.harvard.edu/abs/2019ApJ...873..111I}
}

@article{Nishimichi:2018etk,
    author = "Nishimichi, Takahiro and others",
    title = "{Dark Quest. I. Fast and Accurate Emulation of Halo Clustering Statistics and Its Application to Galaxy Clustering}",
    eprint = "1811.09504",
    archivePrefix = "arXiv",
    primaryClass = "astro-ph.CO",
    reportNumber = "YITP-19-73",
    doi = "10.3847/1538-4357/ab3719",
    journal = "Astrophys. J.",
    volume = "884",
    pages = "29",
    year = "2019"
}

@article{Miyatake:2020uhg,
    author = "Miyatake, Hironao and Kobayashi, Yosuke and Takada, Masahiro and Nishimichi, Takahiro and Shirasaki, Masato and Sugiyama, Sunao and Takahashi, Ryuichi and Osato, Ken and More, Surhud and Park, Youngsoo",
    title = "{Cosmological inference from an emulator based halo model. I. Validation tests with HSC and SDSS mock catalogs}",
    eprint = "2101.00113",
    archivePrefix = "arXiv",
    primaryClass = "astro-ph.CO",
    reportNumber = "IPMU21-0002, YITP-21-03",
    doi = "10.1103/PhysRevD.106.083519",
    journal = "Phys. Rev. D",
    volume = "106",
    number = "8",
    pages = "083519",
    year = "2022"
}

@article{Mead:2020qdk,
    author = "Mead, A. J. and Verde, L.",
    title = "{Including beyond-linear halo bias in halo models}",
    eprint = "2011.08858",
    archivePrefix = "arXiv",
    primaryClass = "astro-ph.CO",
    doi = "10.1093/mnras/stab748",
    journal = "Mon. Not. Roy. Astron. Soc.",
    volume = "503",
    number = "2",
    pages = "3095--3111",
    year = "2021"
}

@article{Smith:2002dz,
    author = "Smith, R. E. and Peacock, J. A. and Jenkins, A. and White, S. D. M. and Frenk, C. S. and Pearce, F. R. and Thomas, P. A. and Efstathiou, G. and Couchmann, H. M. P.",
    collaboration = "VIRGO Consortium",
    title = "{Stable clustering, the halo model and nonlinear cosmological power spectra}",
    eprint = "astro-ph/0207664",
    archivePrefix = "arXiv",
    doi = "10.1046/j.1365-8711.2003.06503.x",
    journal = "Mon. Not. Roy. Astron. Soc.",
    volume = "341",
    pages = "1311",
    year = "2003"
}

@article{Takahashi:2012em,
    author = "Takahashi, Ryuichi and Sato, Masanori and Nishimichi, Takahiro and Taruya, Atsushi and Oguri, Masamune",
    title = "{Revising the Halofit Model for the Nonlinear Matter Power Spectrum}",
    eprint = "1208.2701",
    archivePrefix = "arXiv",
    primaryClass = "astro-ph.CO",
    doi = "10.1088/0004-637X/761/2/152",
    journal = "Astrophys. J.",
    volume = "761",
    pages = "152",
    year = "2012"
}

@ARTICLE{HMCode,
       author = {{Mead}, A.~J. and {Brieden}, S. and {Tr{\"o}ster}, T. and {Heymans}, C.},
        title = "{HMCODE-2020: improved modelling of non-linear cosmological power spectra with baryonic feedback}",
      journal = {\mnras},
         year = 2021,
        month = mar,
       volume = {502},
       number = {1},
        pages = {1401-1422},
          doi = {10.1093/mnras/stab082},
archivePrefix = {arXiv},
       eprint = {2009.01858},
 primaryClass = {astro-ph.CO},
       adsurl = {https://ui.adsabs.harvard.edu/abs/2021MNRAS.502.1401M}
}

@ARTICLE{contreras2024,
       author = {{Contreras}, Sergio and {Angulo}, Raul E. and {Chaves-Montero}, Jon{\'a}s and {Kugel}, Roi and {Schaller}, Matthieu and {Schaye}, Joop},
        title = "{Validating the clustering predictions of empirical models with the FLAMINGO simulations}",
      journal = {\aap},
         year = 2024,
        month = oct,
       volume = {690},
          eid = {A311},
        pages = {A311},
          doi = {10.1051/0004-6361/202451671},
archivePrefix = {arXiv},
       eprint = {2407.18912},
 primaryClass = {astro-ph.GA},
       adsurl = {https://ui.adsabs.harvard.edu/abs/2024A&A...690A.311C}
}

@ARTICLE{Gao2005,
       author = {{Gao}, Liang and {Springel}, Volker and {White}, Simon D.~M.},
        title = "{The age dependence of halo clustering}",
      journal = {\mnras},
         year = 2005,
        month = oct,
       volume = {363},
       number = {1},
        pages = {L66-L70},
          doi = {10.1111/j.1745-3933.2005.00084.x},
archivePrefix = {arXiv},
       eprint = {astro-ph/0506510},
 primaryClass = {astro-ph},
       adsurl = {https://ui.adsabs.harvard.edu/abs/2005MNRAS.363L..66G}
}

@ARTICLE{Wechsler2006,
       author = {{Wechsler}, Risa H. and {Zentner}, Andrew R. and {Bullock}, James S. and {Kravtsov}, Andrey V. and {Allgood}, Brandon},
        title = "{The Dependence of Halo Clustering on Halo Formation History, Concentration, and Occupation}",
      journal = {\apj},
         year = 2006,
        month = nov,
       volume = {652},
       number = {1},
        pages = {71-84},
          doi = {10.1086/507120},
archivePrefix = {arXiv},
       eprint = {astro-ph/0512416},
 primaryClass = {astro-ph},
       adsurl = {https://ui.adsabs.harvard.edu/abs/2006ApJ...652...71W}
}

@ARTICLE{Dalal2008,
       author = {{Dalal}, Neal and {White}, Martin and {Bond}, J. Richard and {Shirokov}, Alexander},
        title = "{Halo Assembly Bias in Hierarchical Structure Formation}",
      journal = {\apj},
         year = 2008,
        month = nov,
       volume = {687},
       number = {1},
        pages = {12-21},
          doi = {10.1086/591512},
archivePrefix = {arXiv},
       eprint = {0803.3453},
 primaryClass = {astro-ph},
       adsurl = {https://ui.adsabs.harvard.edu/abs/2008ApJ...687...12D}
}

@ARTICLE{Contreras2023_MTNG,
       author = {{Contreras}, Sergio and {Angulo}, Raul E. and {Springel}, Volker and {White}, Simon D.~M. and {Hadzhiyska}, Boryana and {Hernquist}, Lars and {Pakmor}, R{\"u}diger and {Kannan}, Rahul and {Hern{\'a}ndez-Aguayo}, C{\'e}sar and {Barrera}, Monica and {Ferlito}, Fulvio and {Delgado}, Ana Maria and {Bose}, Sownak and {Frenk}, Carlos},
        title = "{The MillenniumTNG Project: inferring cosmology from galaxy clustering with accelerated N-body scaling and subhalo abundance matching}",
      journal = {\mnras},
         year = 2023,
        month = sep,
       volume = {524},
       number = {2},
        pages = {2489-2506},
          doi = {10.1093/mnras/stac3699},
archivePrefix = {arXiv},
       eprint = {2210.10075},
 primaryClass = {astro-ph.GA},
       adsurl = {https://ui.adsabs.harvard.edu/abs/2023MNRAS.524.2489C}
}

@ARTICLE{Contreras2020_Scaling,
       author = {{Contreras}, S. and {Angulo}, R.~E. and {Zennaro}, M. and {Aric{\`o}}, G. and {Pellejero-Iba{\~n}ez}, M.},
        title = "{3 per cent-accurate predictions for the clustering of dark matter, haloes, and subhaloes, over a wide range of cosmologies and scales}",
      journal = {\mnras},
         year = 2020,
        month = dec,
       volume = {499},
       number = {4},
        pages = {4905-4917},
          doi = {10.1093/mnras/staa3117},
archivePrefix = {arXiv},
       eprint = {2001.03176},
 primaryClass = {astro-ph.CO},
       adsurl = {https://ui.adsabs.harvard.edu/abs/2020MNRAS.499.4905C}
}

@ARTICLE{Angulo2016_IC,
       author = {{Angulo}, Raul E. and {Pontzen}, Andrew},
        title = "{Cosmological N-body simulations with suppressed variance}",
      journal = {\mnras},
         year = 2016,
        month = oct,
       volume = {462},
       number = {1},
        pages = {L1-L5},
          doi = {10.1093/mnrasl/slw098},
archivePrefix = {arXiv},
       eprint = {1603.05253},
 primaryClass = {astro-ph.CO},
       adsurl = {https://ui.adsabs.harvard.edu/abs/2016MNRAS.462L...1A}
}

@ARTICLE{Angulo2010_scale,
       author = {{Angulo}, R.~E. and {White}, S.~D.~M.},
        title = "{One simulation to fit them all - changing the background parameters of a cosmological N-body simulation}",
      journal = {\mnras},
         year = 2010,
        month = jun,
       volume = {405},
       number = {1},
        pages = {143-154},
          doi = {10.1111/j.1365-2966.2010.16459.x},
archivePrefix = {arXiv},
       eprint = {0912.4277},
 primaryClass = {astro-ph.CO},
       adsurl = {https://ui.adsabs.harvard.edu/abs/2010MNRAS.405..143A}
}

@ARTICLE{Angulo2021_bacco,
       author = {{Angulo}, Raul E. and {Zennaro}, Matteo and {Contreras}, Sergio and {Aric{\`o}}, Giovanni and {Pellejero-Iba{\~n}ez}, Marcos and {St{\"u}cker}, Jens},
        title = "{The BACCO simulation project: exploiting the full power of large-scale structure for cosmology}",
      journal = {\mnras},
         year = 2021,
        month = nov,
       volume = {507},
       number = {4},
        pages = {5869-5881},
          doi = {10.1093/mnras/stab2018},
archivePrefix = {arXiv},
       eprint = {2004.06245},
 primaryClass = {astro-ph.CO},
       adsurl = {https://ui.adsabs.harvard.edu/abs/2021MNRAS.507.5869A}
}

@misc{tensorflow,
title={ {TensorFlow}: Large-Scale Machine Learning on Heterogeneous Systems},
url={https://www.tensorflow.org/},
note={Software available from tensorflow.org},
author={
    Mart\'{\i}n~Abadi and
    Ashish~Agarwal and
    Paul~Barham and
    Eugene~Brevdo and
    Zhifeng~Chen and
    Craig~Citro and
    Greg~S.~Corrado and
    Andy~Davis and
    Jeffrey~Dean and
    Matthieu~Devin and
    Sanjay~Ghemawat and
    Ian~Goodfellow and
    Andrew~Harp and
    Geoffrey~Irving and
    Michael~Isard and
    Yangqing Jia and
    Rafal~Jozefowicz and
    Lukasz~Kaiser and
    Manjunath~Kudlur and
    Josh~Levenberg and
    Dandelion~Man\'{e} and
    Rajat~Monga and
    Sherry~Moore and
    Derek~Murray and
    Chris~Olah and
    Mike~Schuster and
    Jonathon~Shlens and
    Benoit~Steiner and
    Ilya~Sutskever and
    Kunal~Talwar and
    Paul~Tucker and
    Vincent~Vanhoucke and
    Vijay~Vasudevan and
    Fernanda~Vi\'{e}gas and
    Oriol~Vinyals and
    Pete~Warden and
    Martin~Wattenberg and
    Martin~Wicke and
    Yuan~Yu and
    Xiaoqiang~Zheng},
  year={2015},
}

@ARTICLE{Contreras2021_AssemblyBias,
       author = {{Contreras}, S. and {Angulo}, R.~E. and {Zennaro}, M.},
        title = "{A flexible modelling of galaxy assembly bias}",
      journal = {\mnras},
         year = 2021,
        month = jul,
       volume = {504},
       number = {4},
        pages = {5205-5220},
          doi = {10.1093/mnras/stab1170},
archivePrefix = {arXiv},
       eprint = {2005.03672},
 primaryClass = {astro-ph.GA},
       adsurl = {https://ui.adsabs.harvard.edu/abs/2021MNRAS.504.5205C}
}

@ARTICLE{Paranjape2018,
       author = {{Paranjape}, Aseem and {Hahn}, Oliver and {Sheth}, Ravi K.},
        title = "{Halo assembly bias and the tidal anisotropy of the local halo environment}",
      journal = {\mnras},
         year = 2018,
        month = may,
       volume = {476},
       number = {3},
        pages = {3631-3647},
          doi = {10.1093/mnras/sty496},
archivePrefix = {arXiv},
       eprint = {1706.09906},
 primaryClass = {astro-ph.CO},
       adsurl = {https://ui.adsabs.harvard.edu/abs/2018MNRAS.476.3631P}
}

@ARTICLE{Croton:2007,
   author = {{Croton}, D.~J. and {Gao}, L. and {White}, S.~D.~M.},
    title = "{Halo assembly bias and its effects on galaxy clustering}",
  journal = {\mnras},
   eprint = {astro-ph/0605636},
     year = 2007,
    month = feb,
   volume = 374,
    pages = {1303-1309},
      doi = {10.1111/j.1365-2966.2006.11230.x},
   adsurl = {http://adsabs.harvard.edu/abs/2007MNRAS.374.1303C}
}

@article{Sinha:2019reo,
    author = "Sinha, Manodeep and Garrison, Lehman H.",
    title = "{corrfunc \textendash{} a suite of blazing fast correlation functions on the CPU}",
    eprint = "1911.03545",
    archivePrefix = "arXiv",
    primaryClass = "astro-ph.CO",
    doi = "10.1093/mnras/stz3157",
    journal = "Mon. Not. Roy. Astron. Soc.",
    volume = "491",
    number = "2",
    pages = "3022--3041",
    year = "2020"
}

@ARTICLE{VIKING,
       author = {{Edge}, A. and {Sutherland}, W. and {Kuijken}, K. and {Driver}, S. and {McMahon}, R. and {Eales}, S. and {Emerson}, J.~P.},
        title = "{The VISTA Kilo-degree Infrared Galaxy (VIKING) Survey: Bridging the Gap between Low and High Redshift}",
      journal = {The Messenger},
         year = 2013,
        month = dec,
       volume = {154},
        pages = {32-34},
       adsurl = {https://ui.adsabs.harvard.edu/abs/2013Msngr.154...32E}
}

@ARTICLE{Hildebrandt,
       author = {{Hildebrandt}, H. and {van den Busch}, J.~L. and {Wright}, A.~H. and {Blake}, C. and {Joachimi}, B. and {Kuijken}, K. and {Tr{\"o}ster}, T. and {Asgari}, M. and {Bilicki}, M. and {de Jong}, J.~T.~A. and {Dvornik}, A. and {Erben}, T. and {Getman}, F. and {Giblin}, B. and {Heymans}, C. and {Kannawadi}, A. and {Lin}, C. -A. and {Shan}, H. -Y.},
        title = "{KiDS-1000 catalogue: Redshift distributions and their calibration}",
      journal = {\aap},
         year = 2021,
        month = mar,
       volume = {647},
          eid = {A124},
        pages = {A124},
          doi = {10.1051/0004-6361/202039018},
archivePrefix = {arXiv},
       eprint = {2007.15635},
 primaryClass = {astro-ph.CO},
       adsurl = {https://ui.adsabs.harvard.edu/abs/2021A&A...647A.124H}
}

@ARTICLE{Giblin,
       author = {{Giblin}, Benjamin and {Heymans}, Catherine and {Asgari}, Marika and {Hildebrandt}, Hendrik and {Hoekstra}, Henk and {Joachimi}, Benjamin and {Kannawadi}, Arun and {Kuijken}, Konrad and {Lin}, Chieh-An and {Miller}, Lance and {Tr{\"o}ster}, Tilman and {van den Busch}, Jan Luca and {Wright}, Angus H. and {Bilicki}, Maciej and {Blake}, Chris and {de Jong}, Jelte and {Dvornik}, Andrej and {Erben}, Thomas and {Getman}, Fedor and {Napolitano}, Nicola R. and {Schneider}, Peter and {Shan}, HuanYuan and {Valentijn}, Edwin},
        title = "{KiDS-1000 catalogue: Weak gravitational lensing shear measurements}",
      journal = {\aap},
         year = 2021,
        month = jan,
       volume = {645},
          eid = {A105},
        pages = {A105},
          doi = {10.1051/0004-6361/202038850},
archivePrefix = {arXiv},
       eprint = {2007.01845},
 primaryClass = {astro-ph.CO},
       adsurl = {https://ui.adsabs.harvard.edu/abs/2021A&A...645A.105G}
}

@ARTICLE{LePhare1,
       author = {{Arnouts}, S. and {Cristiani}, S. and {Moscardini}, L. and {Matarrese}, S. and {Lucchin}, F. and {Fontana}, A. and {Giallongo}, E.},
        title = "{Measuring and modelling the redshift evolution of clustering: the Hubble Deep Field North}",
      journal = {\mnras},
         year = 1999,
        month = dec,
       volume = {310},
       number = {2},
        pages = {540-556},
          doi = {10.1046/j.1365-8711.1999.02978.x},
archivePrefix = {arXiv},
       eprint = {astro-ph/9902290},
 primaryClass = {astro-ph},
       adsurl = {https://ui.adsabs.harvard.edu/abs/1999MNRAS.310..540A}
}

@ARTICLE{LePhare2,
       author = {{Ilbert}, O. and {Arnouts}, S. and {McCracken}, H.~J. and {Bolzonella}, M. and {Bertin}, E. and {Le F{\`e}vre}, O. and {Mellier}, Y. and {Zamorani}, G. and {Pell{\`o}}, R. and {Iovino}, A. and {Tresse}, L. and {Le Brun}, V. and {Bottini}, D. and {Garilli}, B. and {Maccagni}, D. and {Picat}, J.~P. and {Scaramella}, R. and {Scodeggio}, M. and {Vettolani}, G. and {Zanichelli}, A. and {Adami}, C. and {Bardelli}, S. and {Cappi}, A. and {Charlot}, S. and {Ciliegi}, P. and {Contini}, T. and {Cucciati}, O. and {Foucaud}, S. and {Franzetti}, P. and {Gavignaud}, I. and {Guzzo}, L. and {Marano}, B. and {Marinoni}, C. and {Mazure}, A. and {Meneux}, B. and {Merighi}, R. and {Paltani}, S. and {Pollo}, A. and {Pozzetti}, L. and {Radovich}, M. and {Zucca}, E. and {Bondi}, M. and {Bongiorno}, A. and {Busarello}, G. and {de La Torre}, S. and {Gregorini}, L. and {Lamareille}, F. and {Mathez}, G. and {Merluzzi}, P. and {Ripepi}, V. and {Rizzo}, D. and {Vergani}, D.},
        title = "{Accurate photometric redshifts for the CFHT legacy survey calibrated using the VIMOS VLT deep survey}",
      journal = {\aap},
         year = 2006,
        month = oct,
       volume = {457},
       number = {3},
        pages = {841-856},
          doi = {10.1051/0004-6361:20065138},
archivePrefix = {arXiv},
       eprint = {astro-ph/0603217},
 primaryClass = {astro-ph},
       adsurl = {https://ui.adsabs.harvard.edu/abs/2006A&A...457..841I}
}

@ARTICLE{GAMA,
       author = {{Driver}, S.~P. and {Hill}, D.~T. and {Kelvin}, L.~S. and {Robotham}, A.~S.~G. and {Liske}, J. and {Norberg}, P. and {Baldry}, I.~K. and {Bamford}, S.~P. and {Hopkins}, A.~M. and {Loveday}, J. and {Peacock}, J.~A. and {Andrae}, E. and {Bland-Hawthorn}, J. and {Brough}, S. and {Brown}, M.~J.~I. and {Cameron}, E. and {Ching}, J.~H.~Y. and {Colless}, M. and {Conselice}, C.~J. and {Croom}, S.~M. and {Cross}, N.~J.~G. and {de Propris}, R. and {Dye}, S. and {Drinkwater}, M.~J. and {Ellis}, S. and {Graham}, Alister W. and {Grootes}, M.~W. and {Gunawardhana}, M. and {Jones}, D.~H. and {van Kampen}, E. and {Maraston}, C. and {Nichol}, R.~C. and {Parkinson}, H.~R. and {Phillipps}, S. and {Pimbblet}, K. and {Popescu}, C.~C. and {Prescott}, M. and {Roseboom}, I.~G. and {Sadler}, E.~M. and {Sansom}, A.~E. and {Sharp}, R.~G. and {Smith}, D.~J.~B. and {Taylor}, E. and {Thomas}, D. and {Tuffs}, R.~J. and {Wijesinghe}, D. and {Dunne}, L. and {Frenk}, C.~S. and {Jarvis}, M.~J. and {Madore}, B.~F. and {Meyer}, M.~J. and {Seibert}, M. and {Staveley-Smith}, L. and {Sutherland}, W.~J. and {Warren}, S.~J.},
        title = "{Galaxy and Mass Assembly (GAMA): survey diagnostics and core data release}",
      journal = {\mnras},
         year = 2011,
        month = may,
       volume = {413},
       number = {2},
        pages = {971-995},
          doi = {10.1111/j.1365-2966.2010.18188.x},
archivePrefix = {arXiv},
       eprint = {1009.0614},
 primaryClass = {astro-ph.CO},
       adsurl = {https://ui.adsabs.harvard.edu/abs/2011MNRAS.413..971D}
}

@ARTICLE{Georgiou,
       author = {{Georgiou}, Christos and {Chisari}, Nora Elisa and {Bilicki}, Maciej and {La Barbera}, Francesco and {Napolitano}, Nicola R. and {Roy}, Nivya and {Tortora}, Crescenzo},
        title = "{Intrinsic galaxy alignments in the KiDS-1000 bright sample: dependence on colour, luminosity, morphology and galaxy scale}",
      journal = {arXiv e-prints},
         year = 2025,
        month = feb,
          eid = {arXiv:2502.09452},
        pages = {arXiv:2502.09452},
          doi = {10.48550/arXiv.2502.09452},
archivePrefix = {arXiv},
       eprint = {2502.09452},
 primaryClass = {astro-ph.CO},
       adsurl = {https://ui.adsabs.harvard.edu/abs/2025arXiv250209452G}
}

@ARTICLE{treecorr,
       author = {{Jarvis}, M. and {Bernstein}, G. and {Jain}, B.},
        title = "{The skewness of the aperture mass statistic}",
      journal = {\mnras},
         year = 2004,
        month = jul,
       volume = {352},
       number = {1},
        pages = {338-352},
          doi = {10.1111/j.1365-2966.2004.07926.x},
archivePrefix = {arXiv},
       eprint = {astro-ph/0307393},
 primaryClass = {astro-ph},
       adsurl = {https://ui.adsabs.harvard.edu/abs/2004MNRAS.352..338J}
}

@ARTICLE{Zennaro2024,
       author = {{Zennaro}, Matteo and {Aric{\`o}}, Giovanni and {Garc{\'\i}a-Garc{\'\i}a}, Carlos and {Angulo}, Ra{\'u}l E. and {Ondaro-Mallea}, Lurdes and {Contreras}, Sergio and {Nicola}, Andrina and {Schaller}, Matthieu and {Schaye}, Joop},
        title = "{A 1\% accurate method to include baryonic effects in galaxy-galaxy lensing models}",
      journal = {arXiv e-prints},
         year = 2024,
        month = dec,
          eid = {arXiv:2412.08623},
        pages = {arXiv:2412.08623},
          doi = {10.48550/arXiv.2412.08623},
archivePrefix = {arXiv},
       eprint = {2412.08623},
 primaryClass = {astro-ph.CO},
       adsurl = {https://ui.adsabs.harvard.edu/abs/2024arXiv241208623Z}
}

@ARTICLE{Arico2021a,
       author = {{Aric{\`o}}, Giovanni and {Angulo}, Raul E. and {Contreras}, Sergio and {Ondaro-Mallea}, Lurdes and {Pellejero-Iba{\~n}ez}, Marcos and {Zennaro}, Matteo},
        title = "{The BACCO simulation project: a baryonification emulator with neural networks}",
      journal = {\mnras},
         year = 2021,
        month = sep,
       volume = {506},
       number = {3},
        pages = {4070-4082},
          doi = {10.1093/mnras/stab1911},
archivePrefix = {arXiv},
       eprint = {2011.15018},
 primaryClass = {astro-ph.CO},
       adsurl = {https://ui.adsabs.harvard.edu/abs/2021MNRAS.506.4070A}
}

@ARTICLE{2023Chavesmontero,
       author = {{Chaves-Montero}, Jon{\'a}s and {Angulo}, Raul E. and {Contreras}, Sergio},
        title = "{The galaxy formation origin of the lensing is low problem}",
      journal = {\mnras},
         year = 2023,
        month = may,
       volume = {521},
       number = {1},
        pages = {937-951},
          doi = {10.1093/mnras/stad243},
archivePrefix = {arXiv},
       eprint = {2211.01744},
 primaryClass = {astro-ph.CO},
       adsurl = {https://ui.adsabs.harvard.edu/abs/2023MNRAS.521..937C}
}

@ARTICLE{Debackere2021MNRAS.505..593D,
       author = {{Debackere}, Stijn N.~B. and {Schaye}, Joop and {Hoekstra}, Henk},
        title = "{How baryons can significantly bias cluster count cosmology}",
      journal = {\mnras},
         year = 2021,
        month = jul,
       volume = {505},
       number = {1},
        pages = {593-609},
          doi = {10.1093/mnras/stab1326},
archivePrefix = {arXiv},
       eprint = {2101.07800},
 primaryClass = {astro-ph.CO},
       adsurl = {https://ui.adsabs.harvard.edu/abs/2021MNRAS.505..593D}
}

@ARTICLE{Amon2023MNRAS.518..477A,
       author = {{Amon}, A. and {Robertson}, N.~C. and {Miyatake}, H. and {Heymans}, C. and {White}, M. and {DeRose}, J. and {Yuan}, S. and {Wechsler}, R.~H. and {Varga}, T.~N. and {Bocquet}, S. and {Dvornik}, A. and {More}, S. and {Ross}, A.~J. and {Hoekstra}, H. and {Alarcon}, A. and {Asgari}, M. and {Blazek}, J. and {Campos}, A. and {Chen}, R. and {Choi}, A. and {Crocce}, M. and {Diehl}, H.~T. and {Doux}, C. and {Eckert}, K. and {Elvin-Poole}, J. and {Everett}, S. and {Fert{\'e}}, A. and {Gatti}, M. and {Giannini}, G. and {Gruen}, D. and {Gruendl}, R.~A. and {Hartley}, W.~G. and {Herner}, K. and {Hildebrandt}, H. and {Huang}, S. and {Huff}, E.~M. and {Joachimi}, B. and {Lee}, S. and {MacCrann}, N. and {Myles}, J. and {Navarro-Alsina}, A. and {Nishimichi}, T. and {Prat}, J. and {Secco}, L.~F. and {Sevilla-Noarbe}, I. and {Sheldon}, E. and {Shin}, T. and {Tr{\"o}ster}, T. and {Troxel}, M.~A. and {Tutusaus}, I. and {Wright}, A.~H. and {Yin}, B. and {Aguena}, M. and {Allam}, S. and {Annis}, J. and {Bacon}, D. and {Bilicki}, M. and {Brooks}, D. and {Burke}, D.~L. and {Carnero Rosell}, A. and {Carretero}, J. and {Castander}, F.~J. and {Cawthon}, R. and {Costanzi}, M. and {da Costa}, L.~N. and {Pereira}, M.~E.~S. and {de Jong}, J. and {De Vicente}, J. and {Desai}, S. and {Dietrich}, J.~P. and {Doel}, P. and {Ferrero}, I. and {Frieman}, J. and {Garc{\'\i}a-Bellido}, J. and {Gerdes}, D.~W. and {Gschwend}, J. and {Gutierrez}, G. and {Hinton}, S.~R. and {Hollowood}, D.~L. and {Honscheid}, K. and {Huterer}, D. and {Kannawadi}, A. and {Kuehn}, K. and {Kuropatkin}, N. and {Lahav}, O. and {Lima}, M. and {Maia}, M.~A.~G. and {Marshall}, J.~L. and {Menanteau}, F. and {Miquel}, R. and {Mohr}, J.~J. and {Morgan}, R. and {Muir}, J. and {Paz-Chinch{\'o}n}, F. and {Pieres}, A. and {Plazas Malag{\'o}n}, A.~A. and {Porredon}, A. and {Rodriguez-Monroy}, M. and {Roodman}, A. and {Sanchez}, E. and {Serrano}, S. and {Shan}, H. and {Suchyta}, E. and {Swanson}, M.~E.~C. and {Tarle}, G. and {Thomas}, D. and {To}, C. and {Zhang}, Y.},
        title = "{Consistent lensing and clustering in a low-S$_{8}$ Universe with BOSS, DES Year 3, HSC Year 1, and KiDS-1000}",
      journal = {\mnras},
         year = 2023,
        month = jan,
       volume = {518},
       number = {1},
        pages = {477-503},
          doi = {10.1093/mnras/stac2938},
archivePrefix = {arXiv},
       eprint = {2202.07440},
 primaryClass = {astro-ph.CO},
       adsurl = {https://ui.adsabs.harvard.edu/abs/2023MNRAS.518..477A}
}

@ARTICLE{Viola2015MNRAS.452.3529V,
       author = {{Viola}, M. and {Cacciato}, M. and {Brouwer}, M. and {Kuijken}, K. and {Hoekstra}, H. and {Norberg}, P. and {Robotham}, A.~S.~G. and {van Uitert}, E. and {Alpaslan}, M. and {Baldry}, I.~K. and {Choi}, A. and {de Jong}, J.~T.~A. and {Driver}, S.~P. and {Erben}, T. and {Grado}, A. and {Graham}, Alister W. and {Heymans}, C. and {Hildebrandt}, H. and {Hopkins}, A.~M. and {Irisarri}, N. and {Joachimi}, B. and {Loveday}, J. and {Miller}, L. and {Nakajima}, R. and {Schneider}, P. and {Sif{\'o}n}, C. and {Verdoes Kleijn}, G.},
        title = "{Dark matter halo properties of GAMA galaxy groups from 100 square degrees of KiDS weak lensing data}",
      journal = {\mnras},
         year = 2015,
        month = oct,
       volume = {452},
       number = {4},
        pages = {3529-3550},
          doi = {10.1093/mnras/stv1447},
archivePrefix = {arXiv},
       eprint = {1507.00735},
 primaryClass = {astro-ph.GA},
       adsurl = {https://ui.adsabs.harvard.edu/abs/2015MNRAS.452.3529V}
}

@ARTICLE{Edo2018MNRAS.476.4662V,
       author = {{van Uitert}, Edo and {Joachimi}, Benjamin and {Joudaki}, Shahab and {Amon}, Alexandra and {Heymans}, Catherine and {K{\"o}hlinger}, Fabian and {Asgari}, Marika and {Blake}, Chris and {Choi}, Ami and {Erben}, Thomas and {Farrow}, Daniel J. and {Harnois-D{\'e}raps}, Joachim and {Hildebrandt}, Hendrik and {Hoekstra}, Henk and {Kitching}, Thomas D. and {Klaes}, Dominik and {Kuijken}, Konrad and {Merten}, Julian and {Miller}, Lance and {Nakajima}, Reiko and {Schneider}, Peter and {Valentijn}, Edwin and {Viola}, Massimo},
        title = "{KiDS+GAMA: cosmology constraints from a joint analysis of cosmic shear, galaxy-galaxy lensing, and angular clustering}",
      journal = {\mnras},
         year = 2018,
        month = jun,
       volume = {476},
       number = {4},
        pages = {4662-4689},
          doi = {10.1093/mnras/sty551},
archivePrefix = {arXiv},
       eprint = {1706.05004},
 primaryClass = {astro-ph.CO},
       adsurl = {https://ui.adsabs.harvard.edu/abs/2018MNRAS.476.4662V}
}

@ARTICLE{Dvornik2018MNRAS.479.1240D,
       author = {{Dvornik}, Andrej and {Hoekstra}, Henk and {Kuijken}, Konrad and {Schneider}, Peter and {Amon}, Alexandra and {Nakajima}, Reiko and {Viola}, Massimo and {Choi}, Ami and {Erben}, Thomas and {Farrow}, Daniel J. and {Heymans}, Catherine and {Hildebrandt}, Hendrik and {Sif{\'o}n}, Crist{\'o}bal and {Wang}, Lingyu},
        title = "{Unveiling galaxy bias via the halo model, KiDS, and GAMA}",
      journal = {\mnras},
         year = 2018,
        month = sep,
       volume = {479},
       number = {1},
        pages = {1240-1259},
          doi = {10.1093/mnras/sty1502},
archivePrefix = {arXiv},
       eprint = {1802.00734},
 primaryClass = {astro-ph.CO},
       adsurl = {https://ui.adsabs.harvard.edu/abs/2018MNRAS.479.1240D}
}

@ARTICLE{Garcia-Garcia2024JCAP...08..024G,
       author = {{Garc{\'\i}a-Garc{\'\i}a}, Carlos and {Zennaro}, Matteo and {Aric{\`o}}, Giovanni and {Alonso}, David and {Angulo}, Raul E.},
        title = "{Cosmic shear with small scales: DES-Y3, KiDS-1000 and HSC-DR1}",
      journal = {\jcap},
         year = 2024,
        month = aug,
       volume = {2024},
       number = {8},
          eid = {024},
        pages = {024},
          doi = {10.1088/1475-7516/2024/08/024},
archivePrefix = {arXiv},
       eprint = {2403.13794},
 primaryClass = {astro-ph.CO},
       adsurl = {https://ui.adsabs.harvard.edu/abs/2024JCAP...08..024G}
}

@ARTICLE{AricoDES2023A&A...678A.109A,
       author = {{Aric{\`o}}, Giovanni and {Angulo}, Raul E. and {Zennaro}, Matteo and {Contreras}, Sergio and {Chen}, Angela and {Hern{\'a}ndez-Monteagudo}, Carlos},
        title = "{DES Y3 cosmic shear down to small scales: Constraints on cosmology and baryons}",
      journal = {\aap},
         year = 2023,
        month = oct,
       volume = {678},
          eid = {A109},
        pages = {A109},
          doi = {10.1051/0004-6361/202346539},
archivePrefix = {arXiv},
       eprint = {2303.05537},
 primaryClass = {astro-ph.CO},
       adsurl = {https://ui.adsabs.harvard.edu/abs/2023A&A...678A.109A}
}

@ARTICLE{Hearin2016MNRAS.460.2552H,
       author = {{Hearin}, Andrew P. and {Zentner}, Andrew R. and {van den Bosch}, Frank C. and {Campbell}, Duncan and {Tollerud}, Erik},
        title = "{Introducing decorated HODs: modelling assembly bias in the galaxy-halo connection}",
      journal = {\mnras},
         year = 2016,
        month = aug,
       volume = {460},
       number = {3},
        pages = {2552-2570},
          doi = {10.1093/mnras/stw840},
archivePrefix = {arXiv},
       eprint = {1512.03050},
 primaryClass = {astro-ph.CO},
       adsurl = {https://ui.adsabs.harvard.edu/abs/2016MNRAS.460.2552H}
}

@ARTICLE{Nicola2024JCAP...02..015N,
       author = {{Nicola}, Andrina and {Hadzhiyska}, Boryana and {Findlay}, Nathan and {Garc{\'\i}a-Garc{\'\i}a}, Carlos and {Alonso}, David and {Slosar}, An{\v{z}}e and {Guo}, Zhiyuan and {Kokron}, Nickolas and {Angulo}, Ra{\'u}l and {Aviles}, Alejandro and {Blazek}, Jonathan and {Dunkley}, Jo and {Jain}, Bhuvnesh and {Pellejero}, Marcos and {Sullivan}, James and {Walter}, Christopher W. and {Zennaro}, Matteo and {LSST Dark Energy Science Collaboration}},
        title = "{Galaxy bias in the era of LSST: perturbative bias expansions}",
      journal = {\jcap},
         year = 2024,
        month = feb,
       volume = {2024},
       number = {2},
          eid = {015},
        pages = {015},
          doi = {10.1088/1475-7516/2024/02/015},
archivePrefix = {arXiv},
       eprint = {2307.03226},
 primaryClass = {astro-ph.CO},
       adsurl = {https://ui.adsabs.harvard.edu/abs/2024JCAP...02..015N}
}

@ARTICLE{Beutler2017MNRAS.466.2242B,
       author = {{Beutler}, Florian and {Seo}, Hee-Jong and {Saito}, Shun and {Chuang}, Chia-Hsun and {Cuesta}, Antonio J. and {Eisenstein}, Daniel J. and {Gil-Mar{\'\i}n}, H{\'e}ctor and {Grieb}, Jan Niklas and {Hand}, Nick and {Kitaura}, Francisco-Shu and {Modi}, Chirag and {Nichol}, Robert C. and {Olmstead}, Matthew D. and {Percival}, Will J. and {Prada}, Francisco and {S{\'a}nchez}, Ariel G. and {Rodriguez-Torres}, Sergio and {Ross}, Ashley J. and {Ross}, Nicholas P. and {Schneider}, Donald P. and {Tinker}, Jeremy and {Tojeiro}, Rita and {Vargas-Maga{\~n}a}, Mariana},
        title = "{The clustering of galaxies in the completed SDSS-III Baryon Oscillation Spectroscopic Survey: anisotropic galaxy clustering in Fourier space}",
      journal = {\mnras},
         year = 2017,
        month = apr,
       volume = {466},
       number = {2},
        pages = {2242-2260},
          doi = {10.1093/mnras/stw3298},
archivePrefix = {arXiv},
       eprint = {1607.03150},
 primaryClass = {astro-ph.CO},
       adsurl = {https://ui.adsabs.harvard.edu/abs/2017MNRAS.466.2242B}
}

@ARTICLE{Amico2020JCAP...05..005D,
       author = {{d'Amico}, Guido and {Gleyzes}, J{\'e}r{\^o}me and {Kokron}, Nickolas and {Markovic}, Katarina and {Senatore}, Leonardo and {Zhang}, Pierre and {Beutler}, Florian and {Gil-Mar{\'\i}n}, H{\'e}ctor},
        title = "{The cosmological analysis of the SDSS/BOSS data from the Effective Field Theory of Large-Scale Structure}",
      journal = {\jcap},
         year = 2020,
        month = may,
       volume = {2020},
       number = {5},
          eid = {005},
        pages = {005},
          doi = {10.1088/1475-7516/2020/05/005},
archivePrefix = {arXiv},
       eprint = {1909.05271},
 primaryClass = {astro-ph.CO},
       adsurl = {https://ui.adsabs.harvard.edu/abs/2020JCAP...05..005D}
}

@ARTICLE{Ivanov2020JCAP...05..042I,
       author = {{Ivanov}, Mikhail M. and {Simonovi{\'c}}, Marko and {Zaldarriaga}, Matias},
        title = "{Cosmological parameters from the BOSS galaxy power spectrum}",
      journal = {\jcap},
         year = 2020,
        month = may,
       volume = {2020},
       number = {5},
          eid = {042},
        pages = {042},
          doi = {10.1088/1475-7516/2020/05/042},
archivePrefix = {arXiv},
       eprint = {1909.05277},
 primaryClass = {astro-ph.CO},
       adsurl = {https://ui.adsabs.harvard.edu/abs/2020JCAP...05..042I}
}

@ARTICLE{Taylor2011,
       author = {{Taylor}, Edward N. and {Hopkins}, Andrew M. and {Baldry}, Ivan K. and {Brown}, Michael J.~I. and {Driver}, Simon P. and {Kelvin}, Lee S. and {Hill}, David T. and {Robotham}, Aaron S.~G. and {Bland-Hawthorn}, Joss and {Jones}, D.~H. and {Sharp}, R.~G. and {Thomas}, Daniel and {Liske}, Jochen and {Loveday}, Jon and {Norberg}, Peder and {Peacock}, J.~A. and {Bamford}, Steven P. and {Brough}, Sarah and {Colless}, Matthew and {Cameron}, Ewan and {Conselice}, Christopher J. and {Croom}, Scott M. and {Frenk}, C.~S. and {Gunawardhana}, Madusha and {Kuijken}, Konrad and {Nichol}, R.~C. and {Parkinson}, H.~R. and {Phillipps}, S. and {Pimbblet}, K.~A. and {Popescu}, C.~C. and {Prescott}, Matthew and {Sutherland}, W.~J. and {Tuffs}, R.~J. and {van Kampen}, Eelco and {Wijesinghe}, D.},
        title = "{Galaxy And Mass Assembly (GAMA): stellar mass estimates}",
      journal = {\mnras},
         year = 2011,
        month = dec,
       volume = {418},
       number = {3},
        pages = {1587-1620},
          doi = {10.1111/j.1365-2966.2011.19536.x},
archivePrefix = {arXiv},
       eprint = {1108.0635},
 primaryClass = {astro-ph.CO},
       adsurl = {https://ui.adsabs.harvard.edu/abs/2011MNRAS.418.1587T}
}

@ARTICLE{Baldry2012,
       author = {{Baldry}, I.~K. and {Driver}, S.~P. and {Loveday}, J. and {Taylor}, E.~N. and {Kelvin}, L.~S. and {Liske}, J. and {Norberg}, P. and {Robotham}, A.~S.~G. and {Brough}, S. and {Hopkins}, A.~M. and {Bamford}, S.~P. and {Peacock}, J.~A. and {Bland-Hawthorn}, J. and {Conselice}, C.~J. and {Croom}, S.~M. and {Jones}, D.~H. and {Parkinson}, H.~R. and {Popescu}, C.~C. and {Prescott}, M. and {Sharp}, R.~G. and {Tuffs}, R.~J.},
        title = "{Galaxy And Mass Assembly (GAMA): the galaxy stellar mass function at z < 0.06}",
      journal = {\mnras},
         year = 2012,
        month = mar,
       volume = {421},
       number = {1},
        pages = {621-634},
          doi = {10.1111/j.1365-2966.2012.20340.x},
archivePrefix = {arXiv},
       eprint = {1111.5707},
 primaryClass = {astro-ph.CO},
       adsurl = {https://ui.adsabs.harvard.edu/abs/2012MNRAS.421..621B}
}

@ARTICLE{GAMARandoms,
       author = {{Farrow}, D.~J. and {Cole}, Shaun and {Norberg}, Peder and {Metcalfe}, N. and {Baldry}, I. and {Bland-Hawthorn}, Joss and {Brown}, Michael J.~I. and {Hopkins}, A.~M. and {Lacey}, Cedric G. and {Liske}, J. and {Loveday}, Jon and {Palamara}, David P. and {Robotham}, A.~S.~G. and {Sridhar}, Srivatsan},
        title = "{Galaxy and mass assembly (GAMA): projected galaxy clustering}",
      journal = {\mnras},
         year = 2015,
        month = dec,
       volume = {454},
       number = {2},
        pages = {2120-2145},
          doi = {10.1093/mnras/stv2075},
archivePrefix = {arXiv},
       eprint = {1509.02159},
 primaryClass = {astro-ph.GA},
       adsurl = {https://ui.adsabs.harvard.edu/abs/2015MNRAS.454.2120F}
}




\appendix
\section{Internal inconsistency in clustering data}\label{app:clustering_inconsistency}

When analysing GAMA clustering alone to check the internal consistency of our data we found a strong constraint on $S_8$, which is not expected from clustering data alone and was inconsistent with our fiducial result. This constraint is shown in Figure \ref{app:clustering_inconsistency} (blue contour). To determine where this spurious constraint was coming from we first analysed the clustering data removing each of the four stellar mass bins in turn. We found that if we removed bin 3, where $10.5 < \mathrm{{log}}(M_{{\star}}/h^{{-2}}M_{{\odot}}) \leq 10.7$, the strong constraint on $S_8$ disappeared (orange contour). Next we varied the scales included in bin 3 and found that if we removed large scales, where $r_{\mathrm{max}}>1.5 \ h^{-1}\mathrm{Mpc}$, the strong constraint on $S_8$ also disappeared (green contour). This led us to conclude that there is likely an uncontrolled systematic effect impacting the large scales of clustering bin 3. When we removed stellar mass bin 3 from the clustering in our fiducial joint analysis of GAMA galaxy clustering and KiDS-1000 galaxy-galaxy lensing we found very little impact on our results (see Figure \ref{fig:S8_summary_analysis_variations}).
\begin{figure}
    \centering
    \includegraphics[width=1.0\linewidth]{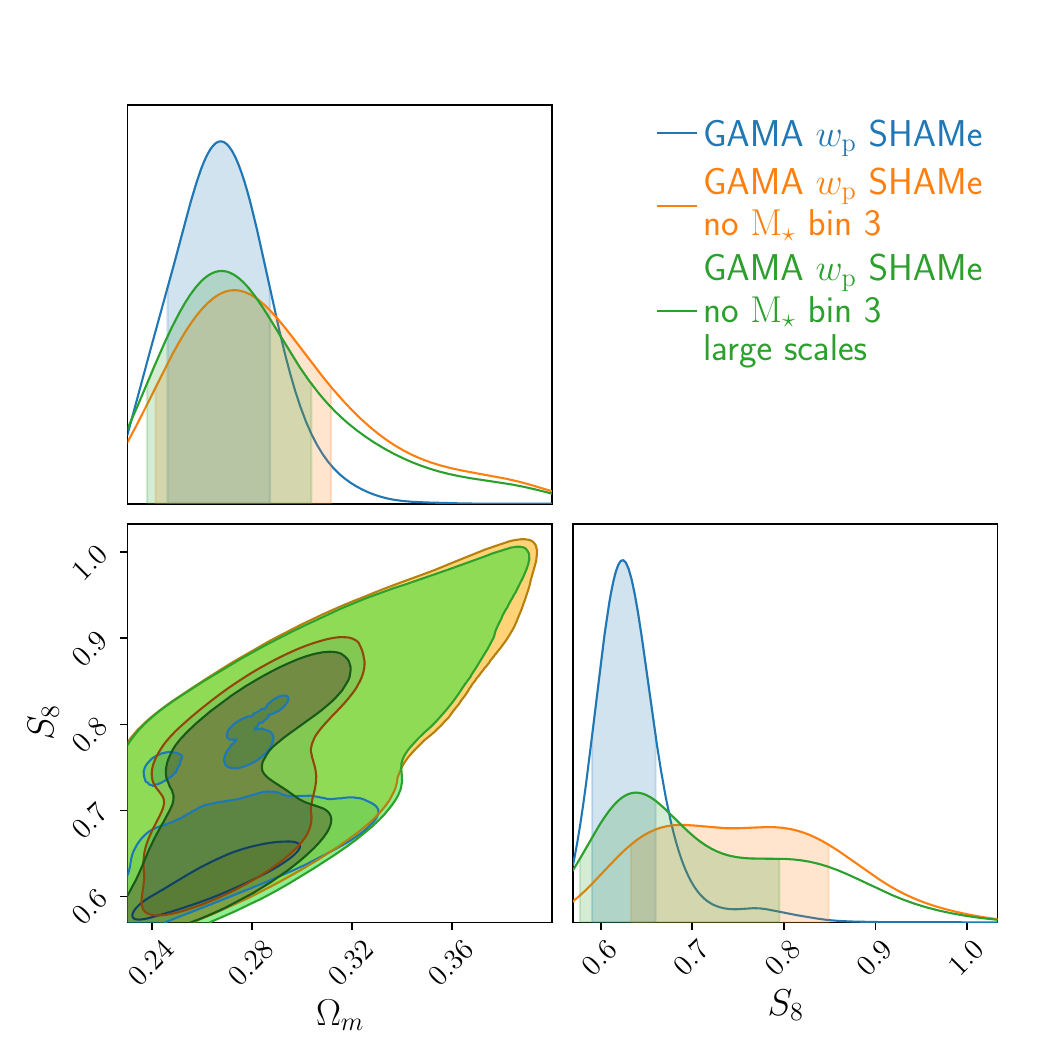}
    \caption{Marginalised posterior constraints on $S_8$ and $\Omega_\mathrm{m}$ for GAMA galaxy clustering alone using SHAMe with stellar mass bin 3 (blue), without stellar mass bin 3 (orange), and without the large scales of stellar mass bin 3 (green). Stellar mass bin 3 includes $10.5 < \mathrm{{log}}(M_{{\star}}/h^{{-2}}M_{{\odot}}) \leq 10.7$, and we define large scales as $r_{\mathrm{max}}=1.5 \ h^{-1}\mathrm{Mpc}$. Including bin 3 we find a strong constraint on $S_8$, which is inconsistent with our fiducial combined analysis. When we remove bin 3, or even just the large scales of bin 3, this internal inconsistency disappears.}
    \label{fig:wp_with_and_without_bin5}
\end{figure}

\section{Analytic prediction of the correlation function of galaxies selected in a stellar mass bin}\label{sec:derivation}

We derived equation 7,

\begin{equation}
\begin{split}
    w_{\mathrm{p}}({M_\star}_1 < M_\star < {M_\star}_2) \simeq 
&w_{\mathrm{p}1} \left( \frac{n_1}{n_1 - n_2} \right)^2 + w_{\mathrm{p}2} \left( \frac{n_2}{n_1 - n_2} \right)^2 \\
&- 2 \sqrt{w_{\mathrm{p}1} w_{\mathrm{p}2} } \left( \frac{n_1 n_2}{(n_1 - n_2)^2} \right),
\end{split}
\end{equation}

 \noindent by assuming that the galaxies of the sample are selected using a number density cut of $n_1$, but not $n_2$. The amplitude of the clustering signal scales as $n_2$, and the number of objects in the new sample is $(n_1 - n_2)$. Assuming the correlation function goes as the $\delta$, we can define the amplitude of the bin (from the galaxies that are in sample 1 but not sample 2) as:
\begin{equation}
\delta_{\rm bin}(\mathbf{x}) = \frac{n_1\,\delta_1(\mathbf{x}) - n_2\,\delta_2(\mathbf{x})}{n_1 - n_2}.
\end{equation}

And since the correlation function between a sample `a' and a sample `b' goes as $\langle \delta_a \delta_b \rangle$:
\begin{equation}
\begin{split}
w_{p,\rm bin}(r_p) = 
\left(\frac{n_1}{n_1 - n_2}\right)^2 w_{p,11}(r_p)
+ \left(\frac{n_2}{n_1 - n_2}\right)^2 w_{p,22}(r_p) \\
- 2\,\frac{n_1 n_2}{(n_1 - n_2)^2} \, w_{p,12}(r_p).
\end{split}
\end{equation}

From these terms, the only one we do not have from our emulator is the last term, but we can approximate it to:
\begin{equation}
w_{p,12}(r_p) \approx \alpha\,\sqrt{w_{p,11}(r_p)\,w_{p,22}(r_p)}, 
\qquad \alpha \simeq 1.
\end{equation}

With $\alpha$ affected by the (non)correlation of the samples and the differences in bias. For the bins in mass we used in this work, we have a difference of $\sim 1\%$. We perform a similar analysis for equation 8.

\section{Cosmological constraints excluding stellar mass bin 4}\label{sec:NoBin4}

\begin{figure}
    \centering
    \includegraphics[width=1.0\linewidth]{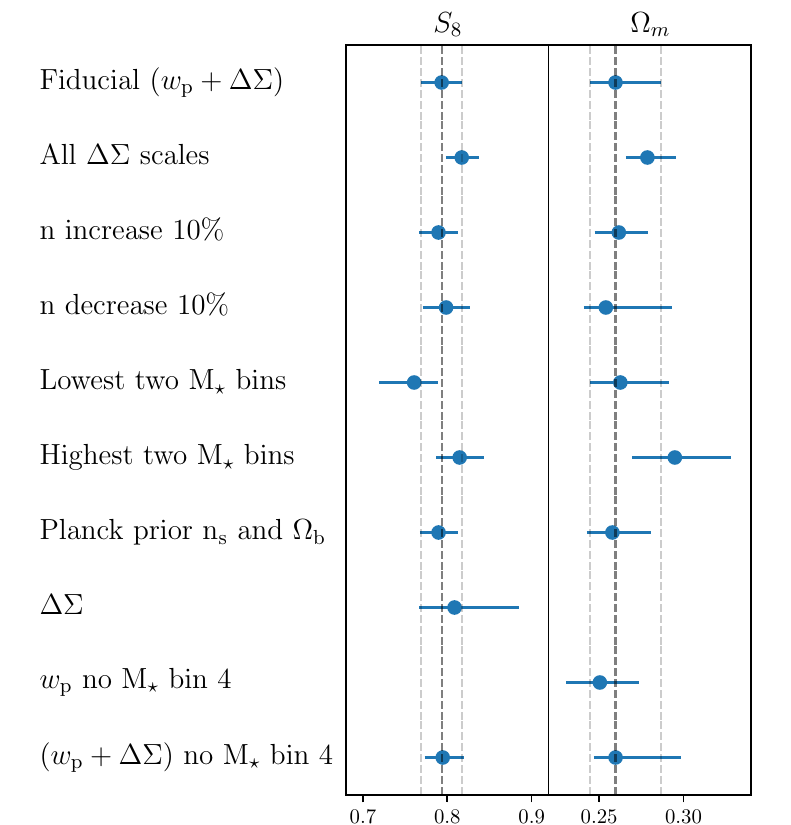}
    \caption{{\bf Similar to Fig. \ref{fig:S8_summary_analysis_variations}, but excluding the stellar mass bin 4 instead of the stellar mass bin 3.}}
    \label{fig:NoBin4}
\end{figure}

As mentioned in Section \ref{sec:fiducial analysis results}, the galaxy selection is performed using cuts in number density rather than stellar mass. This approach mitigates the impact of discrepancies in the stellar mass function, which become substantial at the high–mass end, on the clustering measurements. In Fig. \ref{fig:NoBin4} we examine the cosmological constraints obtained after removing this bin and find that they remain in good agreement with those derived from the main sample. Consequently, we conclude that our implementation adequately accounts for the differences in the stellar mass function for this subsample.


\bsp	
\label{lastpage}
\end{document}